\documentclass[final]{siamltex}

\usepackage{graphicx}
\usepackage{float}
\usepackage{tikz}
\usetikzlibrary{matrix}
\usetikzlibrary{calc}
\usetikzlibrary{patterns,fadings}
\usepackage{colortbl}
\usepackage{xkeyval}   

\usepackage{graphicx,color}
\usepackage{amsmath,amsfonts,amssymb}
\usepackage{array,multirow}
\usepackage{algorithm}
\usepackage{algorithmicx}
\usepackage{algpseudocode}
\usepackage{cite,url}
\usepackage{xspace}
\usepackage{xstring}

\usepackage{fp}
\usepackage{booktabs}
\usepackage{siunitx}

\newcommand{\comment}[1]{\textcolor{red}{[{\bf #1}]}\xspace}

\newcommand{\clawpack}{{\sc Clawpack}\xspace}

\newcommand{\forestclaw}{{\sc ForestClaw}\xspace}
\newcommand{\paramesh}{{\sc Paramesh}\xspace}
\newcommand{\samrai}{{\sc Samrai}\xspace}
\newcommand{\boxlib}{{\sc Boxlib}\xspace}
\newcommand{\amrclaw}{{\sc AMRClaw}\xspace}

\newcommand{\amroc}{{\sc Amroc}\xspace}

\newcommand{\chombo}{{\sc Chombo}\xspace}
\newcommand{\ebchombo}{{\sc EBChombo}\xspace}
\newcommand{\pforest}{\texttt{p4est}\xspace}

\newcommand{\mpi}{{\sc MPI}\xspace}
\newcommand{\pyclaw}{{\sc PyClaw}\xspace}

\newcommand{\flash}{{\sc FLASH}\xspace}
\newcommand{\uintah}{{\sc Uintah}\xspace}

\newcommand{\eqn}[1]{(\ref{eqn:#1})}

\newcommand{\Sect}[1]{Section~\ref{sec:#1}}
\newcommand{\Fig}[1]{Figure~\ref{fig:#1}}
\newcommand{\Tab}[1]{Table~\ref{tab:#1}}
\newcommand{\Alg}[1]{Algorithm~\ref{alg:#1}}

\newcommand{\lmax}{\ell_\mathrm{max}}
\newcommand{\lmin}{\ell_\mathrm{min}}

\newcommand{\mrate}{multirate\xspace}

\newcommand{\mblock}{multiblock\xspace}
\newcommand{\Mblock}{Multiblock\xspace}

\newcommand{\mdata}{metadata\xspace}

\newcommand{\ignore}[1]{}

\newtheorem{thm}{Theorem}[section]

\newtheorem{prp}[thm]{Proposition}

\newcommand{\plotbox}[1]{#1}

\newcommand{\dx}{\ensuremath{\Delta x}\xspace}

\newcommand{\dt}{\ensuremath{\Delta t}\xspace}

\newcommand{\cO}{\mathcal O}

\newcommand{\adaptrun}[1]{\ensuremath{#1\times#1}}

\newcommand{\gpp}{grids-per-process\xspace}
\newcommand{\wallclock}{wall-clock\xspace}
\newcommand{\Wallclock}{Wall-clock\xspace}

\newcommand{\dash}{\textemdash}

\makeatletter
\define@key{results}{w}{\gdef\wall{#1}}
\define@key{results}{a}{\gdef\adv{#1}}
\define@key{results}{gf}{\gdef\gfill{#1}}
\define@key{results}{gc}{\gdef\gcomm{#1}}
\define@key{results}{rg}{\gdef\rg{#1}}
\define@key{results}{asteps}{\gdef\advsteps{#1}}
\define@key{results}{gpp}{\gdef\g{#1}}
\define@key{results}{p}{\gdef\p{#1}}
\define@key{results}{pp}{\gdef\pp{#1}}   
\define@key{results}{lmax}{\gdef\lmax{#1}}
\define@key{results}{mx}{\gdef\mx{#1}}
\define@key{results}{smooth}{\gdef\smooth{#1}}

\define@key{temp}{x}{\gdef\x{#1}}   
\define@key{temp}{star}{\gdef\star{#1}}

\NewDocumentCommand{\rowsbfc}{O{x=-1}m}{%
  \setkeys{temp}{#1}\setkeys{results}{#2} &
  \mx    & {(\smooth)}   &
  \wall  &
  \adv   & {\FPeval{\x}{100*( \adv)/(\wall)}(\num[round-precision=1,round-mode=places]{\x})} &
  \gfill & {\FPeval{\x}{100*(\gfill)/(\wall)}(\num[round-precision=1,round-mode=places]{\x})} &
  \gcomm & {\FPeval{\x}{100*(\gcomm)/(\wall)}(\num[round-precision=1,round-mode=places]{\x})} &
  \rg    & {\FPeval{\x}{100*(\rg)/(\wall)}(\num[round-precision=1,round-mode=places]{\x})} &
  {\FPeval{\x}{(\mx)*(\mx)*(\advsteps)/(\wall)}\x}
}

{\bottomrule
\end{tabular}%
\end{center}%
}

{\bottomrule
\end{tabular}%
\end{center}%
}

\NewDocumentCommand{\rowadv}{O{x=-1,star={}}m}{%
  \setkeys{temp}{#1}\setkeys{results}{#2} &
  \p \star  &
  \wall  &
  \adv   & {\FPeval{\x}{100*( \adv)/(\wall)}(\num[round-precision=1,round-mode=places]{\x})} &
  \gfill & {\FPeval{\x}{100*(\gfill)/(\wall)}(\num[round-precision=1,round-mode=places]{\x})} &
  \gcomm & {\FPeval{\x}{100*(\gcomm)/(\wall)}(\num[round-precision=1,round-mode=places]{\x})} &
  {\FPeval{\x}{(\mx)*(\mx)*(\advsteps)/(\wall)}\num[scientific-notation=fixed,
    fixed-exponent=5,
    table-figures-exponent=1,
    table-figures-decimal=1,
    table-figures-integer=4,
    table-column-width=2.2cm]{\x}}  
}

\newenvironment{advtable}[1]{%
\sisetup{
  round-mode=places,
  round-precision=1,
  table-text-alignment=center,
  table-number-alignment=center,
  table-space-text-post=*,
  input-symbols=()
}
\setlength{\tabcolsep}{-1pt}
\begin{center}
\begin{tabular}{r@{}
  S[table-format=4,table-column-width=1.4cm]      
  S[table-column-width=1.6cm,                     
    table-number-alignment=right]
  *3{S[table-column-width=1.4cm,                  
        table-number-alignment=right]
      S[table-column-width=1.2cm,
        table-text-alignment=left]}
  S[scientific-notation=fixed,                    
    fixed-exponent=5,
    table-figures-exponent=1,
    table-figures-decimal=1,
    table-figures-integer=4,
    table-column-width=2.2cm]
}  
\toprule
\multicolumn{10}{c}{ForestClaw ({#1})} \\
\phantom{} &
{Ranks} &
{Wall} &
\multicolumn{2}{c}{Adv.\ (\%)}   &
\multicolumn{2}{c}{Fill (\%)}   &
\multicolumn{2}{c}{Comm.\ (\%)}  &
\multicolumn{1}{c}{Rate}  \\
\midrule
}
{\bottomrule
\end{tabular}%
\end{center}%
}

\NewDocumentCommand{\rowsphere}{O{x=-1,star={}}m}{%
  \setkeys{temp}{#1}\setkeys{results}{#2} &
  \p   &
  {\FPeval{\x}{round(\g:0)}\x} &
  \lmax  &
  \wall  &
  \adv   & {\FPeval{\x}{100*( \adv)/(\wall)}(\num[round-precision=1,round-mode=places]{\x})} &
  \gfill & {\FPeval{\x}{100*(\gfill)/(\wall)}(\num[round-precision=1,round-mode=places]{\x})} &
  \gcomm & {\FPeval{\x}{100*(\gcomm)/(\wall)}(\num[round-precision=1,round-mode=places]{\x})} &
 {\FPeval{\x}{(\mx)*(\mx)*(\advsteps)/(\wall)}%
 \num[%
    scientific-notation=fixed,                    
    fixed-exponent=4,
    table-figures-exponent=1,
    table-figures-decimal=1,
    table-figures-integer=4,
    table-column-width=2.1cm]%
  {\x}}  
}

\newenvironment{spheretable}[1]{%
\sisetup{
  round-mode=places,
  round-precision=1,
  table-text-alignment=center,
  table-number-alignment=center,
  table-space-text-post=*,
  input-symbols=()
}
\setlength{\tabcolsep}{-1pt}
\begin{center}
\begin{tabular}{r@{}
  S[table-format=3,table-column-width=1cm]      
  S[table-format=3,                               
    table-column-width=1cm,
    table-number-alignment=center]
  S[table-format=2,                               
    table-column-width=0.8cm,
    table-number-alignment=center]
  S[table-column-width=1.7cm,                     
    table-number-alignment=right]
  *3{S[table-column-width=1.2cm,                  
        table-number-alignment=right]
      S[table-column-width=1.2cm,
        table-text-alignment=left]}
  S[scientific-notation=fixed,                    
    fixed-exponent=4,
    table-figures-exponent=1,
    table-figures-decimal=1,
    table-figures-integer=4,
    table-column-width=2.1cm]
}  
\toprule
\multicolumn{12}{c}{ForestClaw ({#1})} \\
\phantom{} &
{Ranks} &
{G} &
{$\lmax$} &
{Wall} &
\multicolumn{2}{c}{Adv.\ (\%)}   &
\multicolumn{2}{c}{Fill (\%)}   &
\multicolumn{2}{c}{Comm.\ (\%)}  &
\multicolumn{1}{c}{Rate}  \\
\midrule
}
{\bottomrule
\end{tabular}%
\end{center}%
}

\NewDocumentCommand{\rowssphere}{O{x=-1}m}{%
  \setkeys{temp}{#1}\setkeys{results}{#2} &
  \p     &
  \gpp   &
  \wall  &
  \adv   & {\FPeval{\x}{100*( \adv)/(\wall)}(\num[round-precision=1,round-mode=places]{\x})} &
  \gfill & {\FPeval{\x}{100*(\gfill)/(\wall)}(\num[round-precision=1,round-mode=places]{\x})} &
  \gcomm & {\FPeval{\x}{100*(\gcomm)/(\wall)}(\num[round-precision=1,round-mode=places]{\x})} &
  \rg    & {\FPeval{\x}{100*(\rg)/(\wall)}(\num[round-precision=1,round-mode=places]{\x})} &
  {\FPeval{\x}{(\mx)*(\mx)*(\advsteps)/(\wall)}\x}  
}

\NewDocumentCommand{\row}{O{x=-1,star={}}m}{%
  \setkeys{temp}{#1}\setkeys{results}{#2} &
  \p \star  &
  \lmax &
  {\FPeval{\x}{round(\g:0)}\x} &
  \wall  &
  \adv   & {\FPeval{\x}{100*( \adv)/(\wall)}(\num[round-precision=1,round-mode=places]{\x})} &
  \gfill & {\FPeval{\x}{100*(\gfill)/(\wall)}(\num[round-precision=1,round-mode=places]{\x})} &
  \gcomm & {\FPeval{\x}{100*(\gcomm)/(\wall)}(\num[round-precision=1,round-mode=places]{\x})} &
  {\FPeval{\x}{(\mx)*(\mx)*(\advsteps)/(\wall)}\x}  
}

\title%
{ForestClaw: A parallel algorithm for patch-based adaptive mesh refinement on a
forest of quadtrees}

\author{%
Donna A.\ Calhoun\thanks{Boise State University, Boise ID, USA (corresponding
author: \texttt{donnacalhoun@boisestate.edu})}%
\and
Carsten Burstedde\thanks{Institut f\"ur Numerische Simulation (INS)
and Hausdorff Center for Mathematics (HCM), Universit\"at Bonn, Germany}%
}

\begin{document}

\maketitle
\begin{abstract}
We describe a parallel, adaptive, \mblock algorithm for
explicit integration of time dependent partial differential equations on
two-dimensional Cartesian grids.  The grid layout we consider consists of a nested
hierarchy of fixed size, non-overlapping, logically Cartesian grids
stored as leaves in a quadtree.  Dynamic grid refinement and parallel
partitioning of the grids is done through the use of the highly
scalable quadtree/octree library \pforest.  Because our concept is
\mblock, we are able to easily solve on a variety of geometries
including the cubed sphere.  In this paper, we pay special attention to
providing details of the parallel ghost-filling algorithm needed to ensure
that both corner and edge ghost regions around each grid hold valid values.

We have implemented this algorithm in the \forestclaw code using single-grid
solvers from \clawpack, a
software package for solving hyperbolic PDEs using finite volumes
methods.  We show weak and strong scalability results for
scalar advection problems on two-dimensional manifold domains on 1 to 64Ki
MPI processes, demonstrating neglible regridding overhead.
\end{abstract}



\pagestyle{myheadings}
\thispagestyle{plain}
\markboth{D.\ Calhoun and C.\ Burstedde}%
         {Parallel, patch-based AMR on quadtrees}

\section{Introduction}
The use of spatial adaptivity is widely recognized as an effective way
to improve the performance of Cartesian grid methods for partial
differential equations (PDEs), and in fact was cited in a recent survey
as the main reason for users to adopt a particular code
\cite{du-an-ga-re-ri-sh-si-we:2009}.  With the ubiquity of multi-core
machines at every level of computing performance, parallel
capabilities are expected for codes running on anything from desktop
machines to petascale supercomputer architectures.  However, building
parallel, adaptive capabilities into solvers is a daunting task, and
often one which is completely orthogonal to the task of improving the
speed and accuracy of single grid solvers for PDEs.  While this
situation presents difficult challenges with respect to numerical
accuracy and parallel performance, we see it as providing a highly
motivating opportunity to investigate a modular strategy to adaptive
solver development that maximizes the reuse of proven algorithms.

We describe a hybrid approach to adaptive mesh refinement (AMR) that
uses the highly scalable quadtree/octree library \pforest
\cite{BursteddeWilcoxGhattas11, IsaacBursteddeGhattas12,
IsaacBursteddeWilcoxEtAl15} to manage a dynamic, multi-resolution
hierarchy of small grids that are distributed in parallel.  This
hierarchy occupies non-overlapping regions of the computational domain
defined by recursively subdividing the domain into quadrants (or
octants in 3D).  Each region is assigned to an owner process,
and the technical issues of parallel mesh management are encapsulated
inside the meshing library.  Our atomic unit of computation is thus a
small uniform grid.  Each such grid is processed by a single grid Cartesian
solver (e.g.\ \clawpack, \cite{be-le:1998}) that mostly
encapsulates the solver details.  Our software implementation,
\forestclaw, provides the central orchestration and coordinates calls
between the mesh management library and solver libraries, including those
AMR tasks related to parallel neighbor communication,
dynamic remeshing (including re-mapping of the solution to newly
created meshes) and time stepping.

There are several existing software frameworks for general patch-based
parallel AMR, including \boxlib, \amrclaw, \samrai, \paramesh, \amroc,
\uintah, \chombo and \ebchombo
\cite{amrclaw,amroc,boxlib,chombo,paramesh,samrai,ebchombo}.  These
codes are typically based on finite volume methods for conservation
laws on logically Cartesian meshes, and so complex geometries are
handled using either mapped grids, or cut-cell approaches, although
only \ebchombo has extensive support for cut cells.  Discussions of
survey and experiences in using several of the high level frameworks
for block-structured AMR described above can be found in
\cite{du-al-be-be-br-br-co-gr-li:2014, du-st:2014}.

With the exception of \paramesh, these software platforms use the
original Berger-Oliger and Berger-Collela block-structured mesh
approach to dynamic mesh refinement.  Such mesh hierarchies consist of
nested, overlapping grids of increasingly finer
resolution.  \paramesh, which is the mesh management library supporting the
\flash multiphysics code, in contrast, consists of a hierarchy of
non-overlapping grids, organized as leaves in a quad- or octree.  The
approach taken by \paramesh is thus most closely related to the
approach we describe here.

What sets the \forestclaw code and project apart from related adaptive
mesh methods, notably the current standard approach described by Berger,
Oliger, Colella, LeVeque and others for solving
conservation laws on Cartesian, finite volume grids \cite{be-ol:1984,
be-co:1989}, is the following.
\begin{itemize}
\item Highly scalable quadtree regridding using a the \pforest parallel mesh
backend that has been demonstrated succesfully at the petascale.
\item A flexible mapped, \mblock infrastructure for solving on a variety
of domains not easily represented by a single Cartesian block.  A key
infrastructure element includes transformations for handling orientation
mis-matches at block boundaries.
\item A detailed description of both a serial and parallel algorithm for filling
ghost regions of grids stored in the leaves of  a quadtree, and:
\item Many customizable options, including a package
handling infrastructure for binding to multiple solvers.
\end{itemize}
Additional features that are based on earlier research on Cartesian
grid methods include built-in support for computing metric terms
needed for solving PDEs on mapped grids that are consistent with
second order finite volume schemes.

Even when not using the adaptive refinement features, one can expect
to benefit from better cache performance for a uniformly refined
domain, more flexible geometric features that the \mblock architecture
provides, and a highly flexible and performant parallel partitioning scheme
based on space-filling curves.

The main focus of this article will be on the algorithmic details,
implemented in the \forestclaw code, associated with orchestrating
parallel communication between grids for second order finite volume
schemes for time dependent hyperbolic PDEs in a two-dimensional
quadtree layout. We highlight our main design goals which were chosen
to facilitate re-use of code involved in grid communication, and the
ease of use for developing new numerical methods.  We provide
performance results on 1 to 64Ki processes, for the solution of a
scalar advection equation, integrated explicitly using a single
globally stable time step.  Detailed numerical accuracy results,
refinement criteria and solutions to more general conservation laws
are discussed in this article, but will be presented in a second paper.

\ignore{%
\comment{Shall we remove this summary paragraph?}
We start by describing our quadtree layout, including how ghost cells
are handled, in \Sect{quadtreelayout}.  We discuss more involved
details concerning parallel algorithms in \Sect{parallel}.  Then we
discuss the time stepping strategy on a the adaptive hierarchy in
\Sect{advance}.  Finally, we show several numerical examples
illustrating the geometric flexibility and parallel performance of the
method.
}

\section{A quadtree layout%
}
\label{sec:quadtreelayout}

\begin{figure}
\begin{center}
\includegraphics[width=0.4\textwidth]{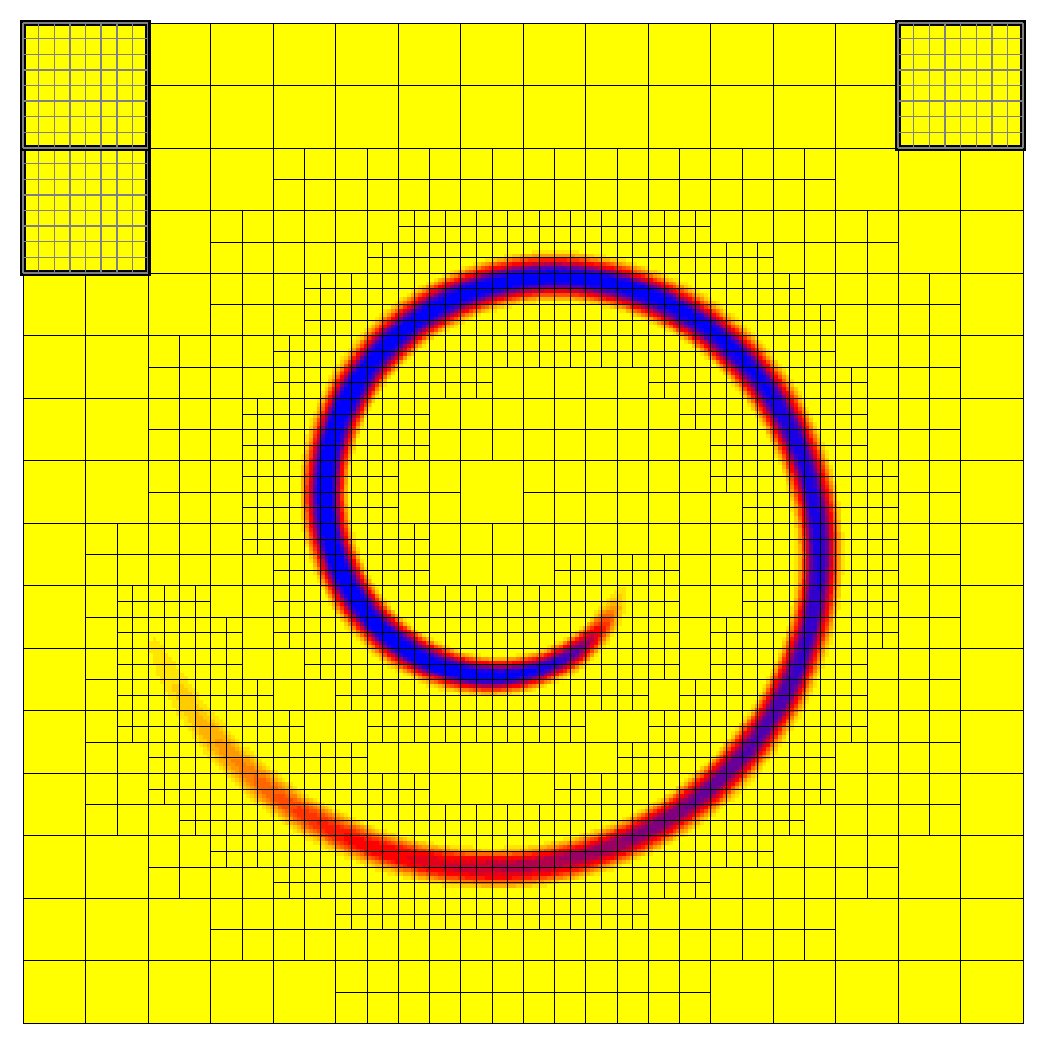}\hfil
\includegraphics[width=0.45\textwidth,clip=true,trim=0cm -1cm  0cm 0cm]{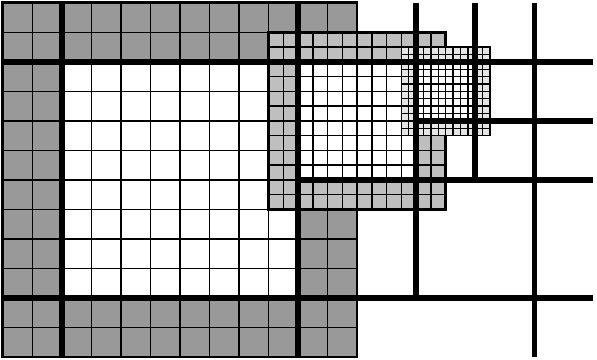}
\end{center}
\caption{The left figure shows quadrants in levels 3 through 6 of a
adaptive quadtree on a single block.  For clarity, we only show grid
lines inside the $8 \times 8$ grids occupying level 3 quadrants
(with darker solid borders).  The right figure shows three $8 \times
8$ computational grids, each with a layer of ghost cells, at three
adjacent levels.  Thick lines indicate quadrant (grid) boundaries.}
\label{fig:grids}
\end{figure}

A \forestclaw domain consists of a static arrangement of one or more
blocks, each of which can be recursively and dynamically
subdivided into quadrants.  When refinement is requested, a
level 0 quadrant, which occupies the same computational space as its
parent block, is partitioned into four equally sized level 1
quadrants. One or more level 1 quadrants can then be partitioned into
four level 2 quadrants each, and so on.  Coarsening proceeds in
reverse.  The collection of all quadrants in a block forms a partition
of a square computational subdomain, with the length of an edge of a
level $\ell$ quadrant being $2^{-\ell}$ times the length of the level
0 edge.  An optional feature of meshes generated using the \pforest
library, and specifically called upon by \forestclaw is that meshes
can be made {\em 2:1 balanced} \cite{SundarSampathBiros08,
IsaacBursteddeGhattas12}.
Any two quadrants that share a face or corner will never be more than
one level apart.  We refer to this arrangement of blocks, each of
which is partitioned into one or more quadrants, as a {\em \mblock
quadtree} layout, or simply a {\em quadtree} layout.
Throughout this paper, we will often use the term ``patch'' to mean a quadrant
together with associated numerical data.
See \Fig{grids} for a typical single block layout generated by \forestclaw.

Each quadrant that makes up the final partitioning of a \forestclaw
quadtree layout is occupied by a fixed-size, logically Cartesian grid.
Each grid has an interior region that fits the quadrant area, made up
of typically $8^2$, $16^2$ or $32^2$
grid cells, and one or more layers of ghost cells that extend
outside in the two coordinate directions.  Solution data on a
computational grid is stored in a contiguous array that includes both
interior and ghost regions, so that a grid with $8^2$
interior cells and two layers of ghost cells stores solution data in a
contiguous array of $12^2$ mesh cells.
The solution at each grid cell stores one or more field variables that make up
the numerical solution, as well as any metric dependent data.
The interior regions of computational grids do not overlap each other,
but the ghost region of one grid will overlap with the interior region
of multiple face-adjacent and corner-adjacent neighbors.  In \forestclaw,
values for the interior grid dimensions and number of ghost cell layers are
the same for all grids, effectively enforcing a constant 2:1 refinement
ratio between grid levels.  The {\em resolution} of a particular grid
is determined by the size of the quadrant it occupies, so a grid
occupying a level $\ell$ quadrant has $2^\ell$ times the resolution of
the same grid in a level 0 quadrant.

When describing numerical schemes, it will
be convenient to refer to the border of the interior region (i.e., the
quadrant) as the grid {\em boundary},
even though this boundary does not enclose the ghost regions belonging to the grid.
When the context is clear, the ``size'' of
a grid should be loosely understood to mean the size of the quadrant
occupied by that grid, although there will also be occasion to
describe a grid using its (fixed) interior dimensions, e.g.\ a $32
\times 32$ grid.  It is also informally understood that the use of the
term ``grid'' often refers to the contiguous array of solution values
associated with the grid, and not just the geometric \mdata needed to
describe the grid.  In this context, a ``coarse grid solution'' or a
``fine grid solution'' is the solution on a coarser or finer grid.  In
the current version of \forestclaw, we store grids (and solution
values) only for those quadrants that make up the final partitioning
of the domain.  If, during refinement, a coarse quadrant is subdivided
into four finer quadrants, the storage for the coarse grid solution
and any coarse grid \mdata is deleted and storage for a finer grid is
allocated in each of the four finer quadrants.  See \Fig{grids} for an
illustration of grids and quadrants.

There are several advantages to tree-based refinement.  One, the
numerical analyst developing methods for an adaptive mesh should find
it relatively simple to work with the quadtree layout, since quadrant
connection patterns appear in only one of three regular arrangements:
a neighboring grid is either the same size, twice the size, or one of
two half sized grids.  Also, it can guaranteed that higher order stencils
will have sufficient data from directly adjacent grids and will never need to
use data from more than two levels of refinement.
Finally, all communication between grids
needed for advancing the solution takes place at grid boundaries,
reducing the reliance on \mdata.  From a performance standpoint,
tree structures have been extensively studied, and so
their performance characteristics in a wide range of scenarios is well
understood.  The information on neighboring quadrants can be cached and exposed
by the meshing library in such a way that they may be located
within a tree traversal by $\cO(1)$-time lookup functions.  Finally,
the grids in a quadtree layout can be enumerated to preserve data
locality using either Morton ordering (as we do in \forestclaw) or
other types of well known space-filling curves \cite{Samet06,
Bader12}.

One potential disadvantage of the quadtree layout is that a single
quadtree may not be appropriate for a general rectangular domain with
a large aspect ratios.  In
\forestclaw, this difficulty can be overcome in at least two ways.
First, one could simply choose fixed size grids with different numbers
of grid cells in each of the two coordinate directions.  For example,
a quadtree in which each leaf contains a $64 \times 16$ grid would
effectively allow one to grid a $4 \times 1$ domain, while maintaining
square grid cells.  This is done for example in the Racoon code
\cite{dr-gr:2005}.  The downside to this approach is that while
individual grid cells have aspect ratios close to 1, the quadrants do
not, making it more difficult to efficiently refine around some
regions. A second approach, and the approach favored in \forestclaw, is
to allow domains to consist of more than one quadtree, or a {\em forest} of
trees.  A $4 \times 1$ domain is naturally divided into four square
blocks, each of which contains a quadtree with square mesh cells.
\forestclaw allows for general arrangements of blocks, with the only
restriction being that face-adjacent blocks must share a complete
face.  One surprisingly useful domain is the ``brick'' domain, an $M
\times N$ arrangement of square blocks in an optimized order.
In \forestclaw, the brick
domain is used for meshing general rectangular domains, the annulus, a
spherical coordinate (e.g.\ latituted/longitude) grid of the mid-latitude region
of the earth, and the torus, all with relatively uniform, square mesh
cells.  Another useful \mblock layout is the cubed sphere grid, widely
used in numerical simulations of weather, climate, geodynamics, etc.%
\ignore{Any
number of blocks can meet at a corner, allowing for a forest whose
blocks are the quads in a general unstructured quadrilateral mesh.
Developing solvers for such mesh layouts is challenging, due to the
presence of metric seams at block boundaries, and so is the topic of
on-going research.}

A quadtree/octree data structure is commonly used in various kinds of numerical
applications, including the fast multipole method, and computer graphics,
making the quadtree a natural choice for interfacing with other libraries.
For this reason, it is well suited as a foundation for more general
libraries doing mesh refinement \cite{BangerthBursteddeHeisterEtAl11,
SampathAdavaniSundarEtAl08}.
A quadrant/octant based approach
has been used in other parallel adaptive frameworks,
including \paramesh, {\sc Nirvana}, {\sc Racoon II},
Peano \cite{WeinzierlMehl11}, and the
Building-Cubes Method \cite{paramesh, dr-gr:2005, zi:2012,
ko-so-eg-ta-ko-ta-sa-ha:2011}.
None of these other codes, however, have general \mblock capabilities,
or documented performance results for adaptive simulations at petascale.
\pforest, on the other
hand, has well-established performance results
\cite{BursteddeGhattasGurnisEtAl08, BursteddeGhattasGurnisEtAl10,
Bui-ThanhBursteddeGhattasEtAl12, RudiMalossiIsaacEtAl15}.

%
The \pforest algorithms unify the design principles underlying the use
of space filling curves for the ordering of elements \cite{Hilbert91,
Lebesgue04, Morton66, GriebelZumbusch99, TuOHallaronGhattas05}, the refinement
one or more tree roots into an adaptive forest \cite{StewartEdwards04,
BangerthHartmannKanschat07}, and the use of linear (i.e.\ leaf-only) octree
storage \cite{SundarSampathBiros08}.  The terms ``leaf''
or``quadrant'' (used interchangeably) refer to an abstract
placeholder for any kind of application data, identified by
discrete tree coordinates and their refinement level.
The quadrants in \pforest can be searched and indexed in a random access
pattern, which we exploit to assemble $\cO(1)$ lookup information on neighbors.
We do not make use of compressed encodings of the leaves that would save
additional memory at the price of
enforcing ordered-only tree traversals
\cite{BungartzMehlWeinzierl06, WeinzierlMehl11}.  The main \pforest
algorithms, including 2:1 balancing and partitioning among the MPI processes,
are similar in spirit to collective MPI commands.  Applications such
as \forestclaw can access the quadrant storage and access all required
neighborhood information without calling MPI directly.

\subsection{Ghost regions}
\label{sec:ghostcells}
As described above, the interior region of each grid (stored as a leaf of a
quadtree) is surrounded by
a layer of one or more ghost cells occupying {\em ghost regions}.  All
communication between grids is facilitated by copying, averaging or
interpolating data from the interior regions of one grid to the ghost
regions of a neighboring grid.  Since we will want to use unsplit
schemes for hyperbolic problems, we also must fill corner ghost regions of
each grid with valid data.

We define grid neighbors as those grids occupying either
face-adjacent or corner-adjacent quadrants.  The coarse grid
neighbors of a level $\ell$ grid are neighboring grids occupying
level $\ell-1$ quadrants, while fine grid neighbors are those
neighboring grids occupying level $\ell+1$ quadrants.  We
use the term {\em coarse ghost regions} to refer
collectively to those ghost regions which are filled using data
copied from the interior of a same-size neighbor or averaged (as described
below) from the interior of a fine grid neighbor.  {\em Fine grid ghost
regions} are those ghost regions which are filled by interpolating
from data in the interior and ghost regions of a
neighboring coarse grid.  A coarse grid is a grid with coarse grid
ghost regions, and a fine grid is a grid with fine grid ghost
regions.  Depending on the context, a grid can be both a coarse grid
and a fine grid.  A ghost region is an interior region if it
is inside the physical domain, and an exterior ghost region otherwise.

The numerical operations used in filling ghost cells are copying between
neighboring grids of the same size, averaging (restricting) data from the
interior of the fine grids to ghost cells of a coarse grid, or interpolating
(prolongating) from a coarse grid to fine grid ghost
cells.  At the boundary of the computational domain, we impose
physical boundary conditions.  For our purposes here, we assume that
we have cell-centered data which represents either a cell average
value (in the finite volume sense) or cell-centered point-wise
values.  For the second order schemes we have implemented, these two
interpretations are interchangeable.

\begin{figure}
\begin{center}
\includegraphics[width=0.6\textwidth]{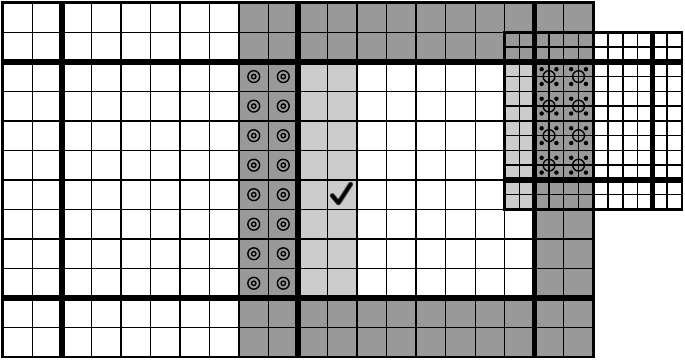}
\end{center}
\caption{Dark-shaded coarse ghost regions (of the center grid)
are filled by copying from a neighboring
same-size grid (left edge) or  averaging from a half-size grid (right edge).  The
open circles are the cell-centered ghost values on the center grid, and smaller black
circles are the values on the neighboring grid that are either copied (same-size neighbor)
 or averaged (finer neighbor).}
\label{fig:coarseghost}
\end{figure}

\begin{figure}
\begin{center}
\includegraphics[width=0.475\textwidth]{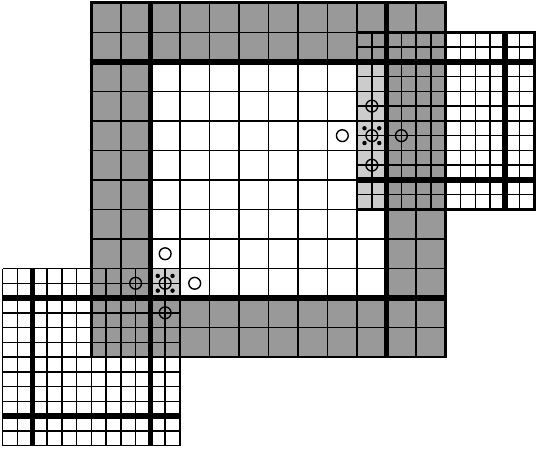}\hfil
\includegraphics[width=0.475\textwidth]{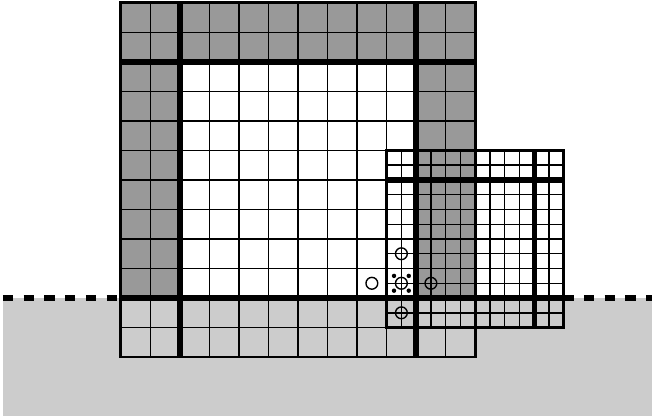}
\end{center}
\caption{Coarse grid interpolation stencils used to fill in fine ghost regions.
The open circles are the coarse grid values used in the stencil
and the filled smaller circles are the fine grid ghost cell values to be filled in. The
stencils used in \forestclaw are applied to a single coarse grid after the coarse grid
ghost regions have been filled.  The right figure shows fine grid ghost cells
at a physical
boundary, where  the coarse grid interpolation stencil requires valid data from the
exterior coarse grid ghost region.
This is fabricated according to the boundary condition.}
\label{fig:interpolation}
\end{figure}

Two key assumptions in our present algorithm for filling in ghost
cells is that copying and averaging only require data from the
interior of a neighboring grid cell, whereas interpolation will in
general require data from both interior and ghost regions.
In \Fig{coarseghost}, we show typical stencils for averaging and copying
from neighboring grids, and
in \Fig{interpolation}, we show a typical 5 point stencil used
to fill fine grid ghost regions.
Because they do not require any data from ghost regions,
the coarse grid regions can be filled first, before filling
fine grid ghost regions.  This ordering of how ghost regions are
filled adds some complexity to the serial and parallel ghost
filling algorithms, but this additional complexity is more than
justified by the greater ease of use obtained with the regular
interpolation stencils, especially when going to higher order.

Filling exterior ghost regions is done using standard methods of copying
or extending data in some way from the neighboring interior cells, depending on
the type of physical boundary condition.  The averaging and
interpolation stencils we use are shown in \Fig{interpolation}.

\subsection{Serial ghost filling algorithm}

Assuming that the values in all interior grid cells are valid, either after
setting initial conditions or running a time step, it remains to compute
correct values in all ghost regions to prepare the following time step.
Because our interpolation stencils rely on data in coarse ghost regions,
a ghost filling algorithm must ensure that all coarse ghost regions
are filled before we fill fine ghost regions.  A relatively
straightforward algorithm that accomplishes this is presented in
\Alg{ghostfill_serial}.  In this algorithm, the expression ``fill
coarse ghost regions'' means to copy or average data from a
neighboring same-size or fine grid neighbor into all edge and corner
coarse ghost regions.  Conversely, ``fill fine ghost regions'' refers to using
interpolation from coarse neighbors to fill all edge and corner fine ghost
regions.
To ensure that exterior corner ghost regions have valid data, physical
boundary conditions are applied to both edge and corner exterior ghost
regions.

The following proposition provides conditions that guarantee that
\Alg{ghostfill_serial} fills all edge and corner coarse and fine ghost
regions in a quadtree layout  with valid data.

\begin{prp}
\label{prop:serial}
Suppose we have a 2:1 balanced quadtree layout, with
grids of fixed size $M\times M$ and $m$ layers of ghost cells each.
Let $w$ be the width
of the stencil used to interpolate from a coarse grid to a fine ghost region.
Assume that the interior regions of all grids contain valid data.
Then, if $m \le M/4$ and $w \le M/2$,
\Alg{ghostfill_serial} is guaranteed to fill in all coarse, fine and
exterior ghost regions with valid data.
\ignore{Moreover, all coarse and fine ghost regions can be filled by visiting
coarse grids only.}
\end{prp}

\begin{figure}
\begin{center}
\includegraphics[width=0.40\textwidth]{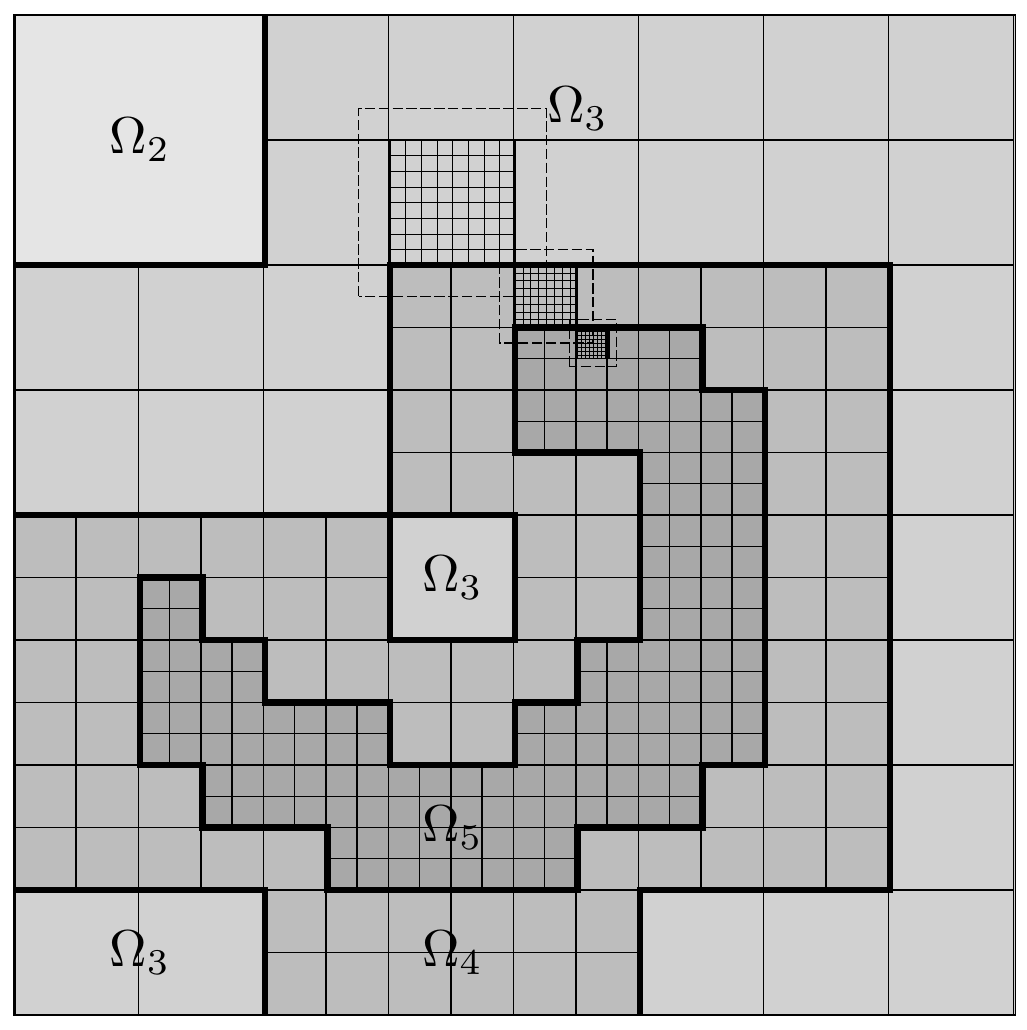}\hfil
\includegraphics[width=0.40\textwidth,clip=true,trim=1.25in 2.5in 1.5in 0.25in]{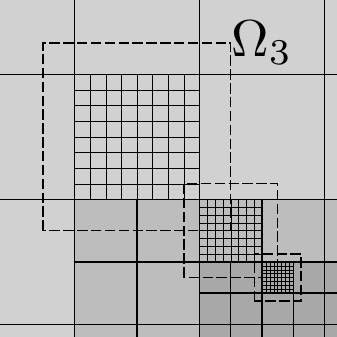}\hfil
\end{center}
\caption{Typical quadtree layout showing non-overlapping refinement regions
$\Omega_{\ell}$ (left),
separated by level curves $\Gamma_{\ell}$ (thick lines).
The dashed lines represent the bounds of the ghost regions of selected grids
with a zoom-in shown in the right hand picture.  By imposing
restrictions on the allowable grid size, the number of ghost cell layers,
and the width of the interpolation stencil, \Alg{ghostfill_serial} is
guaranteed to fill in all corner and edge ghost cell regions with correct values.}
\label{fig:omega_regions}
\end{figure}

To justify this proposition, we only need to demonstrate that an
interpolation stencil can never cross more than one {\em level curve}.
Let a {\em level region} $\Omega_{\ell}$ (\Fig{omega_regions}) be defined
as the polygonal region (possibly multiply-connected) containing the
interiors of all level $\ell$ grids.  We then define a {\em level curve} $\Gamma_{\ell+1}$
as the rectilinear curve (or set of curves) separating $\Omega_{\ell}$ from $\Omega_{\ell+1}$.
In a 2:1 balanced quadtree layout, each curve in $\Gamma_{\ell+1}$ is either a
simple closed curve (the boundary of a rectilinear polygon), or an
open simple rectilinear curve that intersects the physical boundary at
each end.  Furthermore, if $\ell \ne \ell'$, no curve in
$\Gamma_{\ell}$ will intersect a curve in $\Gamma_{\ell'}$ and two
curves in $\Gamma_{\ell}$ can intersect only at a corner point.
From this, and the other conditions laid out in Proposition~\ref{prop:serial},
we conclude that an interpolation stencil cannot cross two level
curves from two distinct levels.  The practical implication of this is
that interpolation stencils will never require data from more than two
adjacent levels, and so an algorithm which first fills all coarse
ghost regions (traversing the coarse grids in any order), and then
fills fine ghost regions (traversing grids in any order), is
guaranteed to fill all ghost regions, without breaking any
data dependency chains between averaging and interpolation.
\begin{algorithm}
\caption{Serial algorithm for updating cells in ghost regions on
all levels $\ell$, $\lmin \le \ell \le \lmax$.  The first application
of physical boundary conditions will in general leave data in exterior corner fine ghost
regions invalid, requiring a second application of physical boundary conditions
after interior fine ghost regions have been filled in an interpolation step.}
\label{alg:ghostfill_serial}
\begin{algorithmic}
\Require Solution on interior of all grids contains valid data for given time $t$.
\Procedure{update\_ghost}{}
\ForAll{levels $\ell$, $\lmin \le \ell \le \lmax$}
\Comment{Copy and average}
\State Fill all interior coarse ghost regions belonging to level $\ell$ grids.
\EndFor
\ForAll{levels $\ell$, $\lmin \le \ell \le \lmax-1$}
\Comment{Apply phys. boundary conditions}
\State Fill exterior coarse ghost regions belong to level $\ell$ grids.
\EndFor
\ForAll{levels $\ell$, $\lmin \le \ell \le \lmax-1$}
\Comment{Interpolate}
\State Fill fine ghost regions belonging to level $\ell+1$ grids.
\EndFor
\ForAll{levels $\ell$, $\lmin+1 \le \ell \le \lmax$}
\Comment{Apply phys. boundary conditions}
\State Fill exterior ghost regions of level $\ell$ grids.
\EndFor
\EndProcedure
\end{algorithmic}
\end{algorithm}


\subsection{\Mblock indexing}

\label{sec:mblock} In the low level routines that implement
\Alg{ghostfill_serial} (and later, the parallel version) in \forestclaw,
we explicitly handle several cases of grid arrangements between pairs of neighboring
grids.

\ignore{
Low level routines which actually carry out the ghost filling
operations (copying, averaging or interpolation) between two
neighboring grids must have information about which grid is the coarse
grid and if the two grids are at different levels, which grid is the
fine grid.  The routine must also know which faces or corners are
involved in the exchange, and possibly the position of a neighboring
fine grid relative to the coarser grid.  In a practical
implementation of these operations between two neighboring grids,
we explicitly handle each of several cases, since the Cartesian
layout of the data in memory makes it inconvenient to abstract the
coarse grid face or corner further, without remapping data in memory.
However, in the \mblock setting, we seek an approach which does not
lead to a prohibitive number of cases between grids with different index orientations.
One of our primary goals is to maximum code re-use by minimizing the
number of cases that must be explicitly handled.
}

For pairs of face or corner adjacent grids on the same block, each
low-level routine handling the ghost-filling designates a coarse grid
(``this'' grid) and a same-size or fine ``neighbor'' grid.  We then
explicitly handle 20 cases.  At each coarse grid face, we have three
cases, one for a same-size neighbor, and two for each fine grid neighbor.
At each corner, we have two cases, one for a same-size neighbor, and
one for a fine grid neighbor.
Considering four faces and four corners per patch, we obtain $4 \times (3 + 2)
= 20$.  There is no advantage in reducing the number of
cases below these 20, since doing so would require an expensive remapping of
coarse and fine grid data in memory and would lead to code which is
hard to read, maintain, or modify.  The case of a double-size
neighbor is not explicitly handled, since each routine is always
written from the perspective of the coarsest grid.


When two neighboring grids are on different blocks, additional
information about the relative orientations of the indices is needed.
Holding the coarse grid
fixed, the face-adjacent neighboring grid can be rotated through one
of four possible positions in the plane, or through two possible
positions out of the plane, so that the z-axes (the directions of
which are determined from a right hand rule) of the coarse grid and
neighboring grid point in opposite directions.  Taking into account the two
possible positions that a fine grid can have relative to a coarse grid
face, the total number of arrangements between a coarse grid and a
neighboring grid at one of four coarse grid faces is 96 (eight
positions for each of three types of grids at each of four faces).
Similarly, the number of possible configurations for an adjacent
corner grid is 32 for both same-size or fine grid neighbors (8 for
each same-size corner-adjacent neighbor at each corner, or 8 for each
fine grid corner adjacent neighbor).

To avoid this combinatorial explosion of possible grid configurations,
\forestclaw makes use of {\em index transforms} for all corner and
face exchanges at grid boundaries, regardless of whether the exchange is
between grids on
the same block or different blocks.  The use of these transforms
effectively reduces the complexity in handling \mblock orientations to
the 20 cases required by the neighboring grids on the same block.

The index transforms from one index space to the index space
of a neighboring same-size grid, possibly at a \mblock boundary,
has the general form
\begin{equation}
{\bf I}_n = A{\bf I}_c + {\bf F}
\label{eqn:transform}
\end{equation}
where $A$ is a $2\times 2$ matrix, ${\bf F}$ is a $2 \times 1$ vector,
and ${\bf I}_c$ and ${\bf I}_n$ are $2\times 1$ vectors of grid
indices $(i_c,j_c)$ and $(i_n,j_n)$.
The matrix $A$ encodes the orientation of indices on one patch
relative to a second patch, and the vector ${\bf F}$ encodes the position
of these patches relative to each other.  For patches on the same
block, the matrix $A$ is the identity matrix, but for patches on
different blocks, $A$ will be a diagonal or anti-diagonal matrix
whose non-zero entries are 1 or -1.  The vector $F$ depends in general
on the fixed grid size $M$.

In what follows, we use $q_c(I_c)$ to indicate a cell-centered value
on the coarse grid at index coordinates $I_c$, defined in coarse grid
coordinates.  By analogy, the neighboring grid values are indicated
using $q_n(I_n)$, where $I_n$ is obtained using \eqn{transform}.
The transform in \eqn{transform} is provide by \pforest as a low-level
routine, which \forestclaw then uses when copying, averaging
and interpolation at patch boundaries, with no distinction made between
patches on the same block or on different blocks.  With these transformations, the
numerical developer can effectively assume all patches have the same
index orientation and can essentially implement operations between pairs of patches as if
both patches were on the same block.

\paragraph{Copying at patch boundaries}
To use \mblock indexing to fill coarse
ghost regions via copying, we transform grid cell coordinates ${\bf
I}_c$ to get index location ${\bf I}_n$ on the neighboring same-size
grid and then make the assignment $q_c({\bf I}_c) = q_n({\bf I}_n)$.

\paragraph{Averaging and interpolation at patch boundaries}
To fill coarse ghost regions via averaging or fine ghost regions via
interpolation, we need to map a single coarse grid index $I_c$ to four
fine grid indices.  We do this by defining four direction vectors on
the coarse grid which we use to find four fine grid locations
contained within a coarse grid cell.  These direction vectors are
given by
\begin{equation}
{\bf d}_0 = (-1,-1), \quad {\bf d}_1 = (1,-1),
\quad {\bf d}_2 = (-1,1), \quad {\bf d}_3 = (1,1).
\end{equation}
The corresponding fine grid locations, in coarse grid coordinates, are then given by
\begin{equation}
{\bf I}_c^k  = {\bf I}_c + \frac{1}{4}{\bf d}_k, \qquad k = 0,1,2,3 .
\end{equation}
A mapping between coarse and fine grid indices has the general form
\begin{equation}
{\bf I}_f = 2 A {\bf I}_c + {\bf F}^f ,
\end{equation}
where $A$ encodes index orientations between neighboring coarse and fine grids (as in the same-size
transforms), and $F^f$ encodes, in fine grid coordinates,
the location of the fine grid relative to the coarse grid.
The four fine grid interpolation points can then be defined as
\begin{equation}
\begin{aligned}
{\bf I}_f^k  & =  2A \left( {\bf I}_c + \frac{1}{4}{\bf d}_k \right) + {\bf F}^f \\
& = 2A{\bf I}_c + \frac{1}{2}A {\bf d}_k + {\bf F}^f, \qquad k = 0,1,2,3. \\
\end{aligned}
\end{equation}
The vector ${\bf F}^f$ encodes the location of the center of the coarse grid
in fine grid coordinates, and so the entries of ${\bf F}^f$ are half-index values.
It follows, then, that entries in the vector $\frac{1}{2}A {\bf d}_k + {\bf F}^f$ are integers,
and the final fine grid location ${\bf I}_f^k$ will be integer coordinates.

\ignore{
The index location ${\bf I}^c_f$ is the center of a coarse grid cell.
Thus, the entries of $\widetilde A$ are multiples of $2$ and $\widetilde F$
contains integers shifted by $\frac12$. \comment{But don't we say that A has
entries 1, -1, or 0, above?  I think this section still needs work.}
In this convention, $\widetilde{\bf d}_k$ adds or subtracts another half which
ensures that the fine grid indices are integers as well.}

Using the above, we can fill in coarse grid ghost cell values (on a
uniform grid) via averaging from a fine grid neighbor as
\begin{equation}
q_c({\bf I}_c) = \frac{1}{4}\sum_{k=0}^3 q_f({\bf I}_f^k) .
\label{eqn:ghost_coarse}
\end{equation}

To fill in fine grid ghost cells, we use interpolation stencils
that are described entirely in the coarse grid index space, but the
index locations of the fine grid ghost cells to be filled in by the
interpolation must be obtained via the transformation.   The interpolation
stencil can be applied as
\begin{equation}
q_f({\bf I}_f^k) = q_c({\bf I}_c) + \frac{h}{4} \widetilde{\nabla} q_c \cdot {\bf d}_k
\label{eqn:ghost_fine}
\end{equation}
where $\widetilde{\nabla} q_c$ is an approximation to the gradient of the coarse grid
solution and computed using one-sided differences and $h$ is the coarse grid mesh width.

\ignore{
To further reduce the complexity of possible grid configurations,
ghost exchange routines in \forestclaw are all written from the
perspective of the coarse grid, and the fine grid perspective is
realized by swapping the coarse and fine grids in subroutine argument
lists.  In this coarse grid perspective, the coarse grid
is assumed to be fixed in the usual Cartesian
orientation (index $i$ increases to the right; index $j$ increases
going up), and grid transforms handle any index orientation mismatches
at the patch boundaries.  With this approach, only four cases
need to be explicitly handled in subroutines for handling face
exchanges, and four cases for handling corner exchanges.
}

\ignore{
When a coarse grid is visited, coarse grid ghost regions are
filled by copying or averaging from a same-size or fine grid neighbor.
In a second interpolation step, coarse grids fill in the fine grid
ghost regions of neighboring fine grids.  While all four grid faces
and four grid corners must be explicitly handled by ghost exchange
routines, the orientation mismatches between grids at block boundaries
are handled by the transforms and are essentially transparent to a
user or developer.
}

\section{Parallel algorithms}
\label{sec:parallel}

\begin{figure}
\begin{center}
\includegraphics[width=0.40\textwidth]{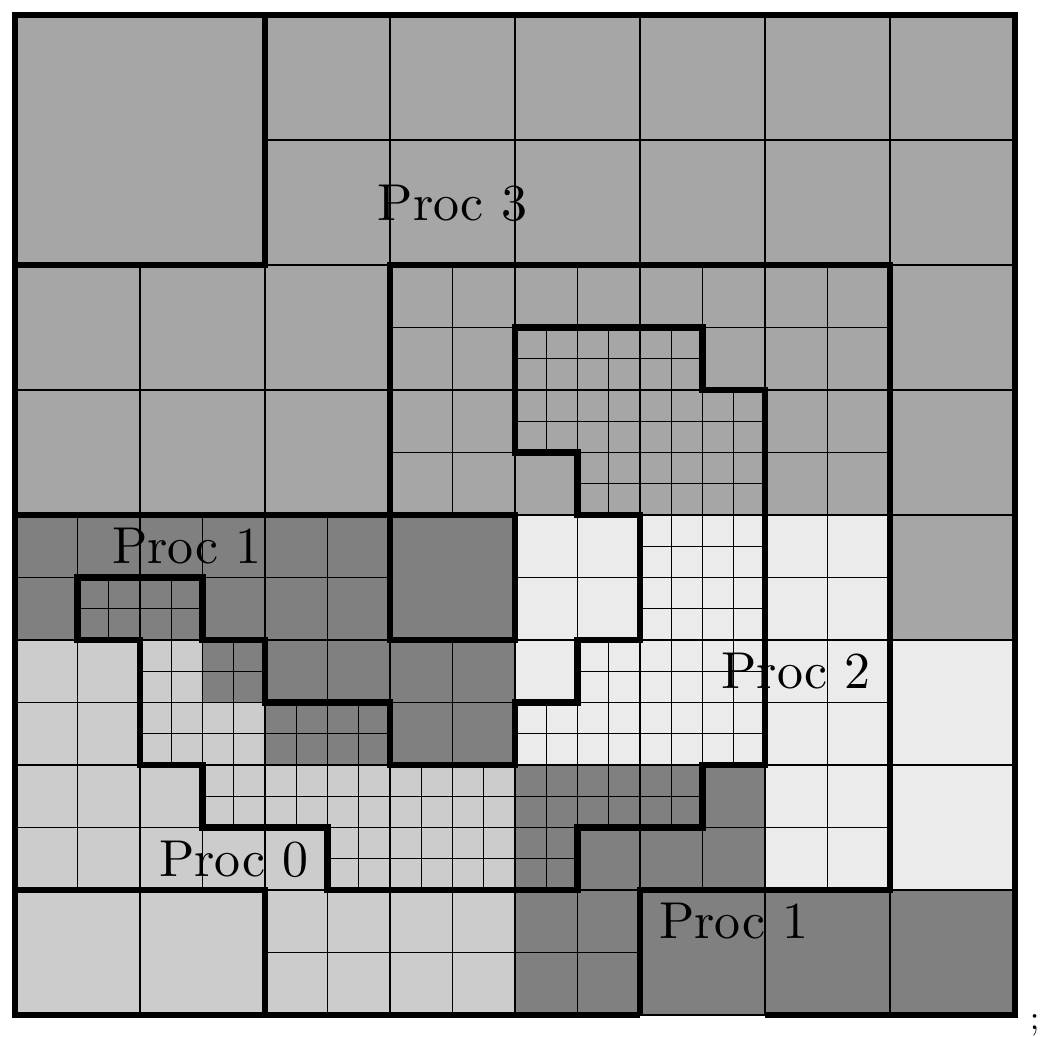}\hfil
\includegraphics[width=0.40\textwidth]
          {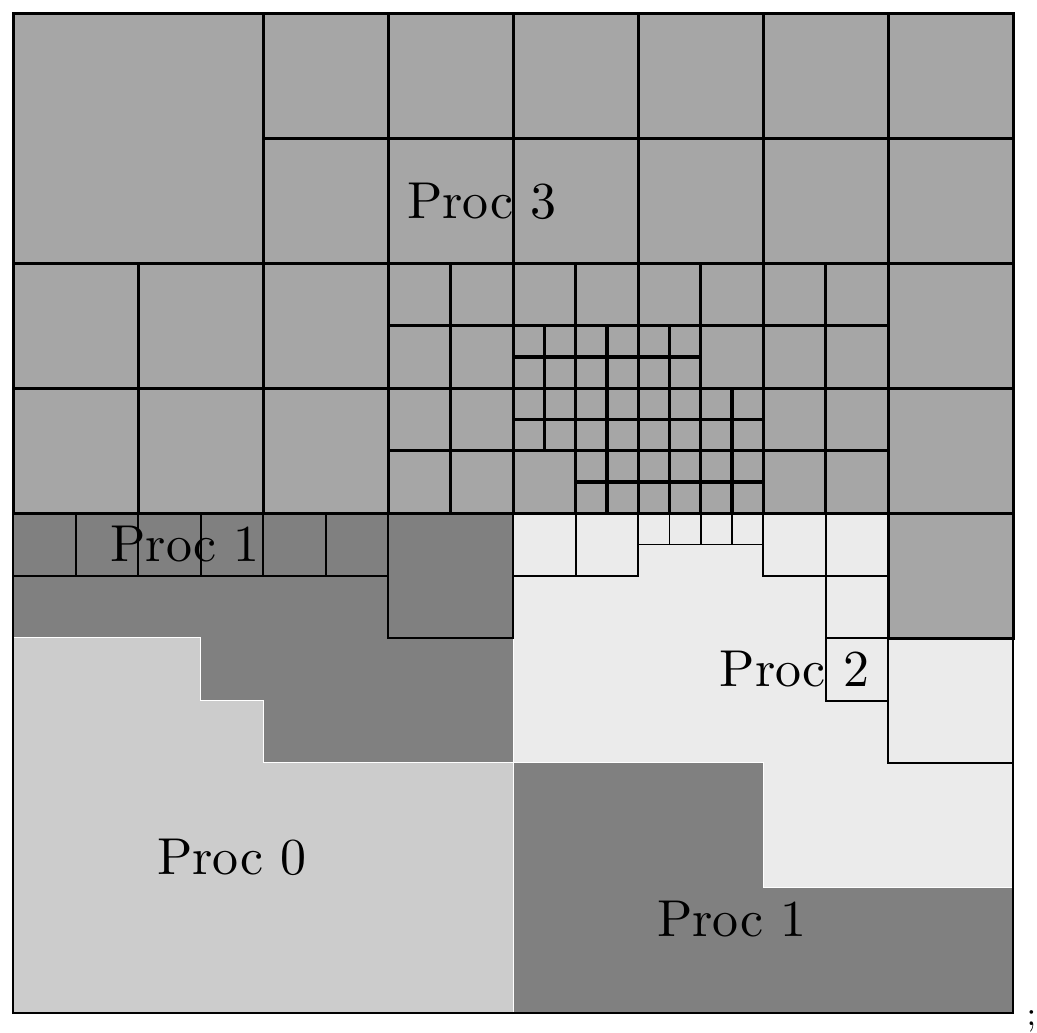}\hfil
\end{center}
\caption{Shaded regions showing processor regions. The left figure
shows patches on all processors, whereas the right figure shows
the patches local to processor 3 (thicker lines) and remote ghost
patches (thinner lines), sent to processor 3 from processors 1 and 2,
and stored as a local copy by processor 3.  Only processor region 1
consists of two disjoint subregions.}
\label{fig:procboundaries}
\end{figure}

Evenly distributing grids to processors using a space-filling curve
results in a partitioning of the quadtree layout into logical {\em
processor regions}.  See \Fig{procboundaries} for a typical
distribution of quadrants to processor regions.  Pairs of processor
regions are separated by simple rectilinear curves, which we call
{\em processor boundaries}.  The {\em parallel boundary} for a
particular processor region is the rectilinear curve made up of all
processor boundaries that separate the particular region from all
other regions.  \pforest (and by extension, \forestclaw) uses Morton
ordering to delineate the space-filling curve.  While this is a
discontinuous ordering, it can be shown that any processor region
defined by this ordering consists of at most two face-connected
sub-regions per tree, and so its parallel boundary will consist of at
most two simple curves for each tree \cite{bu-is:2015}.

Using the single-instruction-multiple-data (SIMD) paradigm underlying
the \mpi model, we describe our parallel exchange and  parallel
ghost filling algorithm
in terms of a {\em local} processor region and {\em remote} processor
regions.  We refer to {\em local} grids as those grids whose
interiors are in the local processor region.  By analogy, remote grids
are those grids whose interiors are in remote processor regions.  {\em
Local ghost regions} and {\em remote ghost regions} are ghost regions
in local or remote processor regions, respectively. To avoid confusion,
we will use terms {\em
local} or {\em remote} ghost regions for those local or remote regions
belonging to local grids only, unless otherwise explicitly stated.
Interior regions and ghost regions of a local or remote grid are said
to be ``on the parallel boundary'' if these regions share a corner or
face with the parallel boundary.

If we allow processor boundaries to extend beyond the physical domain
in an obvious way, we can also designate exterior ghost regions of local grids as
either {\em local} or {\em remote}.

\subsection{Parallel exchange of ghost cell data}
To exchange data across processor boundaries, \forestclaw uses the
data transfer mechanisms available in \pforest.  First,
each processor packs local patches on the parallel boundary
into a sender communication buffer.  This packing exploits the fact
that only the outermost layers of interior cells of a patch need to be
sent, while the center region of a patch is never needed by remote
processors.  Remote patches on the parallel boundary are received by
the processor in receiver communication buffers, unpacked
accordingly, and stored in a ghost patch array.  These local
copies of parallel ghost patches are not stored as part of the local
tree hierarchy and have limited meta-data, but for the purposes of
filling ghost regions, parallel patches can be used just like
local neighboring grid patches.

The \pforest abstract ghost exchange algorithm is based on its
internal knowledge of patches on the parallel boundary and can
optionally be split into an \mpi send phase and an \mpi receive phase.  This
design allows \forestclaw to process the local patches not on the
parallel boundary between calls to send and receive parallel patches,
effectively overlapping communication and computation.  Our approach
to smoothly grade the refinement described in \Sect{smoothrefine}
also makes use of this abstract \pforest ghost exchange facility by
sending and receiving the target refinement level for each patch.
After the target levels for remote patches on the parallel boundary are
received, each local patch can compute its target level as the maximum over
itself and its direct neighbors without further communication.

\subsection{Parallel ghost filling algorithm}
\label{sec:ghostfill_parallel}

Our parallel exchange algorithm needs to manage the sequence of data
transfers and mathematical operations between local and remote
patches.  The challenge in implementing our parallel ghost-filling
algorithm is imposing the correct order on the interleaving of steps
to fill coarse ghost regions and steps to fill fine ghost regions.
Because interpolation stencils may cross multiple parallel boundaries,
our algorithm requires multi-way ghost filling between patches received from
different processors.

In the adaptive setting, parallel boundaries will in general cross
boundaries between levels and fine ghost regions belonging to local
processors may fall on the parallel boundary.  This means that both
remote coarse ghost regions belonging to local patches and local
coarse ghost regions belonging to remote patches may be required to
have valid data before the fine ghost region can be properly filled.
For this reason, a pre-processing step on local patches and a
post-processing step on remote patches must be carried out before and
after parallel communication.  It will also be advantageous (even in
the uniformly refined case) to hide latency associated with parallel
communication.  These pre- and post-processing steps and the send and
recieve calls split the serial algorithm into three distinct steps.
These are labeled \textbf{Steps 1, 2 and 3} and detailed in the parallel
\Alg{ghostfill_parallel}.

\begin{figure}
\begin{center}
\includegraphics[width=0.40\textwidth,clip=true,
trim=7.25cm 4.9cm 2.25cm 4.59cm]{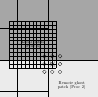}\hfil
\includegraphics[width=0.40\textwidth,clip=true,
trim=4.25cm 4.25cm 4.25cm 4.25cm]{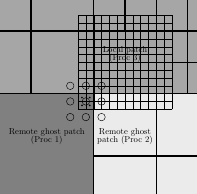}
\end{center}
\caption{%
These figures illustrate the challenge in filling fine corner
ghost regions on the parallel boundary.  In the left figure, the
interpolation stencil needed to fill in the corner fine ghost region
on the local patch crosses a single processor boundary and is applied
to the remote patch on processor 2.  To ensure that the required
coarse ghost region on the remote patch has been filled, a
pre-processing step on all processors fills coarse ghost regions on
the parallel boundary before patches are sent to the remote
processors.  In the right figure, the interpolation stencil crosses
two processor boundaries and the pre-processing step is not sufficient
to ensure that the required coarse ghost regions on the remote patch
on processor 1 are filled.
Therefore, a second ghost filling between locally stored
remote patches from processors 1 and 2 is required.}
\label{fig:procboundaries_zoom}
\end{figure}

\begin{algorithm}
\caption{%
 Parallel interleaved ghost cell update.}
\label{alg:ghostfill_parallel}
\begin{algorithmic}
\Require{Grids on all levels must be time synchronized}
\ForAll{coarse grids {\bf on} the parallel boundary}
\Comment{Step 1}
\State Fill in local coarse ghost regions
\State Apply physical boundary conditions to exterior ghost regions
\EndFor
\State {\bf Send local patches at parallel boundary to remote processors}
\ForAll{coarse grids {\bf not on} the parallel boundary}
\Comment{Step 2}
\State Fill in local coarse regions
\State Apply physical boundary conditions to exterior ghost regions
\State Fill in fine grid neighbors' ghost regions using this grid's interior
\EndFor
\State {\bf Receive patches from remote processors}
\State Fill coarse ghost regions between remote ghost patches from
  different processors.
\ForAll{coarse grids {\bf on} the parallel boundary}
\Comment{Step 3}
\State Fill in remote coarse grid ghost regions, using remote grids
\State Apply physical boundary conditions to exterior ghost regions
\EndFor
\ForAll{{\bf remote} coarse grids}
\State Fill local coarse ghost regions using local grids.
\EndFor
\ForAll{grids {\bf on} the parallel boundary}
\State Fill local coarse ghost regions belonging to remote grids
\State Apply physical boundary conditions to exterior ghost regions
\EndFor
\ForAll{fine grids {\bf on} the parallel boundary}
\State Fill in remote fine  ghost regions, using remote coarse grids.
\EndFor
\ForAll{grids {\bf on} the parallel boundary}
\State Apply physical boundary conditions to exterior ghost regions
\EndFor
\end{algorithmic}
\end{algorithm}


\begin{prp}
Assume we have a 2:1 balanced quadtree layout partitioned to processors
using a space-filling curve.  Let $M$ be the fixed grid size, let $m$
be the number of ghost layers, and $w$ be the width of the stencil
used to interpolate from the coarse grid to fine grid ghost cells.
Assume that the interior regions of all grids contain valid data.
Then, if $m \le M/4$ and $w \le M/2$, \Alg{ghostfill_parallel} is
guaranteed to fill in all coarse, fine and exterior ghost regions
belonging to grids on the local processor with valid data.
\end{prp}

We justify the proposition by describing in detail the three steps of
\Alg{ghostfill_parallel}.

Steps 1 and 2 of our parallel ghost filling algorithm are responsible for
filling the coarse grid regions between any two local neighbor grids,
and all local fine ghost regions not on the parallel boundary.
All local coarse ghost regions can be filled
using data from the interior of local
patches, or by applying physical boundary conditions.  Similarly, all
local fine ghost regions not on the parallel boundary can be
filled using interpolation stencils that rely only on local coarse
grid ghost regions. To overlap communication, Step 1 only fills
coarse grid ghost regions on the parallel boundary.  Following this
step, all grids on the parallel boundary are packed into communication
buffers and sent to remote processors.  Step 2 then continues by filling
all coarse and fine grid regions not on the parallel boundary.

What remains are remote coarse grid regions belonging to local grids
and remote and local fine grid ghost regions on the parallel boundary.
These remaining ghost regions are filled in Step 3, after the
communication step.  In the first loop in Step 3, remote coarse grid
regions belonging local grids on the parallel boundary are filled by
copying or averaging from the locally stored remote parallel patches.
To show that local and remote fine grid regions belonging to local
grids on the parallel boundary are filled by our algorithm, we have to
demonstrate that the stencils involved will have valid data.

In the simplest case, an interpolation stencil needed to fill a remote
or local fine grid region on the parallel boundary does not cross the
parallel boundary.  In this case, the coarse grid ghost data needed
for the stencil will have been filled from the averaging step in Step
1 and so the stencil will have valid data.

A slightly more complicated situation occurs when the interpolation
stencil crosses a single parallel boundary.  In this situation, the
fine grid region to be filled may be either local or remote, but the
interpolation stencil will at least partially depend on data from a
remote coarse grid ghost region.  If the fine grid region is local,
then the coarse grid used for the stencil is local, but since it
crosses the parallel boundary, the stencil will depend on a remote
ghost region belonging to the local coarse grid.  These remote ghost regions
are filled in the first loop of Step 3.  In a second case, the fine
grid region is itself a remote region (but belonging to a local grid)
and the coarse grid used for the
interpolation is itself a remote parallel patch.  In this situation,
the parallel patch must have valid data in any coarse grid ghost
region on the parallel boundary.  These regions are filled in the
second loop of Step 3.  For an illustration of this second case, see
the left plot in \Fig{procboundaries_zoom}.

Finally, the case that requires extra handling is the one in which an
interpolation stencil crosses two or more processor boundaries.  This
situation can arise, for example, when filling corner fine ghost
regions that lie in a remote processor region (see the right plot in
\Fig{procboundaries_zoom}). In this case, the remote (but locally
stored) parallel ghost patch is not guaranteed to have valid data in
all coarse ghost regions on the parallel boundary.  If the coarse grid
region lies in the local processor region, then this coarse grid ghost region
will have been filled in the second loop of Step 3.  But if the coarse
grid ghost region lies in a third processor's region, then we need a mechanism
for filling coarse ghost regions between parallel patches originating from different
processors.  This is included in \Alg{ghostfill_parallel} between Step 2 and
Step 3.

%


\section{Dynamic grid adaptation}
\label{sec:amr}
One of the defining features of \forestclaw, and other adaptive grid
codes, is that grids are dynamically adapted to follow solution
features of interest.  The general refinement strategy in \forestclaw
involves applying refinement and coarsening criteria at regular
regridding intervals to determine if the solution in a particular
quadrant should be replaced with four sibling grids (``refined''), or
if the solution on four sibling grids should be replaced by a single
parent grid (``coarsened'').  Once the refinement and coarsening
criteria have been applied, the quadtree mesh is regenerated and
the numerical solution is adapted to newly created coarse or fine
quadrants.  In a parallel setting, this regridding step is followed by
a parallel partitioning step that re-distributes grids evenly to
processors, correcting potential load imbalances caused by the
regridding.

\ignore{It many cases, it is desirable to smoothly grade transitions between
refinement levels and to effectively extend a tagged region occupied by
one level into the next coarser level. To achieve this, we
follow the initial tagging phase of the regridding process
by a post-processing step that tags additional grids (not initially
tagged based on the solution) to provide buffer grids around specified
levels.  The left plot in \Fig{amrgrid} shows an illustration of smooth
refinement between levels 5 and 6 to ensure that the tracer quantity remains
on the finest level grids.}

\ignore{
\comment{A sentence or two on how cells are tagged}
}

\subsection{Dynamic refinement algorithm}
\label{sec:adapt}
The \forestclaw regridding algorithm proceeds in three
basic steps.  First, each grid in a quadtree layout $Q$ is either
tagged for refinement, tagged, along with any sibling grids, for
coarsening, or left untagged.  A common criterion for tagging cells is
a ``feature'' based refinement, identifying, for example, a sharp jump in
the computed solution or a steep gradient.  Once tagging has been
completed, a quadtree adaptation step creates a new quadtree layout
$Q'$.  A 2:1 balancing step, needed to enforce proper nesting of
refinement levels, may tag additional grids for refinement, and so
coarsening is not always guaranteed to occur based on user-defined
coarsening criteria.  Grids that are tagged for refinement are
guaranteed to be refined.

%
%
After the new mesh is generated, we interpolate the solution from
coarse grids to the newly created fine grids using the same monotone
interpolation scheme we use for filling ghost cells.  Sibling grids
are averaged to a coarser grid using an averaging stencil.
Grids that were neither refined or coarsened are reassigned
unchanged from $Q$ to $Q'$.  Once all grids in $Q'$ are populated, grids
are re-partitioned to processors to ensure proper load balancing.
Again, repartitioning is delegated to the \pforest library.
To support general refinement criteria (examining differences between
neighboring grid cells, for example), the tagging algorithm should run after
a call to the ghost filling algorithm.  Ghost-filling is required once again
on the new and re-partitioned domain $Q'$ to ensure that the adapted solution
is well-defined.

The generation of the new adaptive mesh layout $Q'$ and the 2:1 balancing
are delegated to the \pforest library, which operates largely independent of
the \forestclaw layer.

\ignore{
\begin{algorithm}
\caption{Regridding algorithm for quadtree layout $Q$ to produce new
layout $Q'$. A quadtree is regridded every user-specified
$N$ coarse grid time steps.  In practice, regridding is done every coarse grid
time step ($N=1$). }
\label{alg:regridding}
\begin{algorithmic}
\Require{All grids in $Q$ must be time synchronized at level $t_n$ and have
valid ghost cells.}
\Statex
\State Let $K$ be a subset of indices $\{K_1,K_2,\hdots,K_F\}$, $1 \le K_j \le G$
which start a family of sibling grids.
\Statex
\ForAll{families $f_j = \{g_{K_j},g_{K_j + 1},g_{K_j + 2},g_{K_j+3}\}$,
$j = 1,2,\hdots,F$}
\State Compute $\delta_{K_j+i}^n$, $i = 0,1,2,3$ using either \eqn{delta_undivide} or \eqn{delta_grid}
\If{$\delta_{K_j + i}^n < \tau_c$ for all $i = 0,1,2,3$}
\State Tag all grids in family $f_j$ for coarsening.
\EndIf
\EndFor
\ForAll{grids $g_k$, $k = 1,2, \hdots,G$}
\If{grid $g_k$ is not tagged for coarsening}
\State Compute $\delta_{k}^n$ using either \eqn{delta_undivide} or \eqn{delta_grid}
\If{$\delta_k^n > \tau_r$}
\State Tag grid $g_k$ for refinement
\EndIf
\EndIf
\EndFor
\Statex
\State If smooth refinement is on, exchange target level of ghost grids,
       modify tags.
\State Construct new 2:1 balanced quadtree $Q'$ consisting of grids $g'_k$,
$k = 1,2, \hdots,G'$.
\Statex
\If{$Q'$ is different from $Q$}
\State Copy, average or interpolate grids from $Q$ to $Q'$
\State Repartition grids across processors
\State \Call{update\_ghost}{$\lmin$}
\EndIf
\end{algorithmic}
\end{algorithm}
}%

\ignore{
For example, one could exit the computation of
$\delta_k^n$ as soon as it is detected that for some $(i,j)$, the
tagging criteria in \eqn{delta_undivide} or \eqn{delta_grid} exceed
$\tau_r$ or $\tau_c$.  If the tagging criteria exceeds $\tau_r$ for
some $(i,j)$ the grid should be tagged for refinement, and there is no
need to check the remaining $(i,j)$.  If the criteria exceeds
$\tau_c$, the grid will not be coarsened, and there is no need to
apply the coarsening criteria to the remainder of that grid, or the
remaining sibling grids. With more detailed knowledge of how many
grids are likely to be refined or coarsened during any regridding
step, one could probably implement more optimizations.  However, in
practice, regridding takes such a small fraction of the overall
wall-clock time, that such optimizations would likely have negligible
impact on the total time to solution.
}

\subsection{Smooth refinement}
\label{sec:smoothrefine}

When dynamically adapting grid resolution to follow solution features
of interest, one wishes to ensure that such features are not
too close to coarse-fine boundaries.  To provide a buffer region
around grids that have been tagged for refinement, \forestclaw can
optionally smoothly refine from one region to the next.  This
effectively adds an additional layer of tagged grids around the finest
levels and avoids the situation in which sharp solution features are
just barely contained by the finest level grids.

%
%

After all grids have been tagged for coarsening or refinement,
refinement levels are smoothly graded as follows.  Each grid stores
its current level and a {\em target level} which is either equal to
the current level (i.e.\ the grid is not tagged for refinement or
coarsening), one greater than the current level, or one level less.
Then, for each patch, we compute the maximum of the target level over
that patch and all neighboring patches and pass this to the \pforest
adapt/balance routines.  The 2:1 balancing algorithm then ensures that
neighboring levels will never differ by more than one.


\section{Scaling results}
\label{sec:examples}
We demonstrate our ghost-filling algorithm and parallel
communication scheme using the wave propagation algorithms available in
\clawpack, a software package for solving hyperbolic problems using
high resolution, second order finite volume schemes on logically
Cartesian meshes \cite{ma-ah-be-ca-ge-ha-ke-le-le:2016, le:1996,
le:1997, le:2002, be-le:1998}.
Incorporating the \clawpack Fortran library routines into \forestclaw
required only minimal changes to a few of those routines.  Both
\clawpack 4.6 and \clawpack 5 solvers, along with most \clawpack
applications from those packages, are all available as part of
\forestclaw.

In the following study, we focus  on how the choice of
fixed grid size $M$ affects the efficiency of the adaptive algorithms in
\forestclaw.  The model problems we consider are scalar advection of a tracer field in
a square, replicated domain and on a sphere.
The scalar advection problem is ideally suited for performance
scaling studies because we can easily choose a fixed time step size that
remains stable throughout a simulation and across a wide range of
resolutions.   The Riemann problem for scalar advection has very low
arithmetic intensity, so any overhead associated with communication
and dynamic regridding cannot be easily hidden by the cost of
advancing the solution.  Also, because of the low memory requirements
of the scalar advection problem, we can run problems that are large
enough to maintain appropriate granularity at high processor counts
without becoming memory-bound on lower counts.






\subsection{Constant velocity in a square, replicated domain}
\label{sec:parallel_advection}
\label{sec:scaling}
\begin{figure}
\begin{center}
\includegraphics[width=0.5\textwidth,clip=true,trim=0cm 0.2cm 0cm 0cm]
{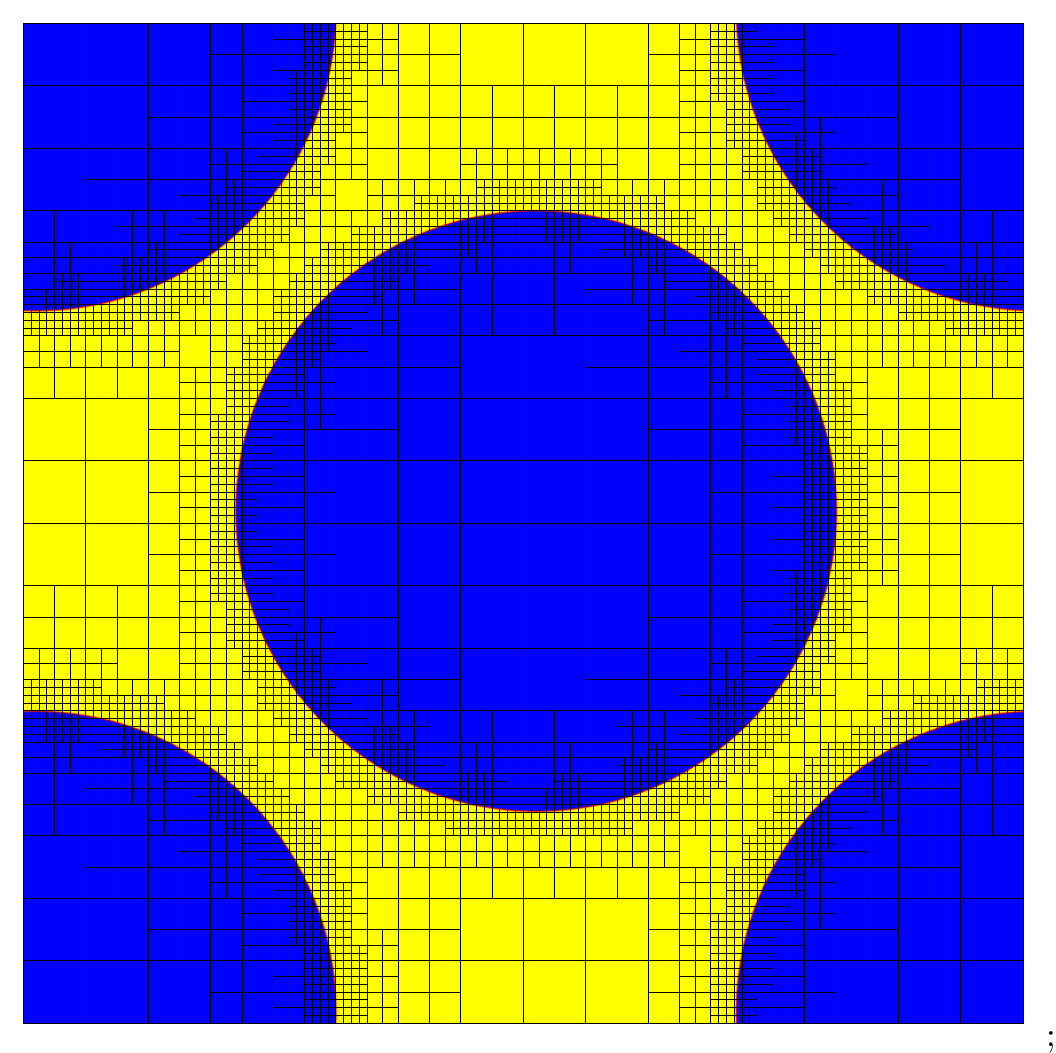}
\includegraphics[width=0.48\textwidth]{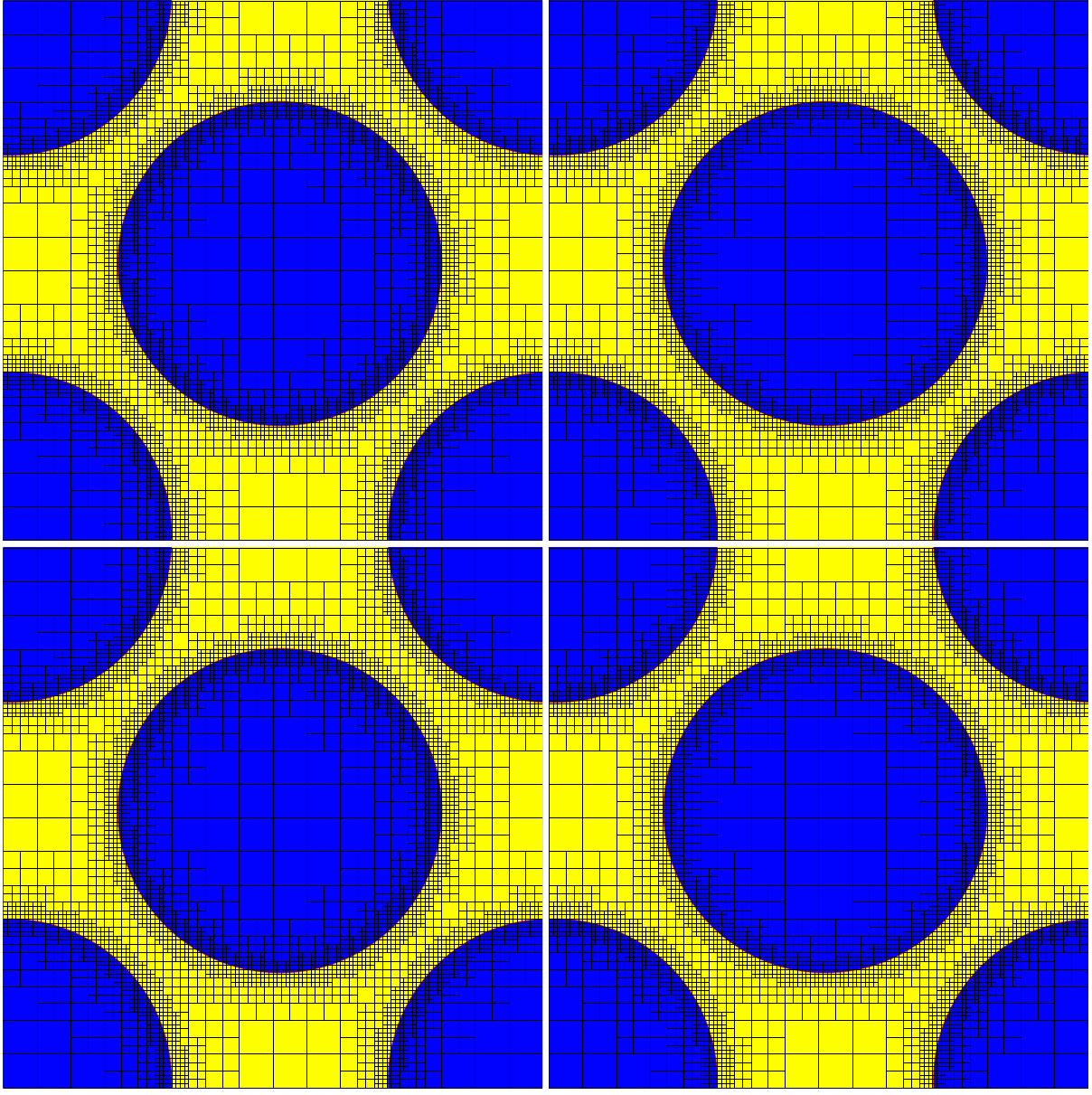}
\end{center}
\caption{Initial tracer field for the replicated
scalar advection problem (left).
The mesh is adapted to the tracer field, which is set to 1 inside of one of
five disks of radius 0.3 and 0 outside.
We use refinement levels 4 through 7 (grid lines not shown).
The domain on the right shows the unit quadtree replicated
four times on a $[0,2] \times [0,2]$ domain.%
}
\label{fig:tracer_init}
\end{figure}

In this first test, we run the scalar advection problem on a sequence
of replicated domains designed to provide meaningful weak scaling
results for adaptive simulations.  Each domain is a \mblock (or ``brick'')
domain consisting of a $2^n \times 2^n$ arrangement of unit blocks, each of
which is an adaptive quadtree.  The initial tracer field (shown in \Fig{tracer_init})
is replicated on each of the blocks in the \mblock domain, and periodic boundary
conditions are used at the physical boundaries of the domain.  Once
the initial tracer field is replicated across the domain, the
\forestclaw simulation is oblivious to the replication, and all
aspects of the parallel regridding, communication and load balancing
algorithms in \forestclaw are rigorously exercised.

\paragraph{Problem setup}
Each unit block in the tracer field on the replicated domain is
initialized using the piecewise constant initial field shown in
\Fig{tracer_init}. The prescribed flow field is the constant velocity
${\bf u} = (u,v) = (0.5,0.5)$, defined using the streamfunction
$\psi(x,y) = -(x-y)/2$ \cite{ca-le:2000, ca:2002}. The periodic boundary
conditions imposed on the physical boundary ensure that each
block runs the same problem.

For all runs, we fix the CFL number $\alpha$ to 0.64 and adjust the
time step for each run to satisfy $\dt = \alpha\dx$, where $\dx$ is
the mesh width for the finest level grids.  The number of time steps
taken is held fixed so that the final time varies with the resolution
of the fixed size grids.  For the adaptive runs, we run the
simulations for 160 time steps to final times $T=0.2$, $T=0.1$ and
$T=0.05$, corresponding to the three different grid resolutions, as
described in the next paragraph.  The uniform results are run for 20
steps, and resulting timings scaled by 8 to make valid
comparisons with the adaptive runs.

We ran three sets of uniform runs and three sets of adaptive runs,
corresponding to fixed size grids $8 \times 8$, $16 \times 16$ and $32
\times 32$. For the uniform runs, we set $\lmin = \lmax = 7$ so that
the effective resolutions on each block in the uniform case are
$1024\times 1024$, $2048 \times 2048$ and $4096 \times 4096$.  For the
adaptive runs, we set $\lmin=4$ and $\lmax = 7$, corresponding to
initial coarse grid resolutions of $128\times 128$, $256\times 256$
and $512\times 512$.   The adaptive mesh is dynamically regenerated
every $2^{\lmax-\lmin} = 2^3 = 8$ times steps.  For the uniform runs,
we disabled the dynamic regridding and only use the initial mesh
created by \pforest. A grid is tagged for refinement if the
difference between its largest value and smallest value exceeds a
refinement threshold $\tau_r=0.25$.  Four sibling grids are tagged for
coarsening if the difference between the largest and smallest value on
each sibling grid does not exceed a coarsening threshold of $\tau_c =
0.001$.  The finest level is smoothly graded for all adaptive runs.

\paragraph{Parallel setup}
Within each of the six sets of runs described above, we vary the
number of processors used and dimensions of the replicated \mblock
domain. For example, on the $32\times32$ runs, we run on replicated domains
ranging from a single block to $256\times 256$ blocks, while
corresponding MPI process counts vary from 1 to 65,536 (shown in
\Tab{scale_gpp}).
All of our parallel runs were done on JUQUEEN, the
BlueGene/Q system at the J\"ulich Supercomputing Centre
(Forschungszentrum, J\"ulich, Germany).  Each node on JUQUEEN has one
16-core PowerPC A2 processor running at 1.6 \si{\giga\hertz} with 16
\si{\giga\byte} RAM.  Unless otherwise stated, we ran 32 ranks (\mpi
processes) on each JUQUEEN node, making 0.5 \si{\giga\byte} of memory
available for each process.

\begin{table}
\footnotesize
\caption{Average number of \gpp for the \adaptrun{32} set of runs for the
replicated, \mblock scalar advection problem.  The leftmost column
shows number of \mpi ranks used for the run and the top row shows the
dimensions of the \mblock brick domain used, i.e.\ $1\times1$, $2
\times 2$, $4 \times 4$ and so on.  Weak scaling results are taken
from runs along diagonals, where the work per process remains fixed.
We carried out similar sets of runs for $8\times8$ and $16\times 16$
fixed size grids, for both adaptive and uniform cases (not shown here).}
\label{tab:scale_gpp}
\begin{center}
\begin{tabular}{r*{9}{S[table-format=4]}}
\toprule
Ranks  &  1 & 2 & 4 &  8 & 16 & 32 & 64 &  128  & 256\\
\midrule
       1  &     5059    &   \dash  &     \dash   &  \dash   &   \dash  & \dash &    \dash  &    \dash & \dash \\
       4  &     1264    &    5059  &     \dash   &  \dash   &   \dash  & \dash &    \dash  &    \dash & \dash \\
      16  &      316    &    1264  &      5059   &  \dash   &   \dash  & \dash &    \dash  &    \dash & \dash \\
      64  &       79    &     316  &      1264   &   5059   &   \dash  & \dash &    \dash  &    \dash & \dash \\
     256  &       19    &      79  &       316   &   1264   &    5059  & \dash &    \dash  &    \dash & \dash \\
    1024  &      \dash  &      19  &        79   &    316   &    1264  &  5059 &    \dash  &    \dash & \dash \\
    4096  &      \dash  &   \dash  &        19   &     79   &     316  &  1264 &     5059  &    \dash & \dash \\
   16384  &      \dash  &   \dash  &     \dash   &     19   &      79  &   316 &     1264  &     5059 & \dash \\
   65536  &      \dash  &   \dash  &     \dash   &  \dash   &      19  &    79 &      316  &     1264 & 5059 \\
\bottomrule
\end{tabular}
\end{center}
\end{table}

\begin{figure}
\begin{center}
\plotbox{\includegraphics[width=0.320\textwidth, clip=true,trim=1cm 0cm 1.95cm 0cm]
  {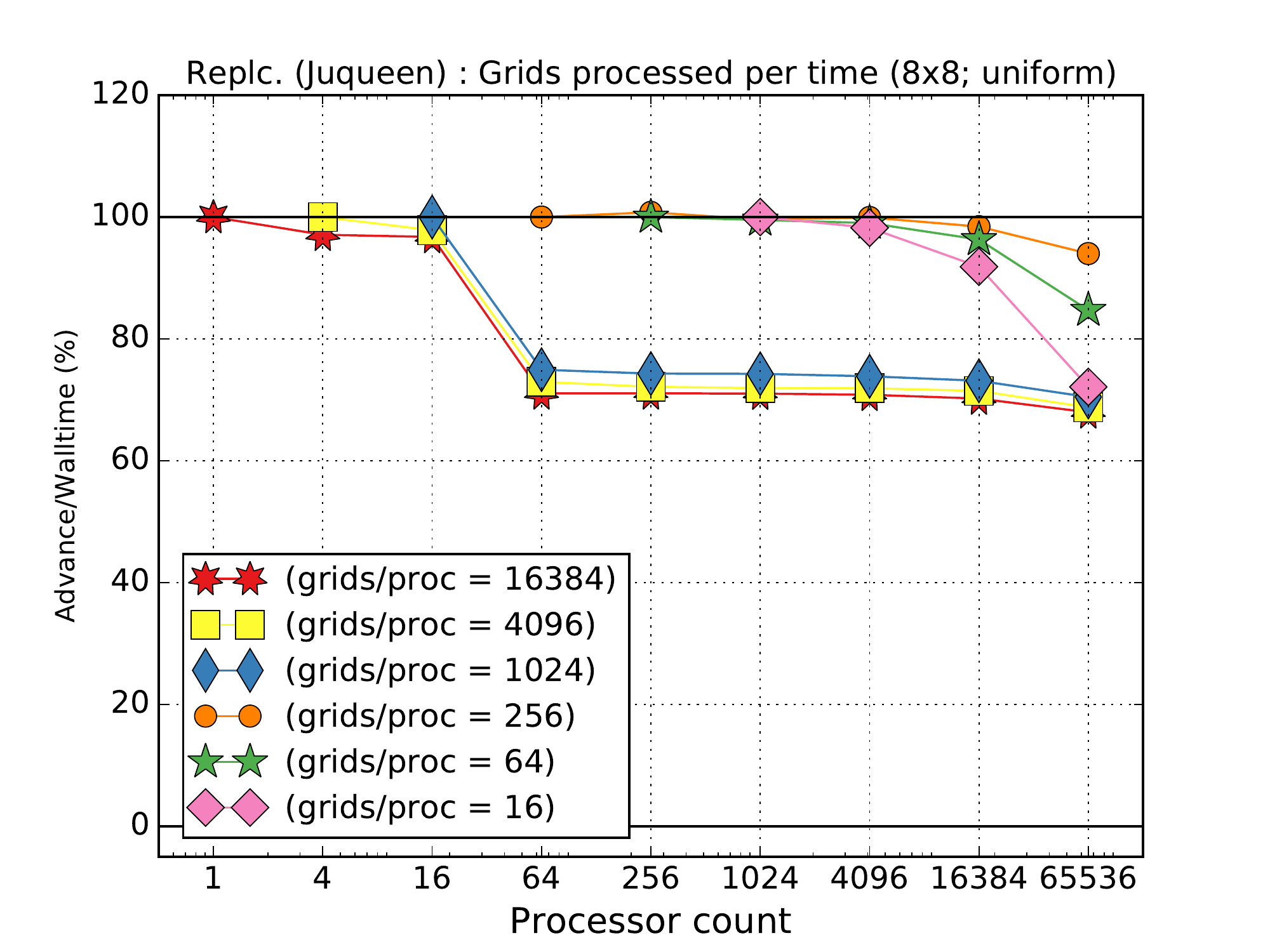}}\hfil
\plotbox{\includegraphics[width=0.320\textwidth, clip=true,trim=1cm 0cm 1.95cm 0cm]
  {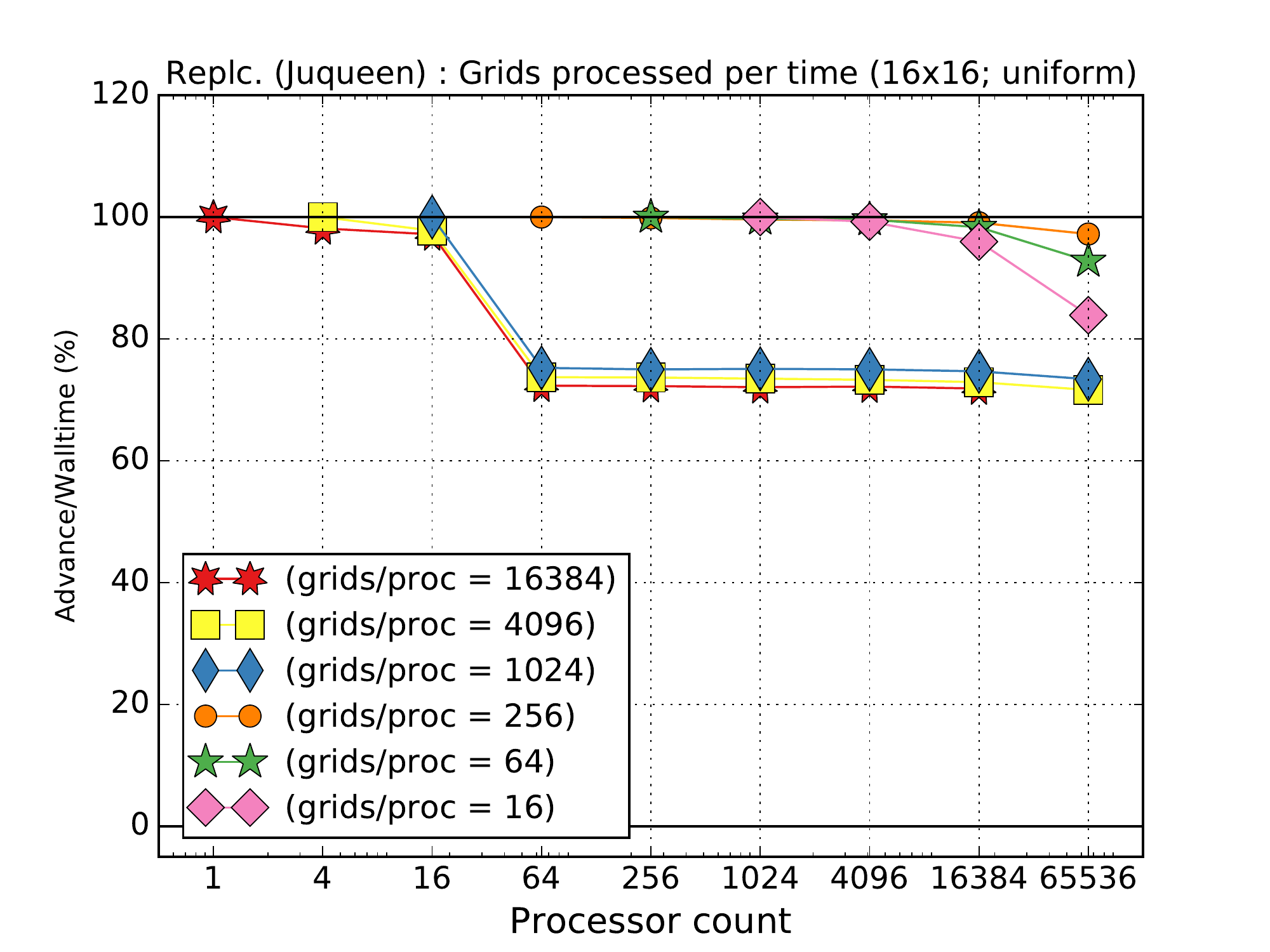}} \hfil
\plotbox{\includegraphics[width=0.320\textwidth, clip=true,trim=1cm 0cm 1.95cm 0cm]
  {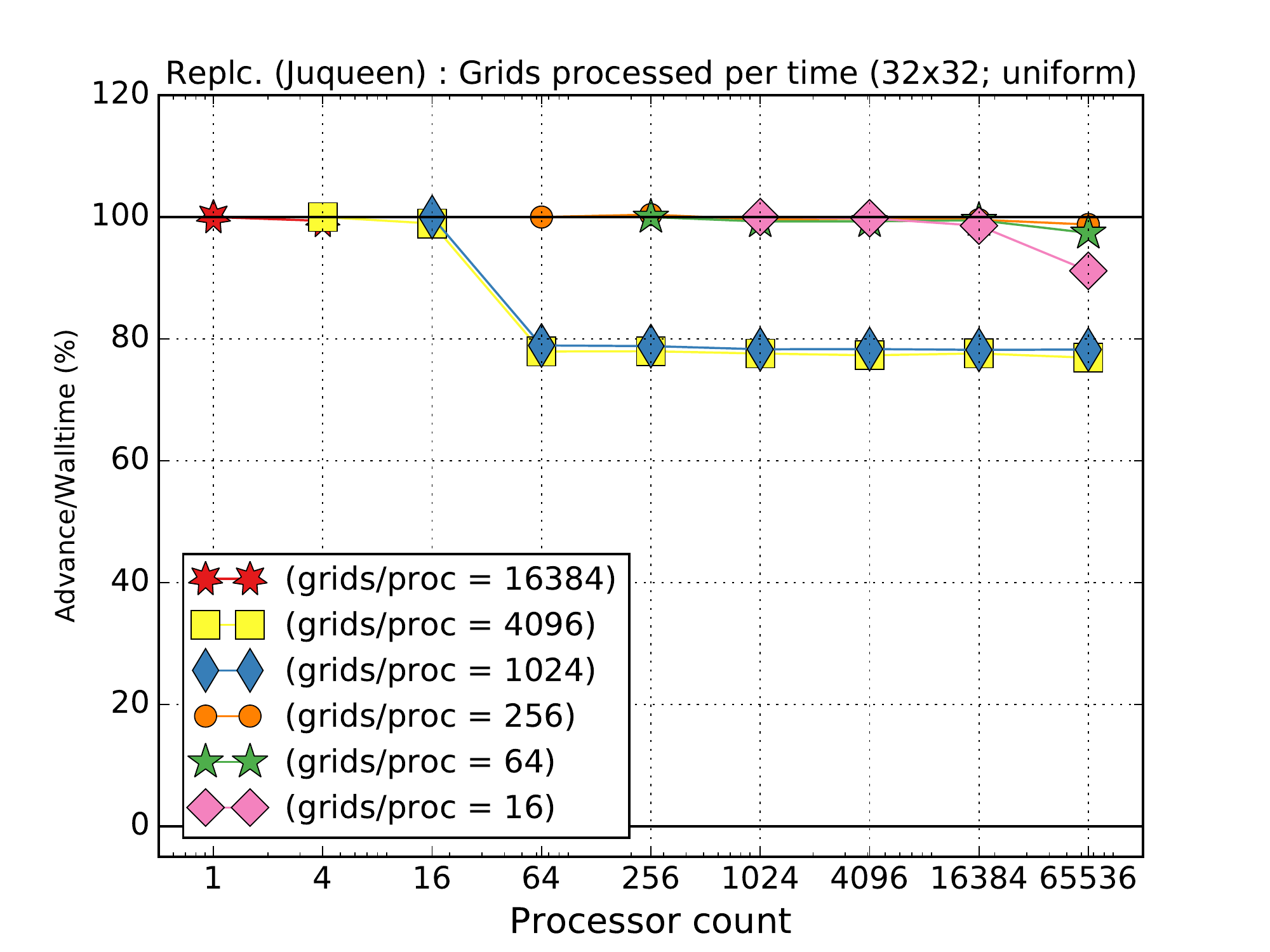}}\\
\plotbox{\includegraphics[width=0.320\textwidth, clip=true,trim=1cm 0cm 1.95cm 0cm]
  {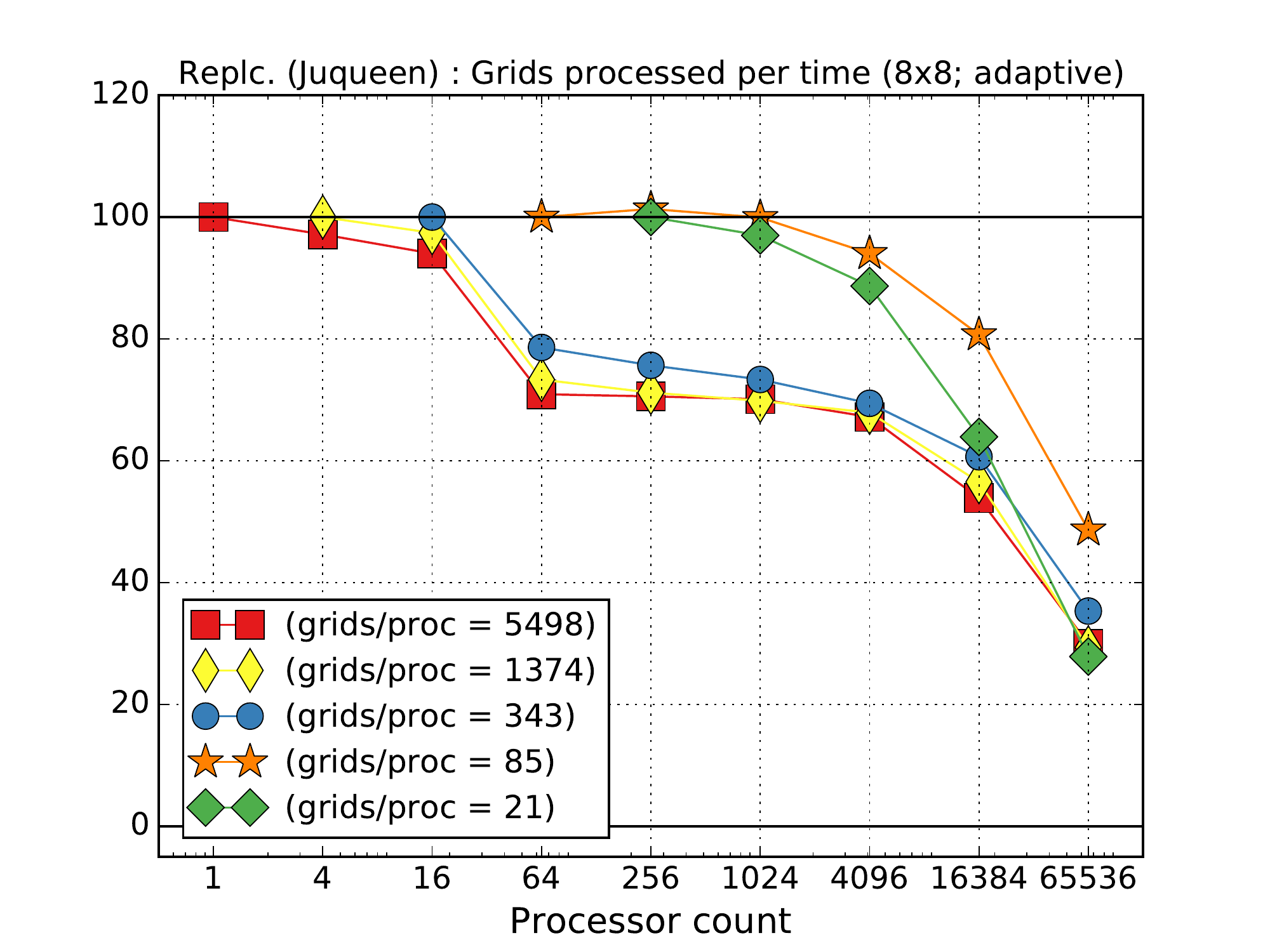}}\hfil
\plotbox{\includegraphics[width=0.320\textwidth, clip=true,trim=1cm 0cm 1.95cm 0cm]
  {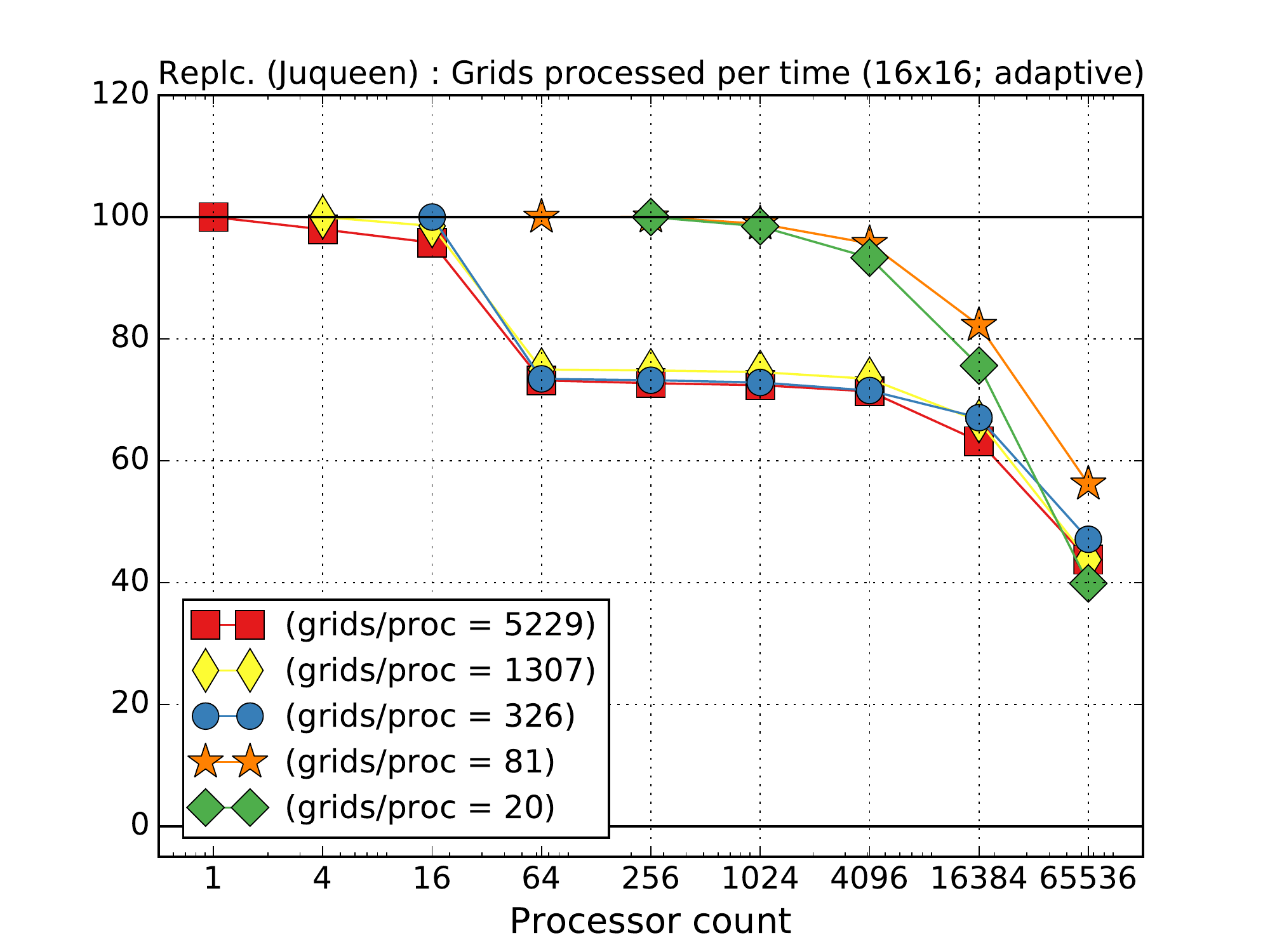}}\hfil
\plotbox{\includegraphics[width=0.320\textwidth, clip=true,trim=1cm 0cm 1.95cm 0cm]
  {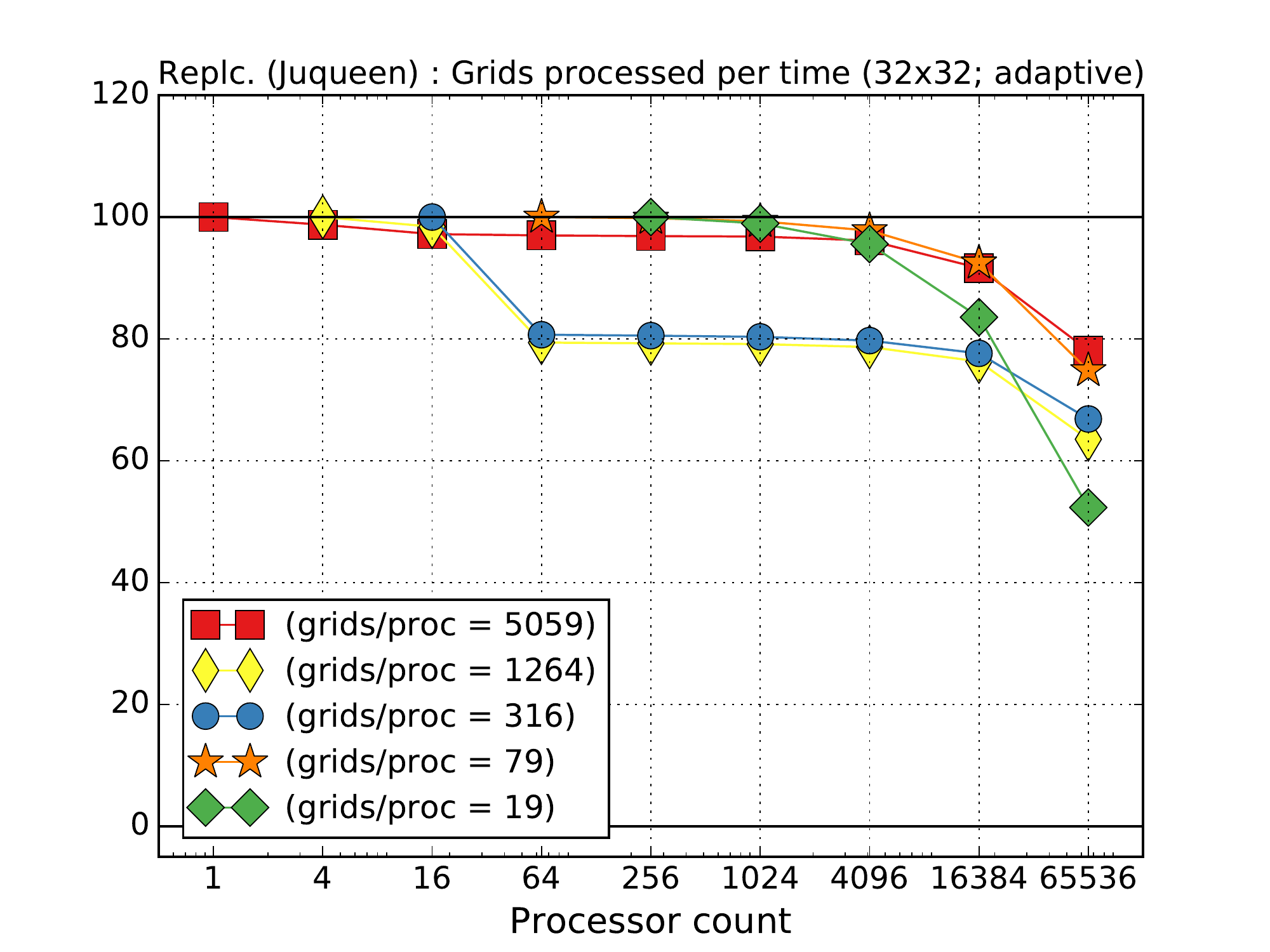}}
\end{center}
\caption{Weak scaling results for the replicated, \mblock scalar
advection problem.  The uniform runs are shown in the top row and the
adaptive runs on the bottom row.  The $8\times8$, $16\times16$ and
$32\times32$ runs are shown in the left, middle and right columns,
respectively.  Each plot shows the efficiency (\%) of the
particular run, where efficiency is computed as the ratio of the
number of grids updated per time for a simulation run on $P$
processes to the number of grids updated per time on 1 process.
The number of \gpp is indicated in the legend and remains constant for
each curve.  The reduction in efficiency going from 16 processes to 64 processes results from running  1 \mpi
process per core on 1,4 and 16 cores to 2 processes per core for 64 and
higher process counts, e.g.\ 32 ranks per node on each JUQUEEN node.}
\label{fig:weak_scaling}
\end{figure}

\begin{figure}
\begin{center}
\plotbox{\includegraphics[width=0.320\textwidth, clip=true,trim=1cm 0cm 1.6cm 0cm]
  {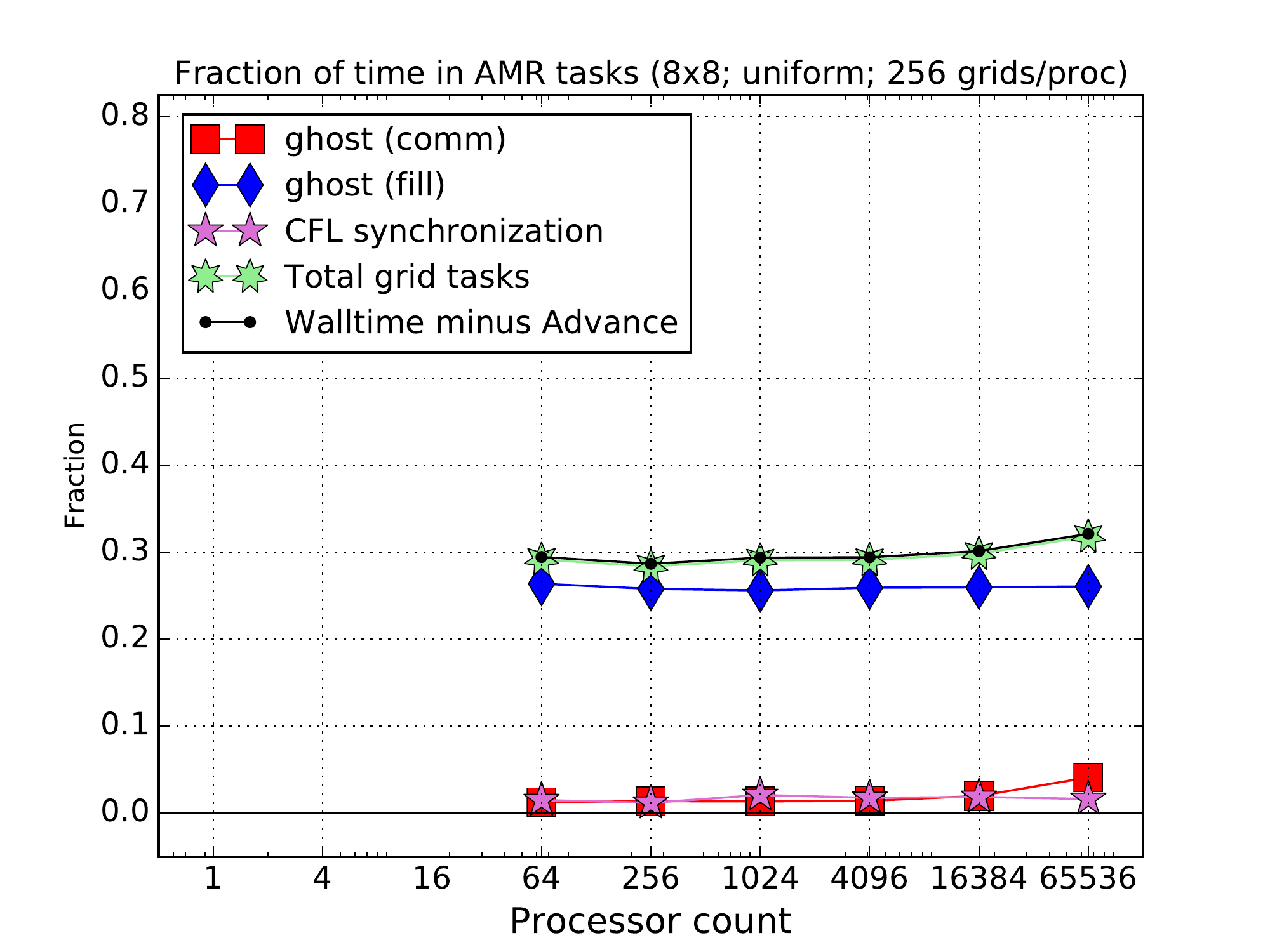}}\hfil
\plotbox{\includegraphics[width=0.320\textwidth, clip=true,trim=1cm 0cm 1.6cm 0cm]
  {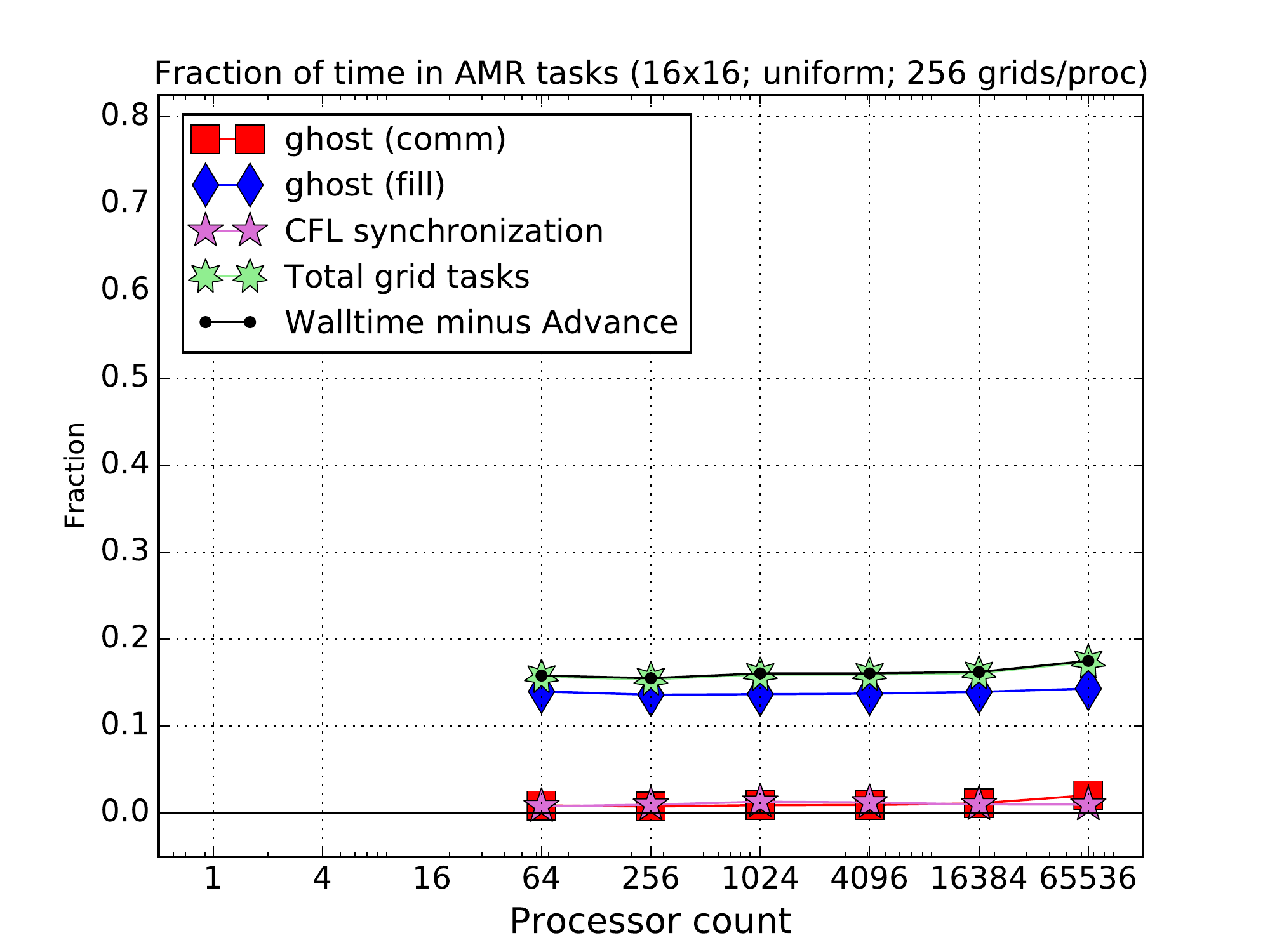}}\hfil
\plotbox{\includegraphics[width=0.320\textwidth, clip=true,trim=1cm 0cm 1.6cm 0cm]
  {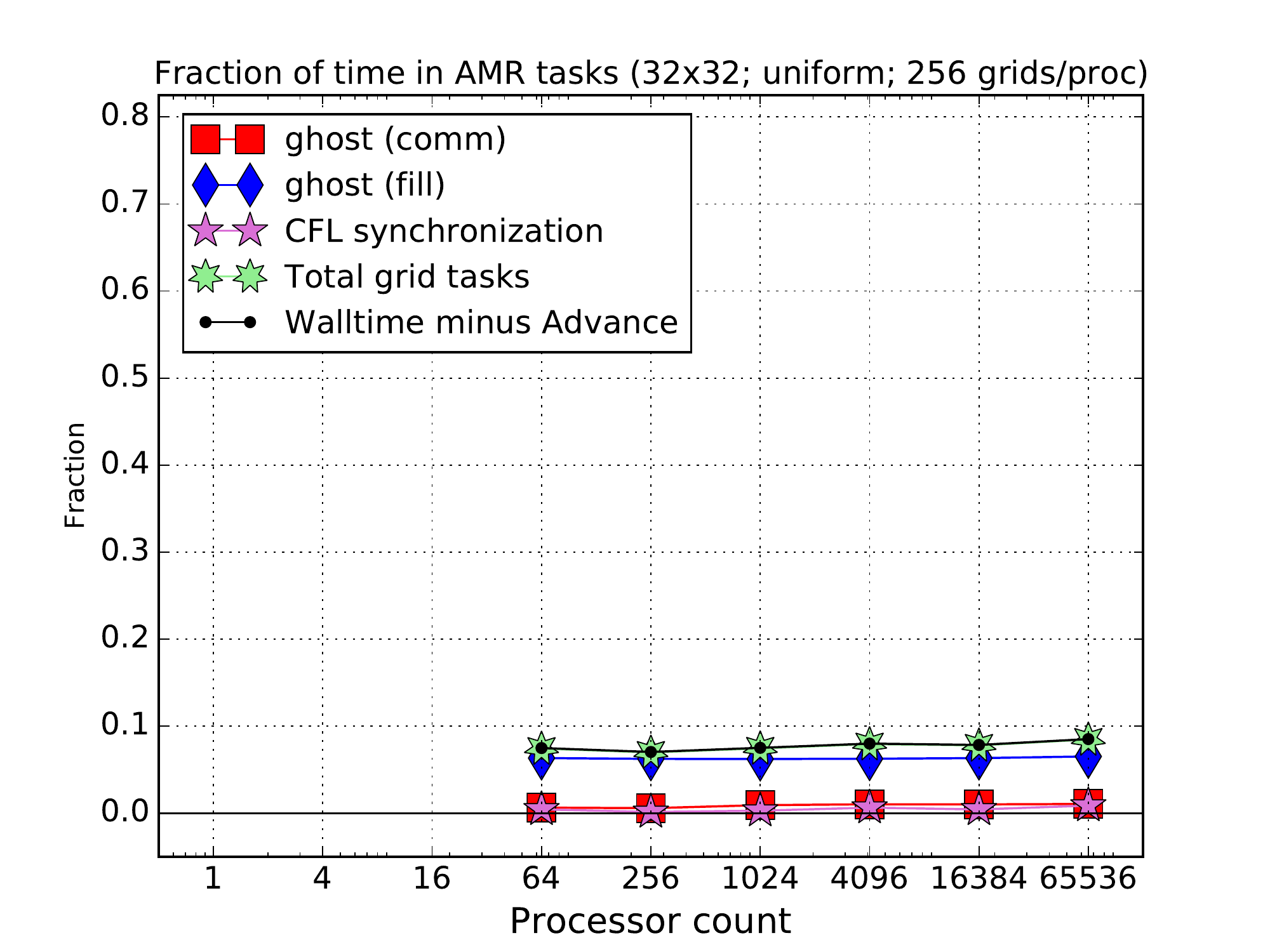}} \\
\plotbox{\includegraphics[width=0.320\textwidth, clip=true,trim=1cm 0cm 1.6cm 0cm]
  {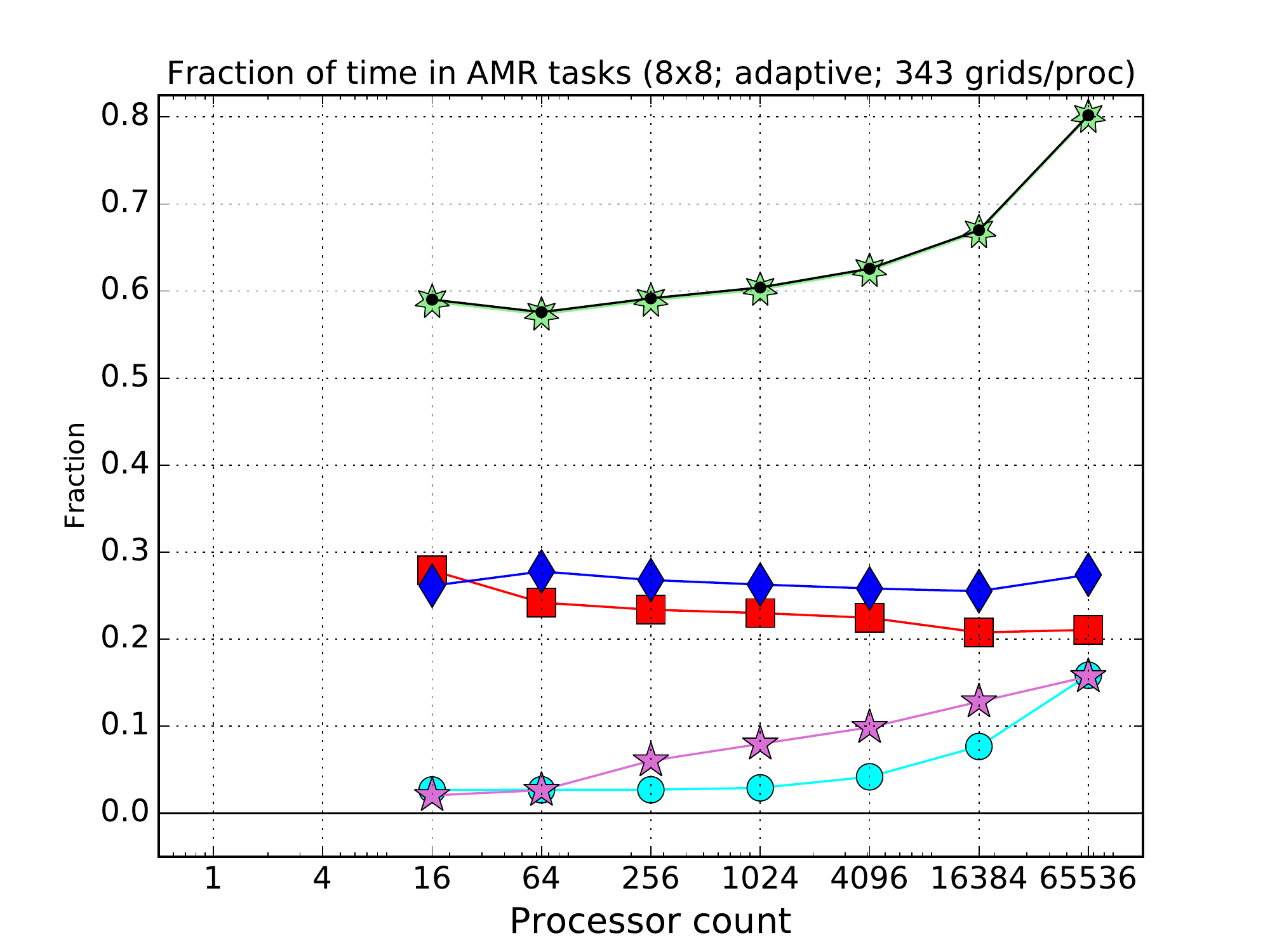}}\hfil
\plotbox{\includegraphics[width=0.320\textwidth, clip=true,trim=1cm 0cm 1.6cm 0cm]
  {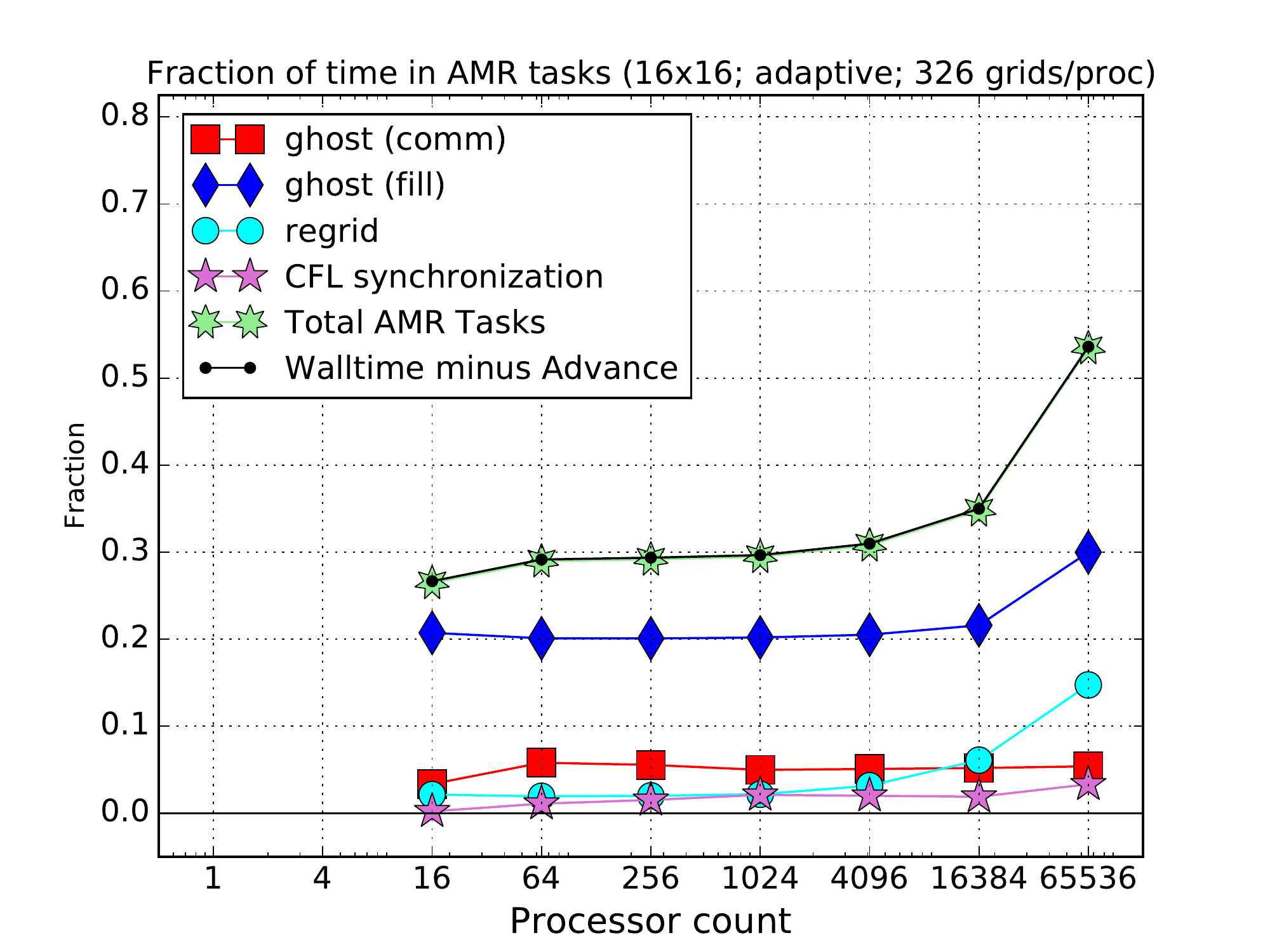}}\hfil
\plotbox{\includegraphics[width=0.320\textwidth, clip=true,trim=1cm 0cm 1.6cm 0cm]
  {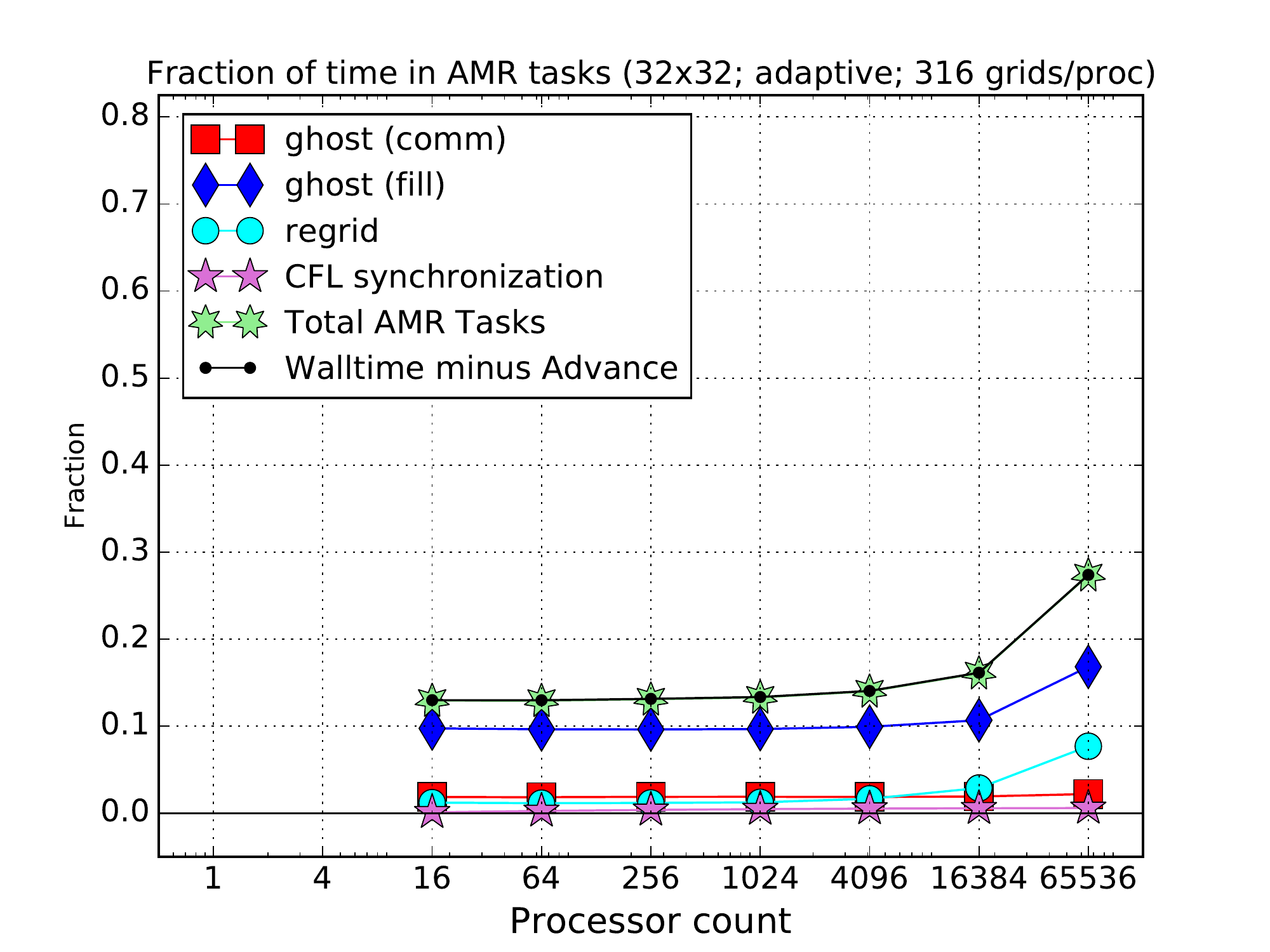}}
\end{center}
\caption{Fraction of time spent in filling ghost cells, ghost cell
communication, CFL synchronization, and (in the adaptive case)
regridding on the replicated \mblock scalar advection problem.
The top row shows fractions for the uniformly refined
simulations, and the bottom row shows fractions for the adaptive
simulations.  Columns show the $8\times8$, $16\times16$ and $32 \times
32$ fixed-size grid runs. Ghost filling tasks include both copying
between grids on the same level and averaging and interpolation at
coarse/fine interfaces.  Regridding tasks include tagging grids,
dynamic mesh regeneration, rebuilding newly coarsened or refined
grids, and partioning to correct load imbalances.  The fractions indicated by the
green stars (total time of all AMR tasks shown in the legends), and black dots (\wallclock time
minus time spent advancing the solution) are nearly identical,
confirming that we have accounted for all significant grid tasks in the indicated
legend entries.}
\label{fig:overhead}
\end{figure}

\begin{figure}
\begin{center}
\plotbox{\includegraphics[width=0.320\textwidth, clip=true,trim=1cm 0cm 1.6cm 0cm]
  {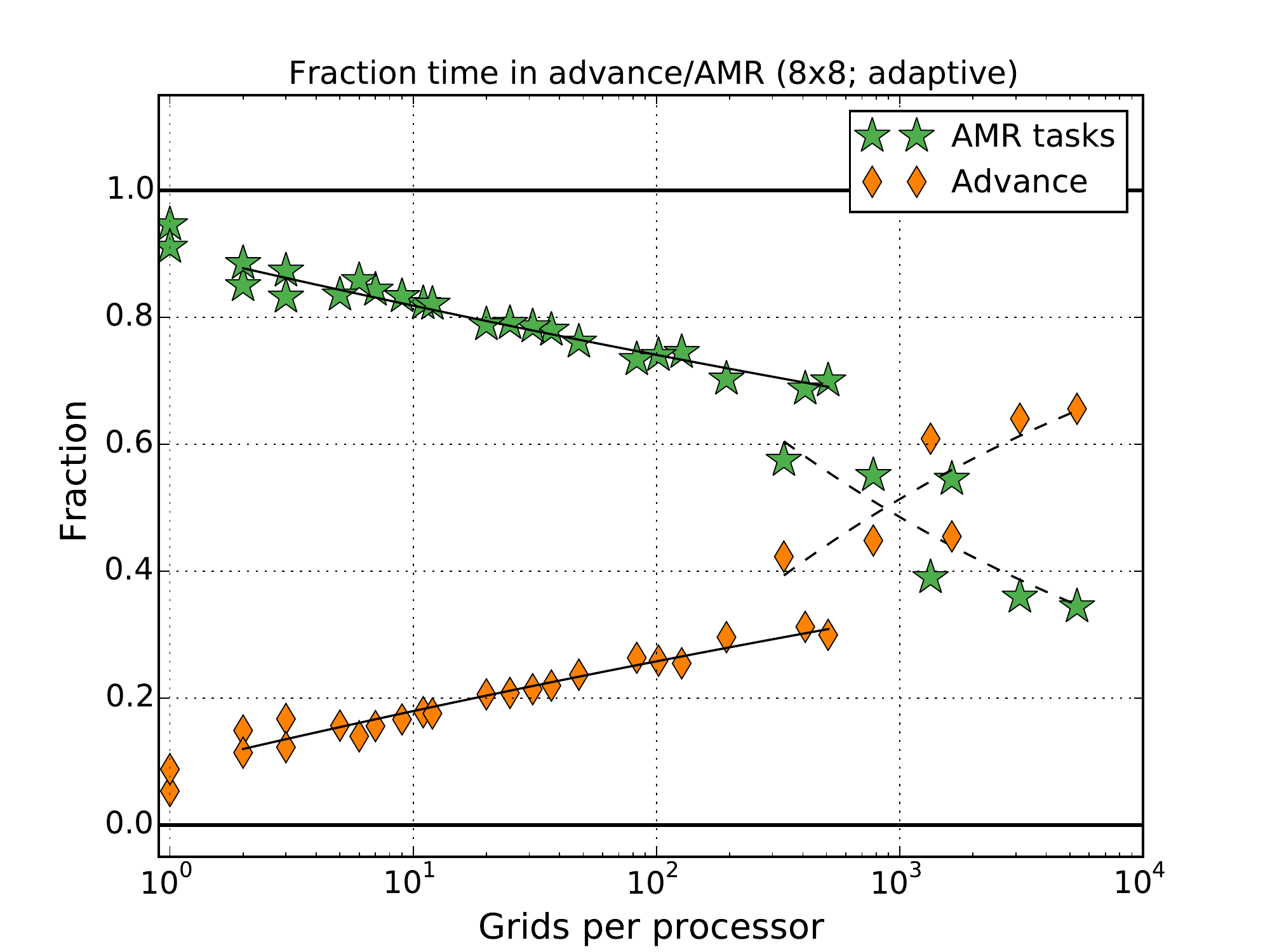}}\hfil
\plotbox{\includegraphics[width=0.320\textwidth, clip=true,trim=1cm 0cm 1.6cm 0cm]
  {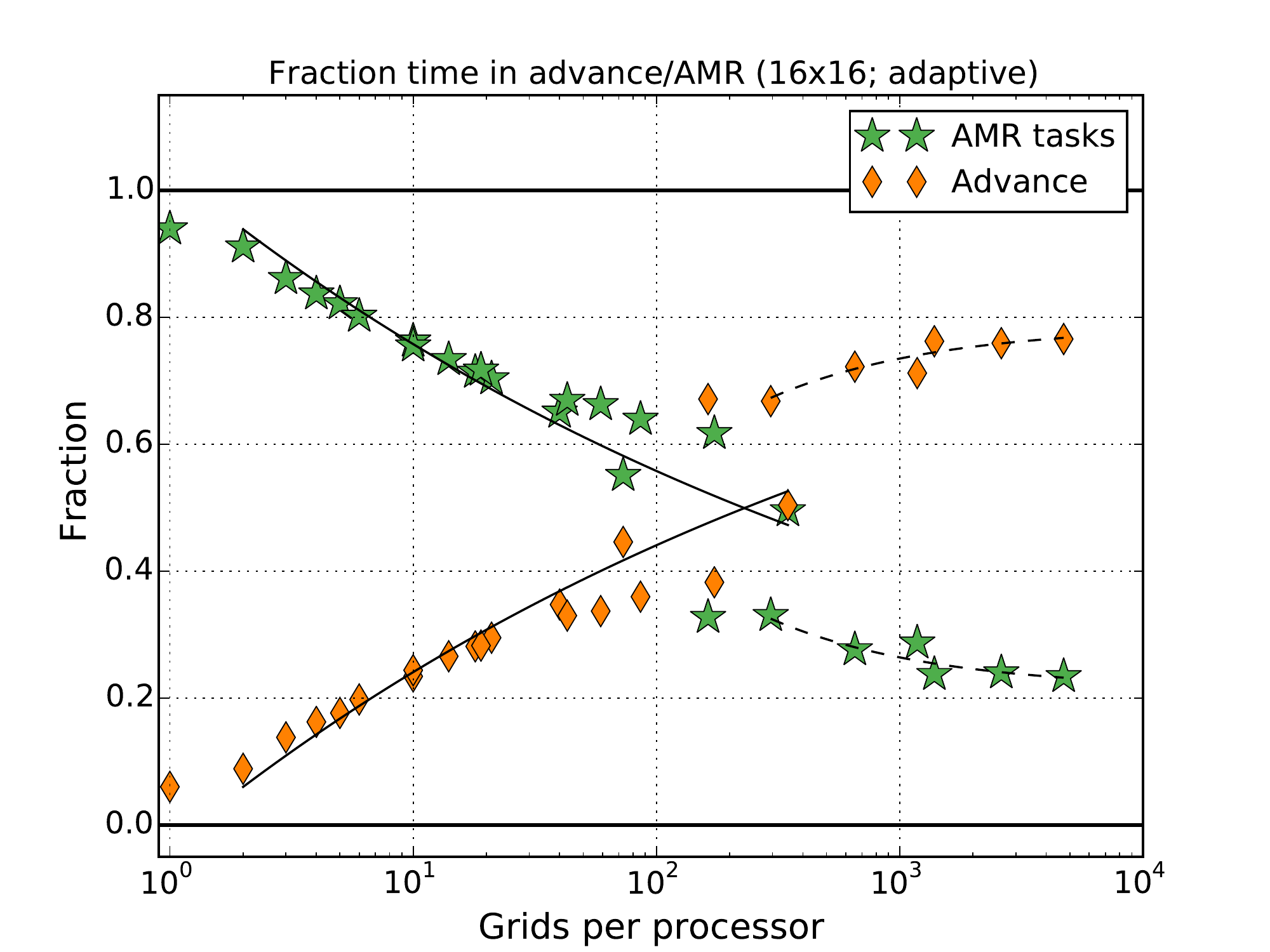}}\hfil
\plotbox{\includegraphics[width=0.320\textwidth, clip=true,trim=1cm 0cm 1.6cm 0cm]
  {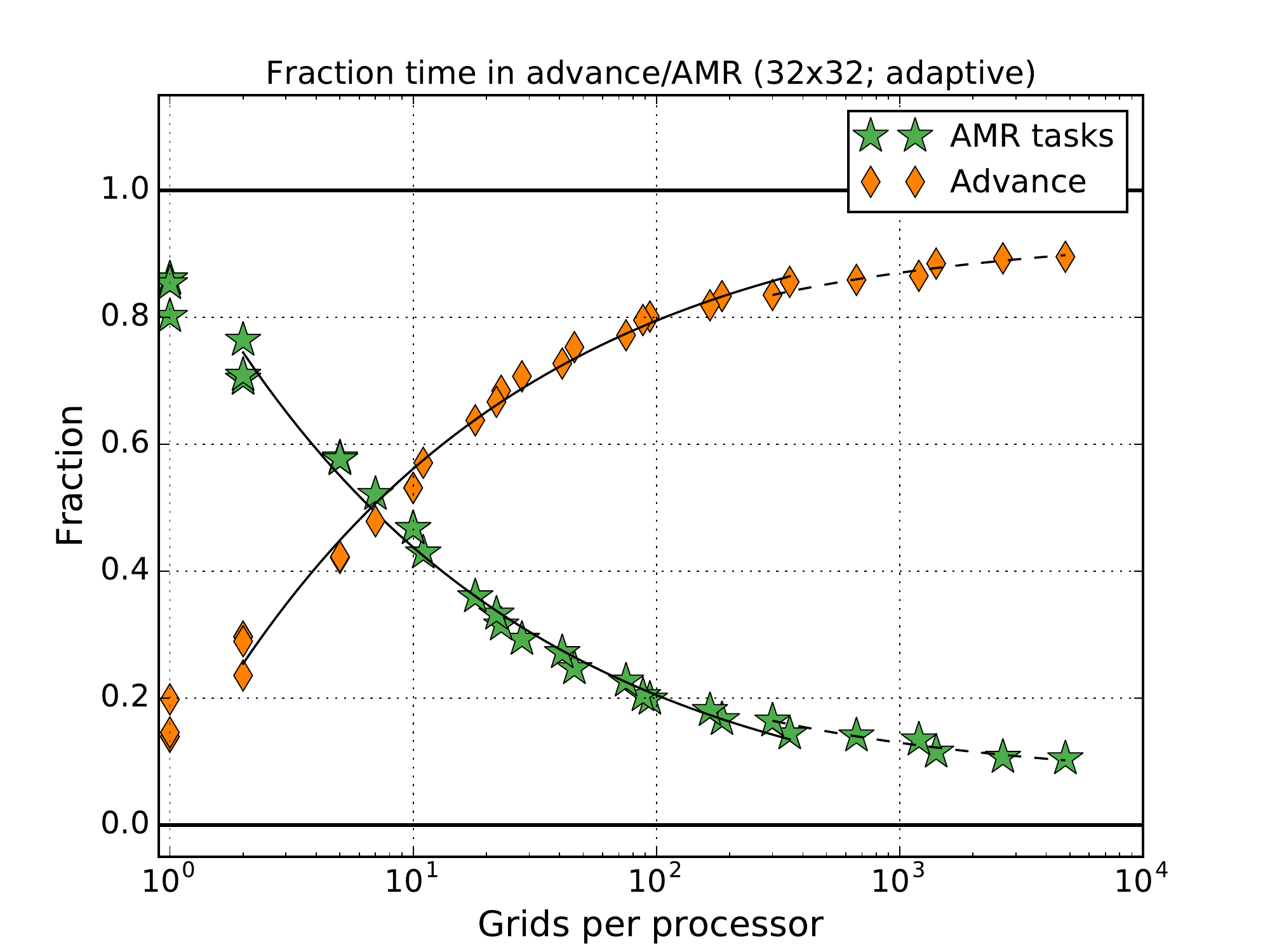}}\\
\end{center}
\caption{Plots showing AMR efficiency as a function of granularity (\gpp), for the scalar advection
problem on a single block.  The solid line shows results for process
counts of 64 or greater, whereas the dashed line shows results for
process counts of 16 or less. }
\label{fig:scale_adaptive}
\end{figure}

\paragraph{Results}
\ignore{
Tables \ref{tab:scale_gpp}, \ref{tab:scale_wallclock} and \ref{tab:scale_rates} report
numerical results from the adaptive $32\times32$ set of runs.   The figures in \Fig{weak_scaling}
show efficiency results from all six sets of runs, and \Fig{overhead} shows the cost of
doing AMR for subsets of runs from each set.  In \Fig{scale_adaptive}, we present a novel
plot showing AMR efficiency as a function of \gpp, or granularity.

In \Tab{scale_gpp}, we see that if we simultaneously increase the size of the replicated domain,
and the process count, we can maintain a fixed number of \gpp (as shown along the diagonals of
the table.  The \wallclock times from \Tab{scale_wallclock} show that when we keep the work per process
fixed, the corresponding times are very nearly constant, at least with sufficient number of \gpp.
Finally, \Tab{scale_rates} shows that for all but the highest process counts, over \%80 of the time
is spent advancing the solution.
}

The weak scaling results for the uniform runs (top row of
\Fig{weak_scaling}) show near perfect scaling with close to 100\%
efficiency on up to 64Ki processes for runs with sufficient
granularity.  Only the $8\times8$ run on 64Ki processes with 16 grids
per process dips below 80\% efficiency. As a point of
comparision, Ketcheson, Mandli et al.\ show 92\% efficiency using
\pyclaw, a massively parallel Python implementation of \clawpack, to
solve the Euler equations on up to 64Ki processes on a uniformly
refined mesh with $400 \times 400$ fixed-sized grids
\cite{ke-ma-ah-al-lu-pa-kn-em:2012}.

\begin{figure}[t]
\begin{center}
\plotbox{\includegraphics[width=0.320\textwidth, clip=true,trim=1cm 0cm 1.6cm 0cm]
  {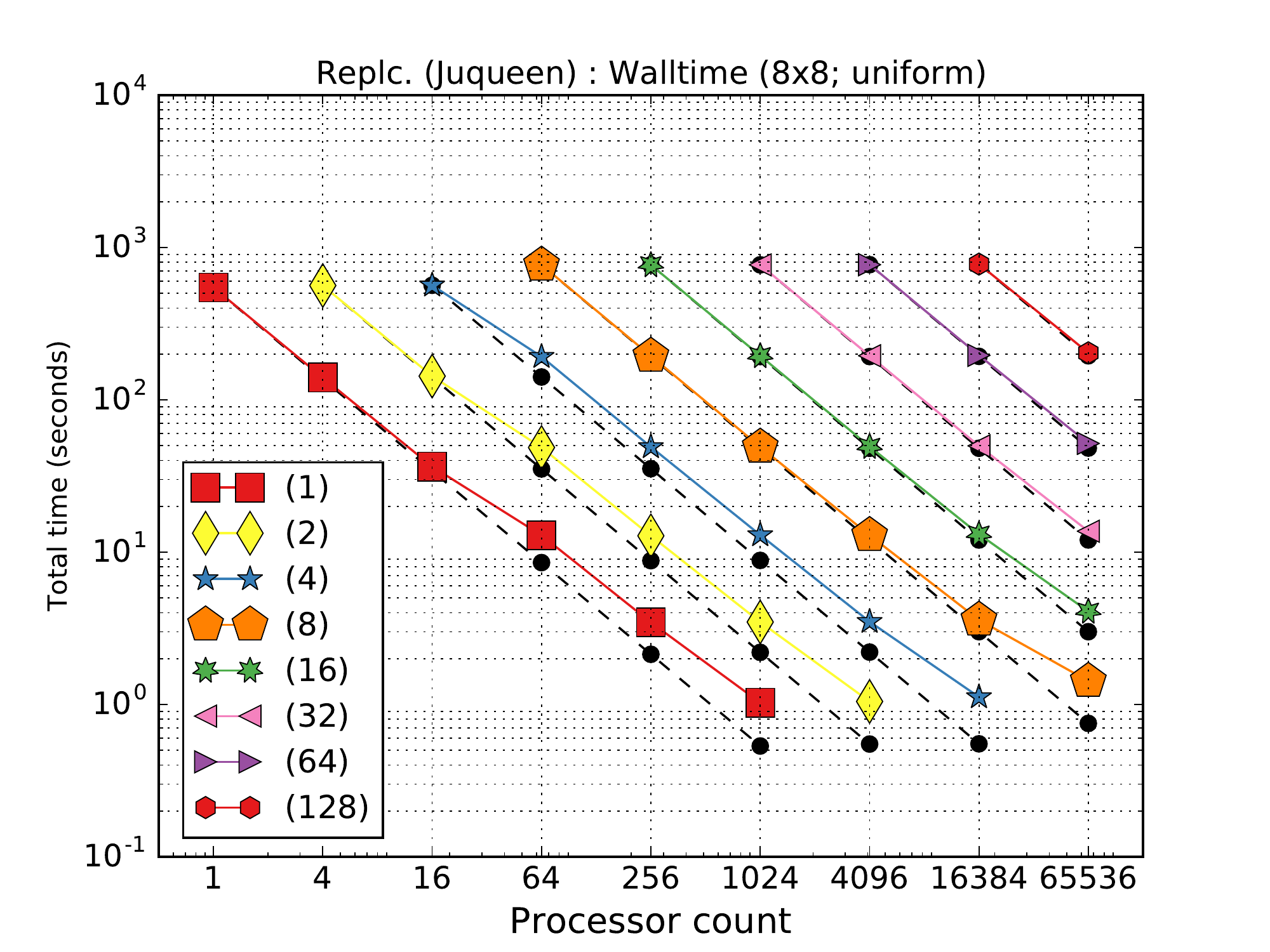}}\hfil
\plotbox{\includegraphics[width=0.320\textwidth, clip=true,trim=1cm 0cm 1.6cm 0cm]
  {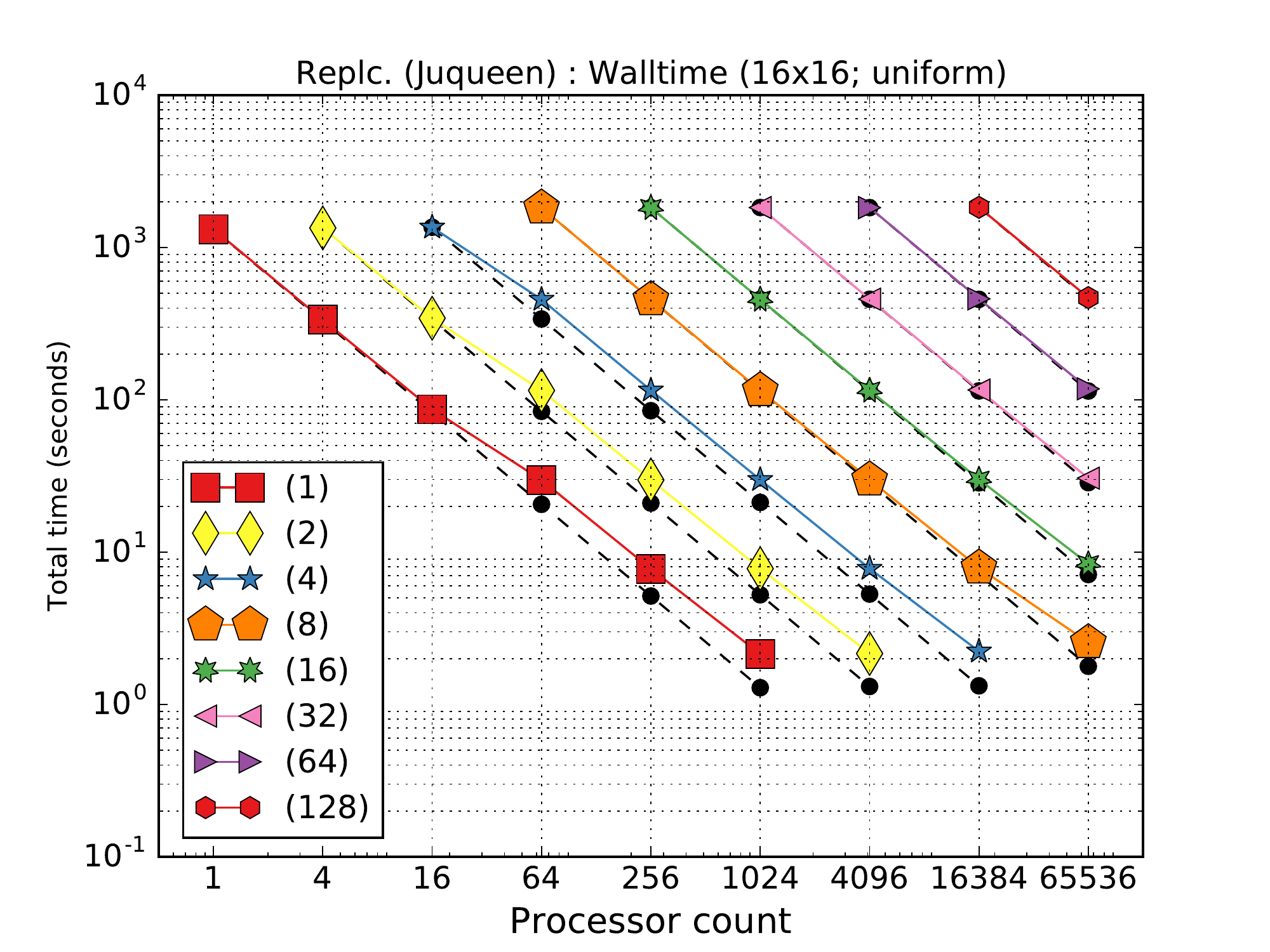}}\hfil
\plotbox{\includegraphics[width=0.320\textwidth, clip=true,trim=1cm 0cm 1.6cm 0cm]
  {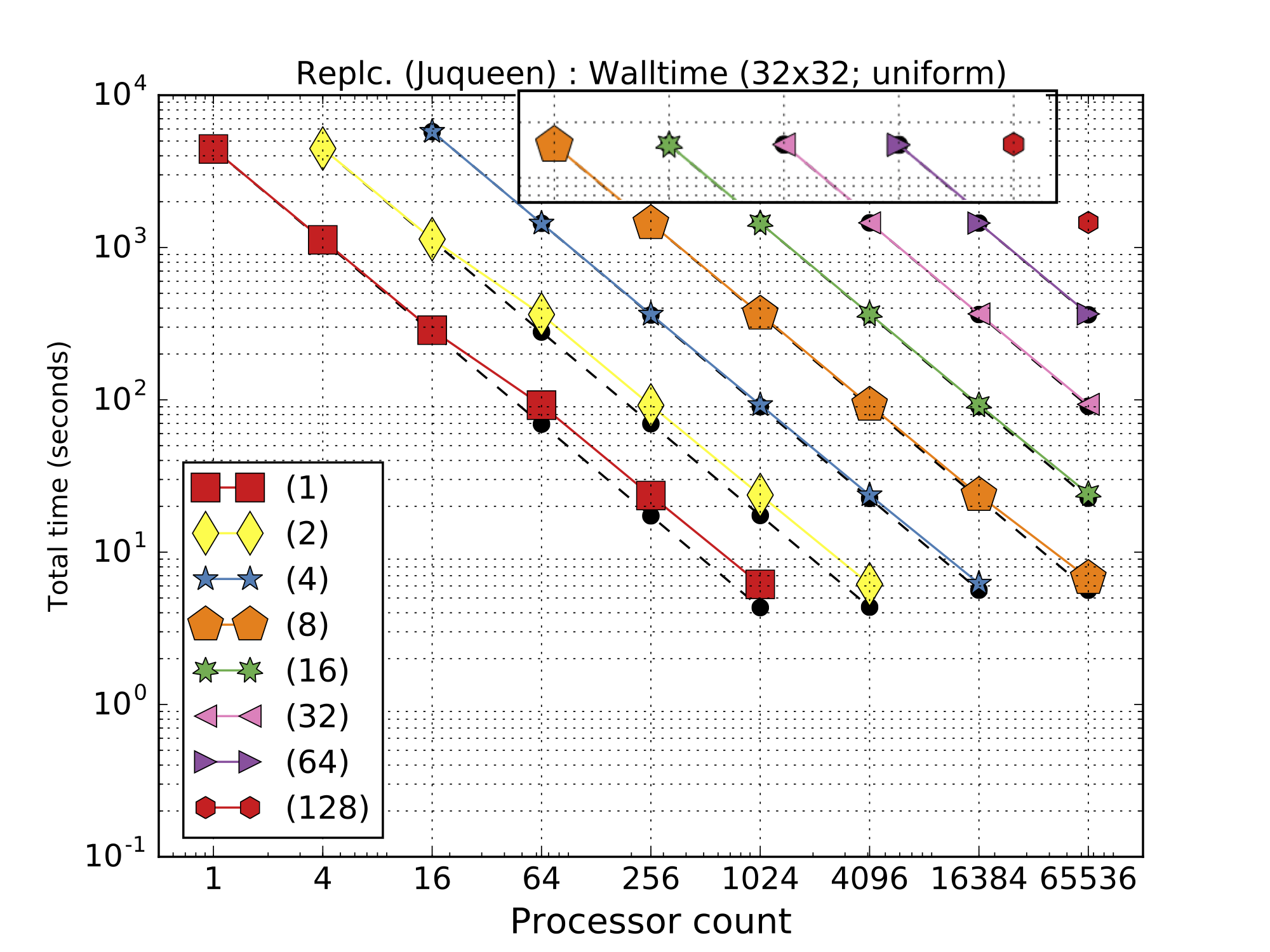}} \\
\plotbox{\includegraphics[width=0.320\textwidth, clip=true,trim=1cm 0cm 1.6cm 0cm]
  {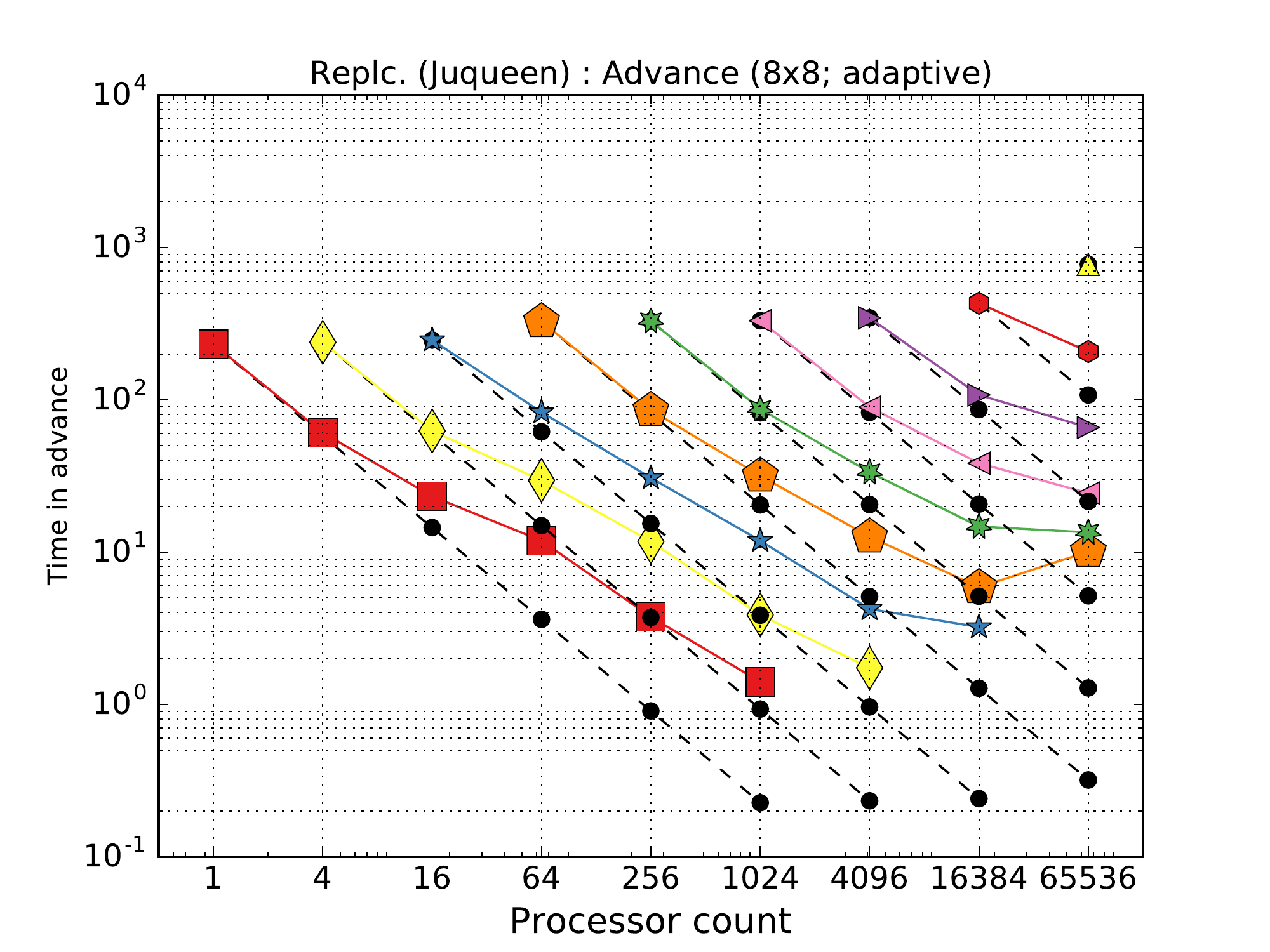}}\hfil
\plotbox{\includegraphics[width=0.320\textwidth, clip=true,trim=1cm 0cm 1.6cm 0cm]
  {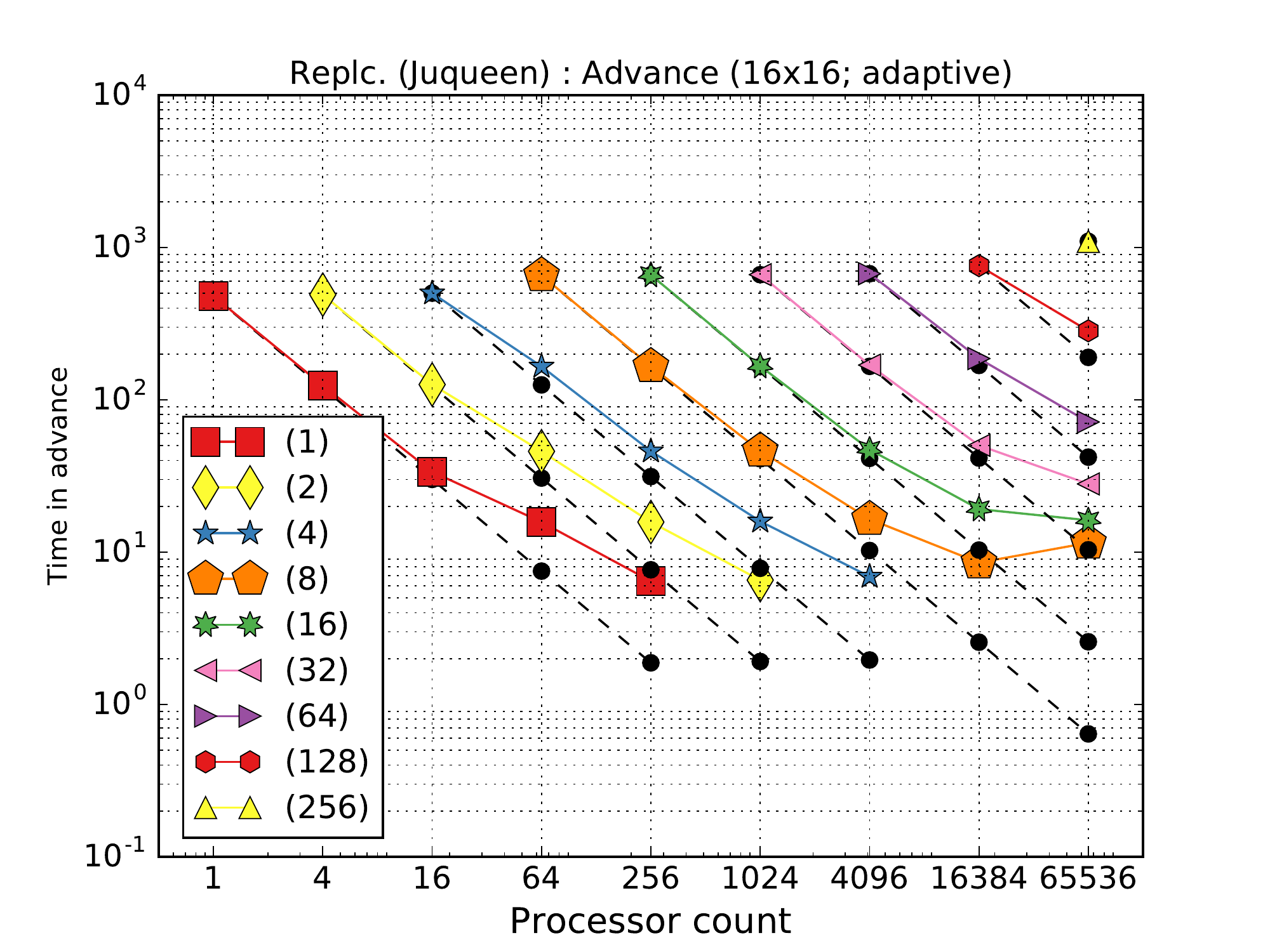}}\hfil
\plotbox{\includegraphics[width=0.320\textwidth, clip=true,trim=1cm 0cm 1.6cm 0cm]
  {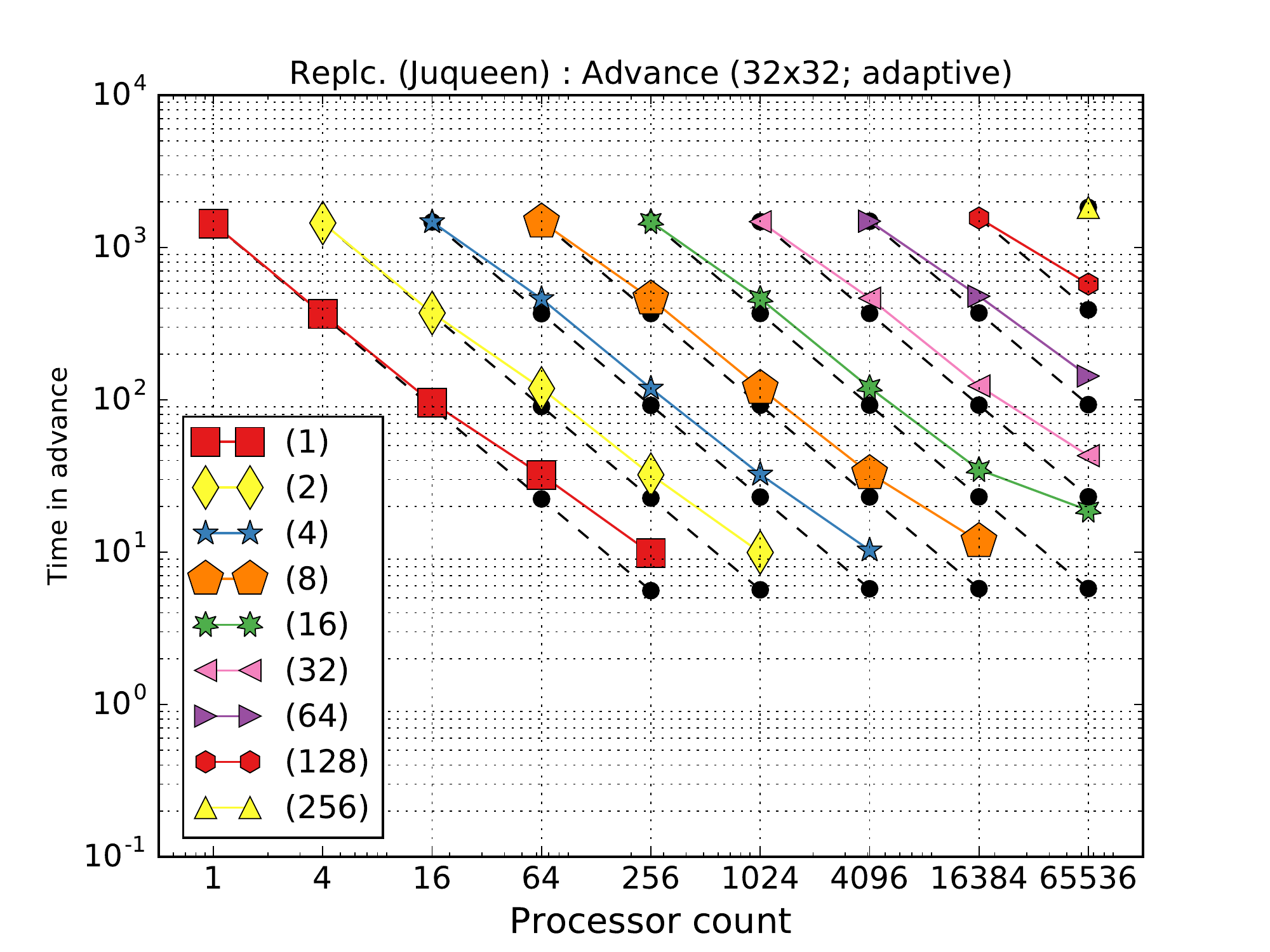}}
\end{center}
\caption{Strong scaling of replicated uniform problem (top) and the replicated
adaptive problem (bottom).  The legend labels indicate the number of blocks in each
direction in the replicated domain.  The values plotted are the \wallclock
time for each run.  The black dashed line is the ideal scaling, i.e.\ slope
$= -2$).  The timing results for the lowest granularity simulations
in the upper right plot
(boxed) could not be computed within allocated time;
we estimated them from runs done on higher process counts to complete the
picture.}
\label{fig:scale_strong}
\end{figure}

In the adaptive case, we see from \Fig{weak_scaling} that we have
close to 90\% or better efficiency on up to 4096 processes for all
grid sizes, even at very high granularity.  For the $32\times 32$
grids, the efficiency on 64Ki only dips below 80\% for simulations with
79 \gpp and below 60\% for 19 \gpp.  Simulations on the $8 \times 8$
grids dips close to or below 60\% on 64Ki processes, regardless of
granularity, but on 16Ki processes are close to 80\% as long as the
granularity exceeds roughly 300 \gpp.  The $16 \times 16$ runs show
close to 80\% or better efficiency on up to 16Ki processes for all
levels of granularity, whereas the 64Ki runs dip below 80\% efficiency,
regardless of granularity.

\Fig{scale_strong} shows strong scaling results for all six sets of
runs.  The data for these results was taken from columns of \Tab{scale_wallclock}
(for the $32\times 32$ adaptive run) and similar tables for the other five
runs.  As with the weak scaling results, the strong scaling results
are nearly perfect for the uniform case and show better efficiency
for higher resolution fixed size grids in the adaptive case.

In \Fig{overhead}, we show a breakdown of overhead costs in managing
both the uniform and adaptive simulations for the case of 256 \gpp in
the uniform case, or 300-350 \gpp for the adaptive case.  In the
uniform case, we see that essentially all overhead is in the filling
of coarse grid ghost regions via copying between grids.  Communication
costs for the uniform case are negligible and for the $32 \times 32$
grids, these communication costs remain essentially flat, even at the
highest process counts.  For the adaptive $8\times8$ runs, the
adaptive overhead is significant, consuming over 50\% of the total
time.  By contrast, the adaptive $32\times32$ runs are nearly as
efficient, in terms of overhead, as the equivalent uniform runs.
Select numerical data from the runs shown in the previous plots are
shown in \Tab{scale_wallclock} and \Tab{scale_rates}.
In the results presented here we see a drop in efficiency when going from 16Ki
to 64Ki processes.
Especially in weak scaling experiments, this is not ideal behavior.
We suspect that the overlap of computation and communication
we implement is not sufficient at this scale to produce optimal scaling.
The precise cause will be investigated using more elaborate profiling.

\ignore{
To achieve the \gpp counts reported in \Fig{overhead} for the adaptive
runs, each unit block was distributed to 16 processes so that the
\pforest mesh was identical for each fixed grid resolution. For the
uniform runs, each unit block was distributed to 64 processes.  At
levels of granularity reported, we are using less than 10\% of the 0.5
\si{\giga\byte} of memory allocated per process.
}

\begin{table}
\footnotesize
\caption{\Wallclock times (seconds) for the \adaptrun{32} set of runs
on the replicated, \mblock
scalar advection problem. The leftmost column shows
number of \mpi ranks used for the run, and the top row shows the
dimensions of the \mblock brick domain used, i.e.\ $1\times1$, $2
\times 2$, $4 \times 4$ and so on. Times along the diagonals remains relatively
fixed, showing good weak scaling results.}
\label{tab:scale_wallclock}
\begin{center}
\begin{tabular}{r*{9}{S[table-format = 4.1]}}
\toprule
Ranks  &  1 & 2 & 4 &  8 & 16 & 32 & 64 &  128 & 256\\
\midrule
       1  &     1430.5  &  \dash   &  \dash   &  \dash   &  \dash  &  \dash   & \dash      & \dash   & \dash \\
       4  &      365.6  &  1450.   &  \dash   &  \dash   &  \dash  &  \dash   & \dash      & \dash  &  \dash \\
      16  &       95.8 &  371.    &  1470.   &  \dash   &  \dash  &  \dash   & \dash      & \dash  &  \dash \\
      64  &       32.2 &  119.    &   460.   &  1470.   & \dash    &   \dash  & \dash     & \dash   &  \dash\\
     256  &       9.85 &  32.2    &   119.   &   461.   &  1480.   &   \dash  & \dash     & \dash   & \dash\\
    1024  &    \dash   &   9.95   &    32.4  &   119.   &   462.   &  1480.   & \dash     & \dash   & \dash\\
    4096  &    \dash   &  \dash   &    10.3  &    32.9  &   120.   &   465.   &   1490.   & \dash   & \dash\\
   16384  &    \dash   &  \dash   &   \dash  &    11.8  &    34.8  &   123.   &    479.   &  1560.  & \dash\\
   65536  &    \dash   &  \dash   &   \dash  &   \dash  &    18.8  &    43.0  &    143.   &   575.  & 1830.\\
\bottomrule
\end{tabular}
\end{center}
\end{table}

\begin{table}
\caption{Details of $32\times 32$ run for entries along the top diagonal in \Tab{scale_gpp} and
\Tab{scale_wallclock}.  The number of \gpp for the runs below is fixed at 5059.  The rate in
the rightmost column is computed as total number of grid cells advanced per time per process.
All results were run using 16 ranks per  JUQUEEN node. }
\label{tab:scale_rates}
\begin{advtable}{$32 \times 32$; Replicated problem}
\rowadv{mx=32,p=1,    pp=1,  w=1430.5, asteps=809584,gpp=5059,a=1292.07,gf=112.474,gc=0.00533591}\\
\rowadv{mx=32,p=4,    pp=2,  w=1448.79,asteps=809584,gpp=5059,a=1299.09,gf=120.585,gc=2.75172}\\
\rowadv{mx=32,p=16,   pp=4,  w=1471.67,asteps=809584,gpp=5059,a=1316.13,gf=123.577,gc=4.89805}\\
\rowadv{mx=32,p=64,   pp=8,  w=1474.89,asteps=809584,gpp=5059,a=1317.3, gf=123.741,gc=6.08375}\\
\rowadv{mx=32,p=256,  pp=16, w=1476.68,asteps=809584,gpp=5059,a=1317.46,gf=124.563,gc=5.93793}\\
\rowadv{mx=32,p=1024, pp=32, w=1477.79,asteps=809584,gpp=5059,a=1316.25,gf=126.114,gc=6.23245}\\
\rowadv{mx=32,p=4096, pp=64, w=1487.81,asteps=809584,gpp=5059,a=1317.17,gf=134.249,gc=5.21293}\\
\rowadv{mx=32,p=16384,pp=128,w=1561.0, asteps=809584,gpp=5059,a=1318.78,gf=204.592,gc=4.40963}\\
\rowadv{mx=32,p=65536,pp=256,w=1831.75,asteps=809584,gpp=5059,a=1329.58,gf=449.497,gc=5.41203}\\
\end{advtable}
\end{table}

Finally, we report results from three sets of single-block adaptive
runs in which we vary the process count and the maximum level of
refinement.  This problem is more typical of how one might allocate
computational resources, in that one would use additional
resources to increase resolution, not run more copies of the same
problem.  We initialize the single block domain using the same initial
conditions and time stepping parameters as in the replicated problem.
The range of refinement levels and process counts we chose are shown
in \Tab{single_gpp}.

The results of these single-block runs allow us to obtain interesting
insight into the balance between AMR efficiency, granularity and
grid size.  In \Fig{scale_adaptive} we show three plots,
corresponding to grid sizes $8\times8$, $16\times16$ and $32
\times 32$.  In these plots, we show the relationship between AMR
efficiency, measured as fraction of time spent in AMR tasks and
advancing the solution, and granularity. Each plot shows a clear
crossover point indicating the minimum granularity needed to
ensure that at least 50\% of computational time is spent advancing the
solution.  What the plots clearly show is that this crossover point
moves left, towards higher granularity, as the fixed-grid size
increases.  For $8\times8$ grids, one needs nearly 1000 \gpp before
one is spending more time advancing the solution than managing the
grids.  But for the $32 \times 32$ grids, allocating roughly 100 \gpp
is enough to ensure that over 80\% of the total time is spent
advancing the solution.  Plotting efficiency data from other
applications and machine architectures should lead to plots with the
same general characteristics as those shown here, and so this novel
approach to illustrating the dependence of AMR efficiency on
granularity and grid size should provide a useful guide in how
to allocate computational resources for adaptive simulations using
\forestclaw.

\begin{table}
\caption{\Wallclock times for the $32\times 32$ adaptive runs on a single
block (quadtree) of the replicated problem.
The minimum level for each run was fixed at $\lmin = 4$
and the maximum levels are listed across the top row.  The
number of \gpp for the topmost run in
each column is listed in parenthesis in the header for that column.
The number of \gpp for remaining runs in the column are roughly one
fourth of the previous entry.}
\label{tab:single_gpp}
\begin{center}
\begin{tabular}{r*{6}{S[table-format=6.2]}}
\toprule
Ranks     & 7 (4798)& 8 (2656) & 9 (1412)& 10 (186)& 11 (94)& 12 (28)  \\
\midrule
       1  & 1380.  & \dash    & \dash   &  \dash   &   \dash  &  \dash   \\
       4  & 358.   &  1570.   & \dash   &  \dash   &   \dash  &  \dash   \\
      16  & 93.9   &   413.   &  1740.  &  \dash   &   \dash  &  \dash   \\
      64  & 31.5   &   134.   &   554.  &  \dash   &   \dash  &  \dash   \\
     256  &  9.54  &    37.7  &   149.  &    606.  &   \dash  &  \dash   \\
    1024  &  \dash &    12.9  &    44.5 &    168.  &    644.  &  \dash   \\
    4096  &  \dash &     5.94 &    17.4 &     55.3 &    188.  &    450.  \\
\bottomrule
\end{tabular}
\end{center}
\end{table}

\ignore{
\begin{figure}
\begin{center}
\plotbox{\includegraphics[width=0.450\textwidth, clip=true,trim=1cm 0cm 1.6cm 0cm]
  {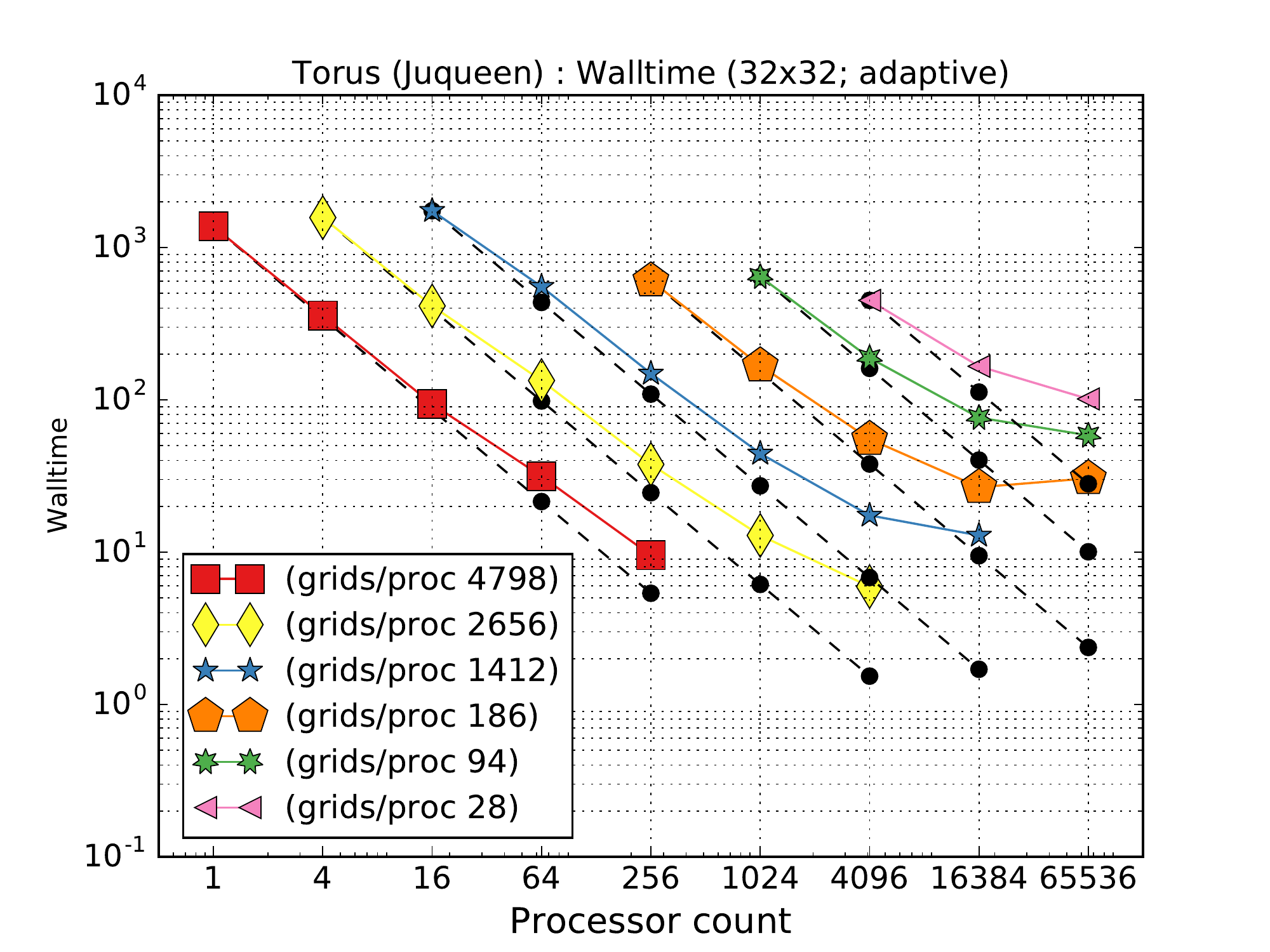}}\hfil
\end{center}
\caption{Strong scaling for \adaptrun{32} simulation on a single block.
See \Tab{single_gpp} for actual timing values for this case and
explanation of the \gpp listed in legend.}
\label{fig:strong_single}
\end{figure}
}

\subsection{Advection on a sphere}
\begin{figure}
\begin{center}
\plotbox{\includegraphics[width=0.3\textwidth,clip=true,trim=5cm 2.5cm  4.5cm 2cm]{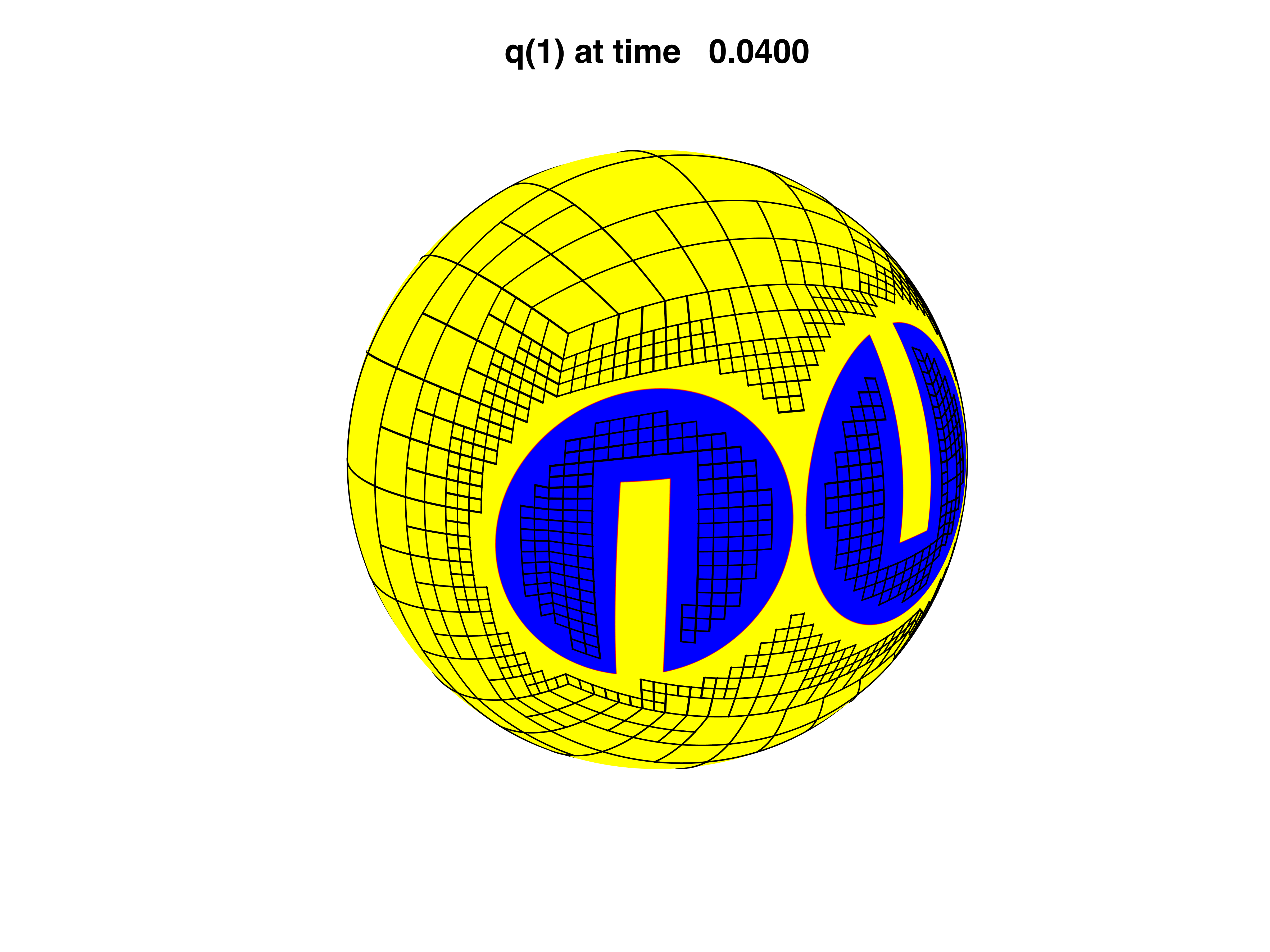}}
\plotbox{\includegraphics[width=0.3\textwidth,clip=true,trim=5cm 2.5cm  4.5cm 2cm]{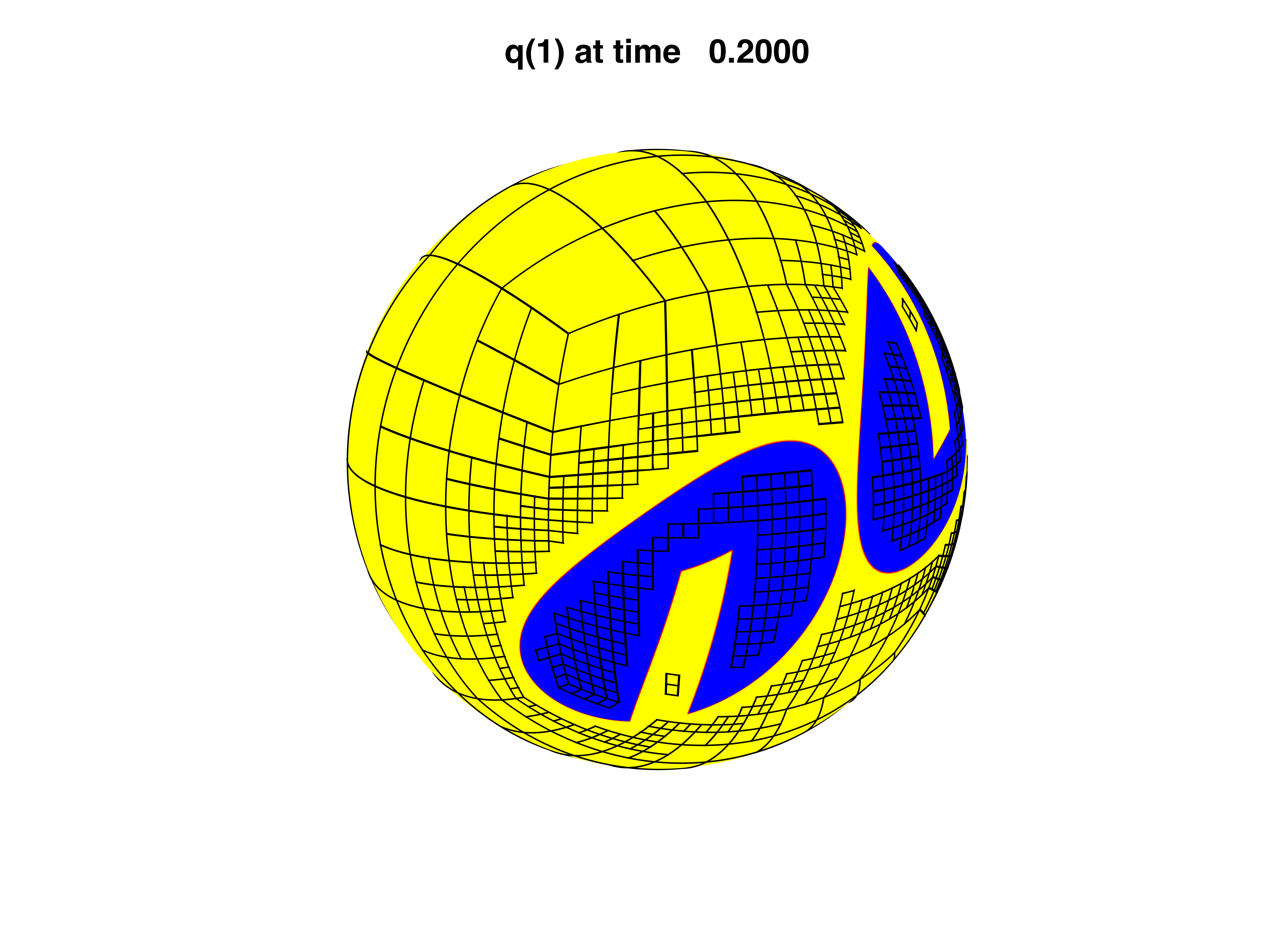}}
\plotbox{\includegraphics[width=0.3\textwidth,clip=true,trim=5cm 2.5cm  4.5cm 2cm]{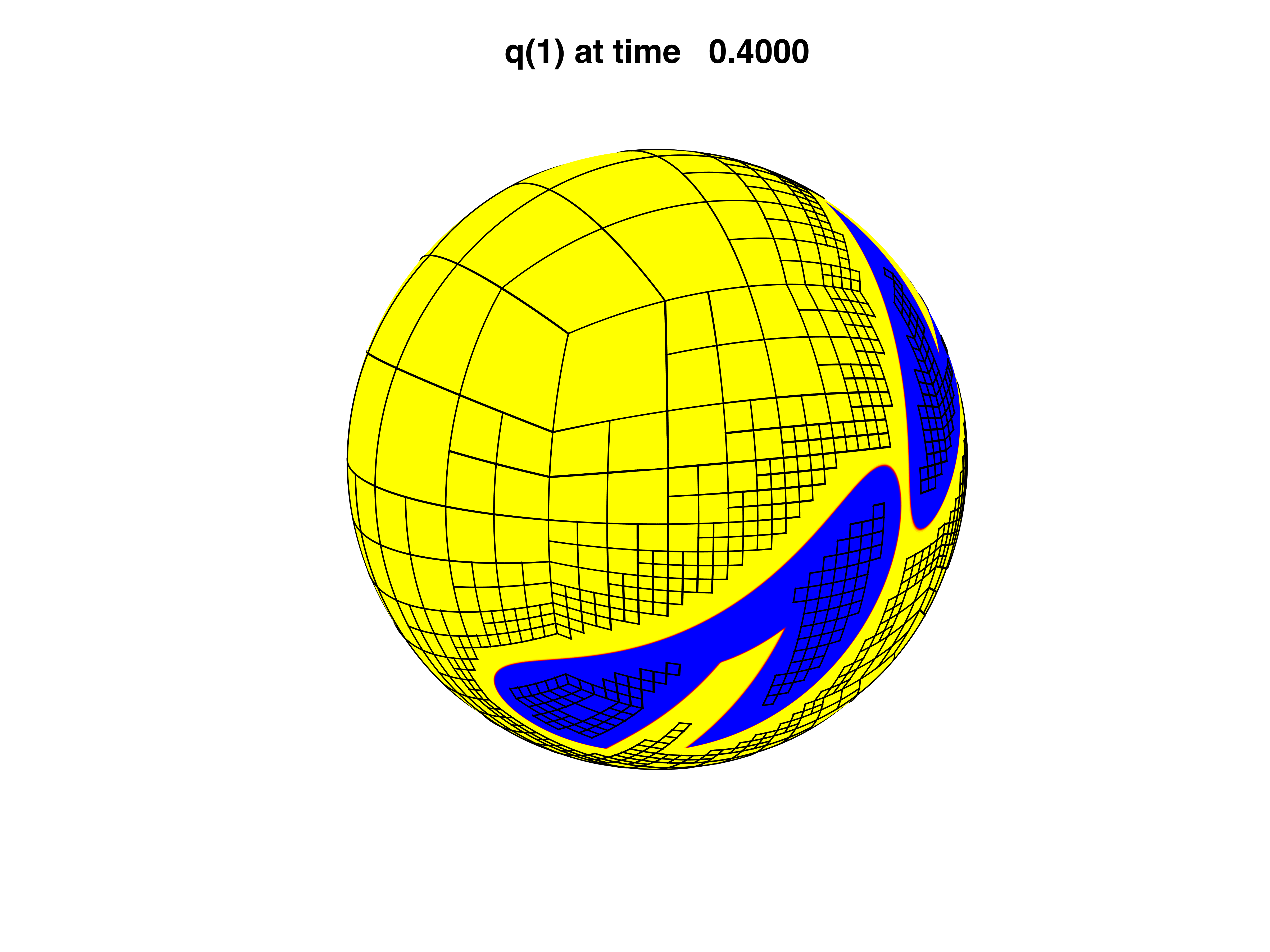}}
\end{center}
\caption{Three views of the slotted disk problem illustrating flow on a cubed sphere grid
for times $t=0.05$ (left), $t = 0.25$ (middle) and $t = 0.4$ (right).
Levels 2--6 are shown (patch borders for level 6 are not shown).}
\label{fig:sphere_map}
\end{figure}

The cubed-sphere grid, shown in \Fig{sphere_map}, has become a popular
alternative to spherical coordinate grids for solving PDEs on the
sphere.  This mapping, and many variants, typically have the properties
that mesh cells are relatively uniform and they do not suffer from the extreme
aspect ratios seen when using spherical coordinates.
The cubed-sphere is also an example of a \mblock domain in which
grid indices at block boundaries do not generally align and so is a
good test case for the \mblock indexing described in \Sect{mblock}.

\ignore{
numerical discretization we use to solve the advection equations on the sphere are
based on the \clawpack solvers available in \forestclaw and described in
\cite{be-ca-he-le:2009, ca-he:2009}.
}

\paragraph{Scalar advection on a manifold}
\label{sec:parallel_sphere}
To  demonstrate the cubed-sphere functionality in \forestclaw, we use the
tracer transport problem proposed by Lauritzen, Skamarock et
al. \cite{la-sk-pr-ta:2012, la-ul-ja-bo-ca-co-en-do-du:2014}.
The test is intended to assess how well tracer transport schemes can
stretch a slotted-disk (see \Fig{sphere_map}) and return it to its initial
shape.
Our goal is to use this example to demonstrate our \mblock mapping
capabilities and to assess how metric terms impact the performance of \forestclaw.

Two slotted disks are initialized
and the
velocity field is given in spherical coordinates $(\rho,\theta)$ as
\begin{equation} 
\Psi(\rho,\theta) = \kappa \sin(\rho-2\pi t/T)^2 \cos(\theta)^2
\cos(\pi t/T) - 2\pi\sin(\theta)/T_f
\end{equation} where $\kappa=2$ and $T_f=5$.  We advect the initial
slotted-disk tracer distribution in this velocity field from the
initial time to final time $T = 0.5$.  For this problem, we test the
fixed size $32\times 32$, and, as with the single block example in the
previous problem, we vary the refinement levels.  A fixed time step on
level $\ell$ is set to $\dt_\ell = (\num{2.5e-3})/2^{\ell - \lmin}$.
We run 5 series of runs, from a uniformly refined run at level 2 (16
$32 \times 32$ patches per block) to adaptive runs which start at
minimum $\lmin=2$ and maximum levels varying from 3 to 6.  The
numerical discretization we use for the sphere is based on \clawpack
and is described in \cite{ca-he:2009,ca-he-le:2008}.

\begin{table}
\caption{\Wallclock times and \gpp (in parenthesis) for the slotted-disk problem on the cubed-sphere.
The header for each column is the maximum level of refinement $\lmax$ on
each block of the six blocks forming the cubed-sphere.  The minimum level
for all runs is $\lmin=2$. }
\label{tab:sphere_wallclock_gpp}
\begin{center}
\begin{tabular}{
  S[table-format=2,table-column-width=1cm]      
  *5{S[table-column-width=2.0cm,table-number-alignment=center]}}
\toprule
Ranks   & 2           & 3             & 4           & 5             & 6            \\
\midrule
   1  &  {274.  (96)} & {1360. (242)} &  \dash       & \dash        & \dash        \\
   4  &  { 72.2 (24)} & { 366. (60)}   &  \dash       & \dash        & \dash        \\
  16  &  { 19.2 (6)}  & { 111. (15)}   & {508. (38)} & \dash        & \dash        \\
  64  & \dash         & \dash         &  {251. (9)}  & {1200. (25)} & \dash        \\
 256  & \dash         & \dash         &  \dash       & { 489. (6)}   & {1700. (16)} \\
\bottomrule
\end{tabular}
\end{center}
\end{table}

\paragraph{Parallel setup}
We increase the refinement level and MPI process counts simultaneously.
For each of the 5 series of runs, we start with
an initial process count and then increase that count by a factor of four as
long as the number of \gpp exceeds a reasonable threshold (10 or so).
In \Tab{sphere_wallclock_gpp} shows the number of processes for each of 11 runs
that we carried out.  All runs were done on JUQUEEN.

\begin{figure}
\begin{center}
\plotbox{\includegraphics[width=0.33\textwidth, clip=true,trim=1cm 0cm 1.6cm 0cm]
  {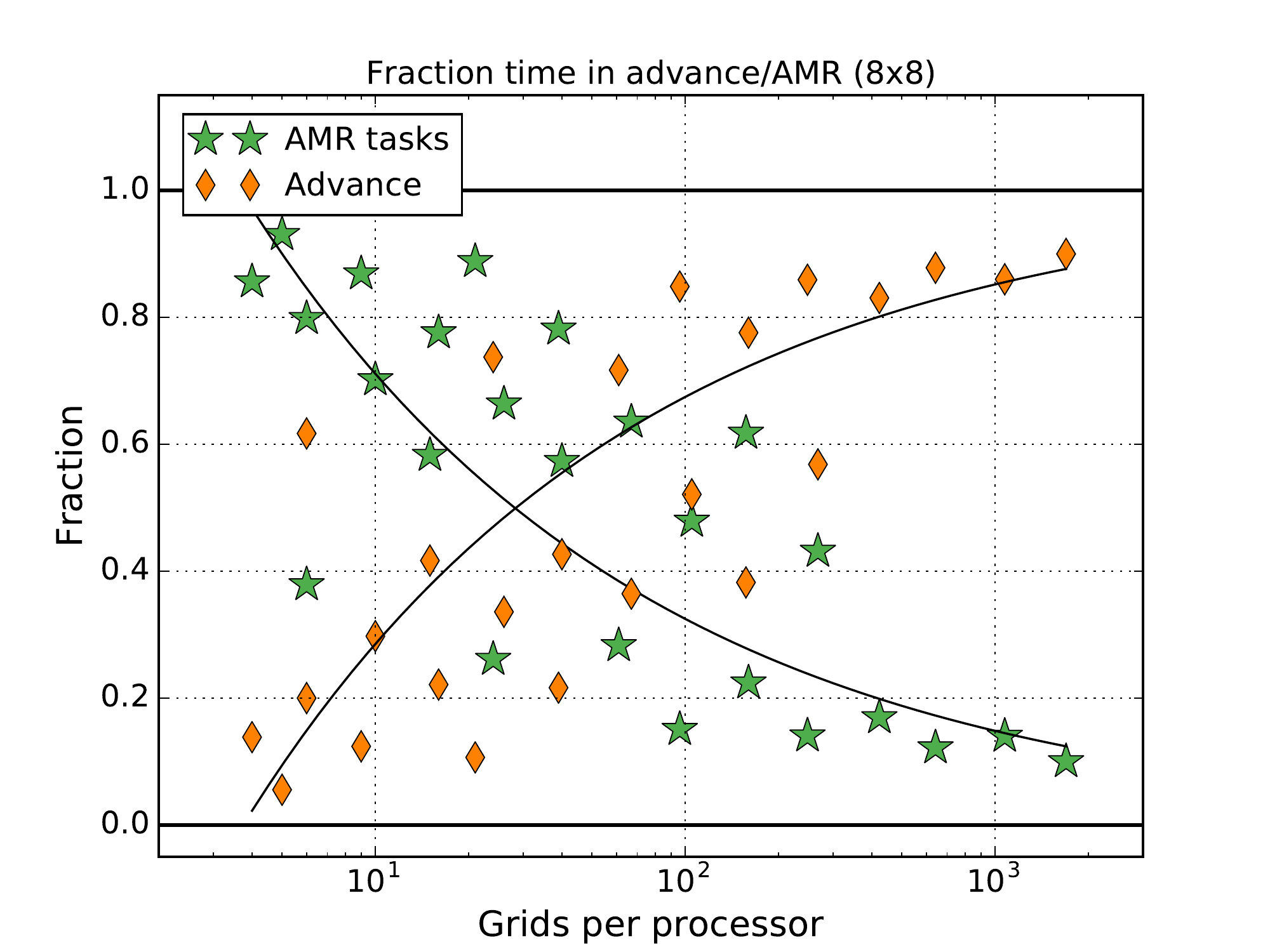}}\hfil
\plotbox{\includegraphics[width=0.33\textwidth, clip=true,trim=1cm 0cm 1.6cm 0cm]
  {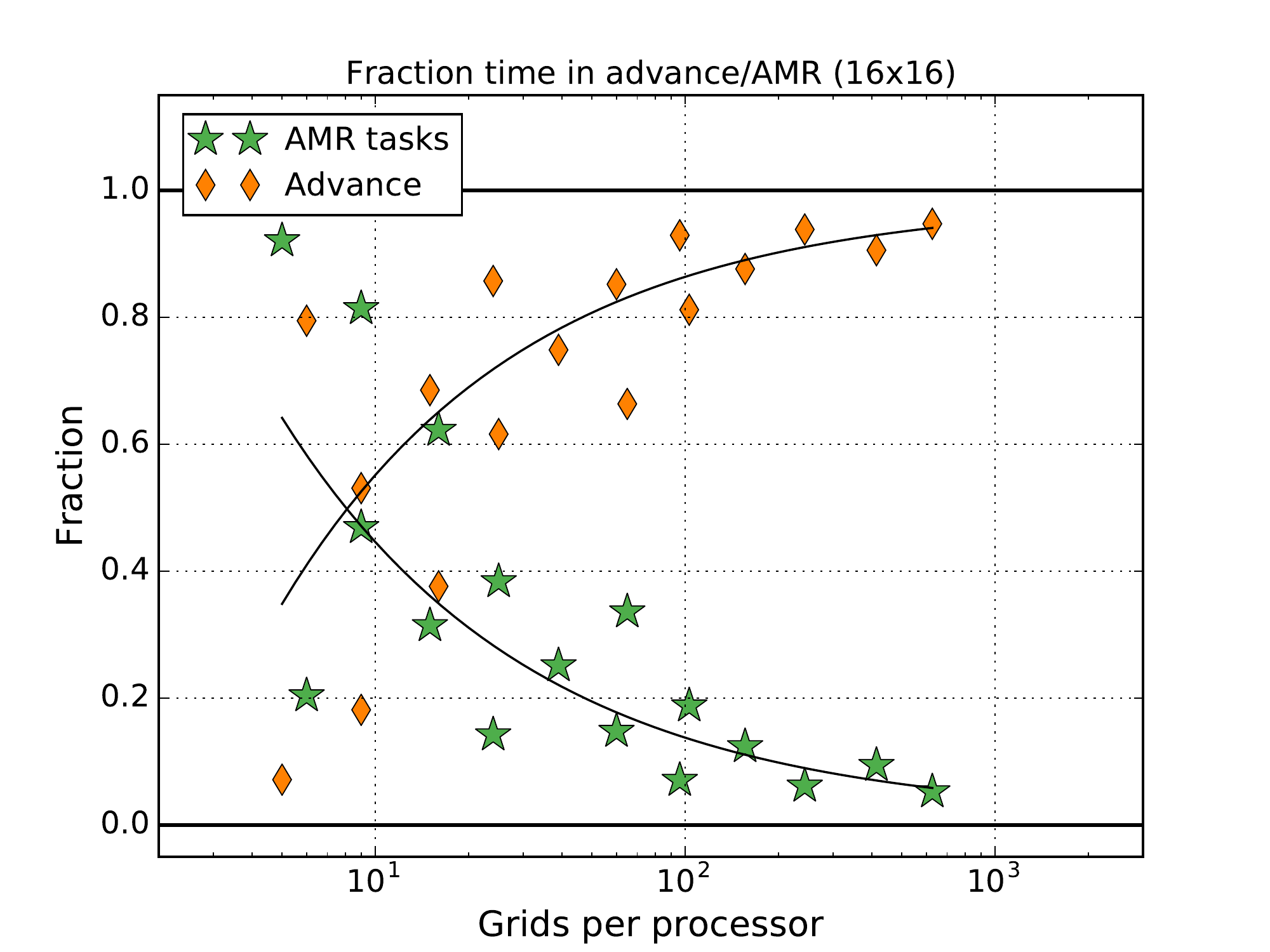}}\hfil
\plotbox{\includegraphics[width=0.33\textwidth, clip=true,trim=1cm 0cm 1.6cm 0cm]
  {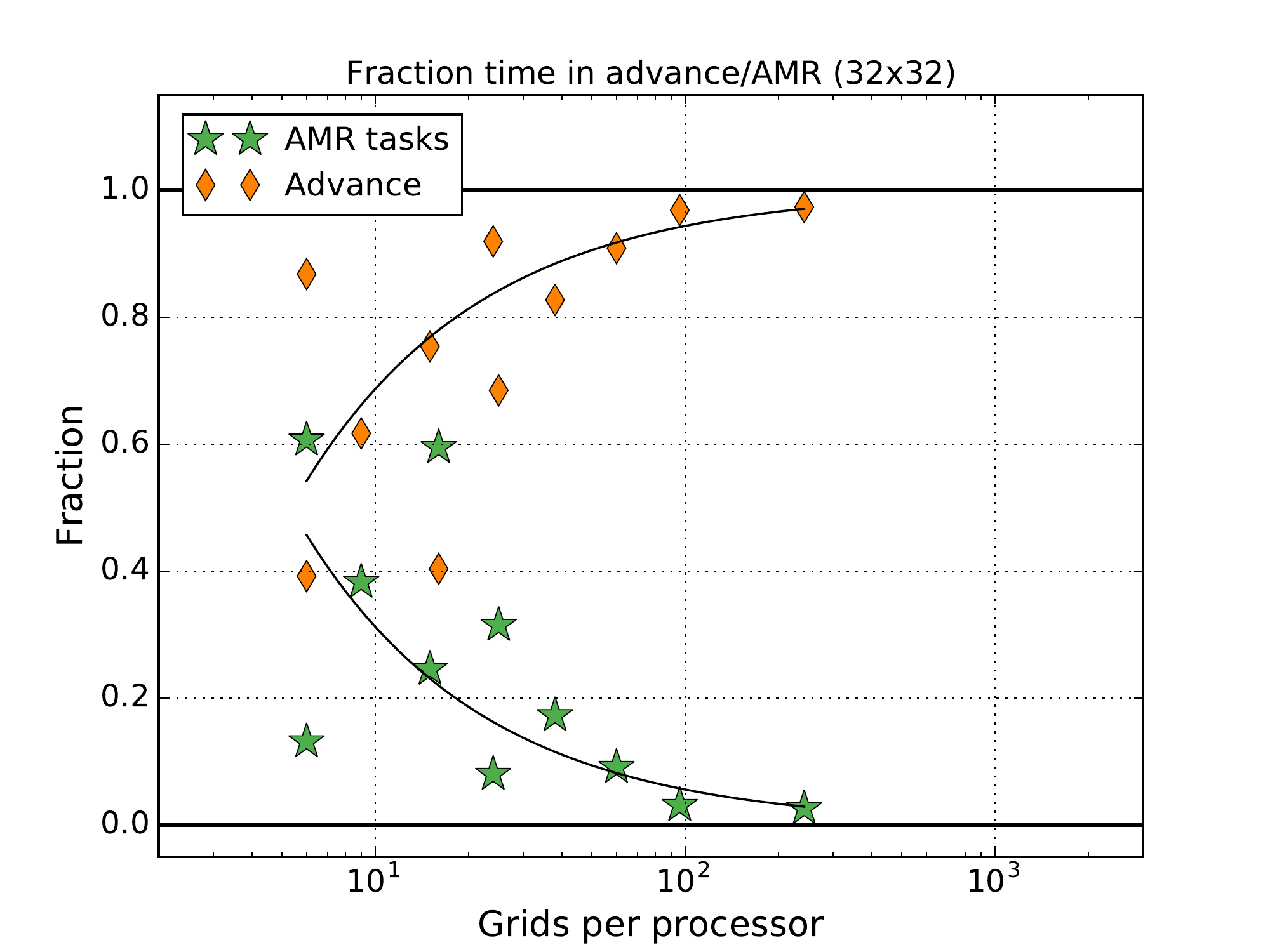}}\\
\end{center}
\caption{Adaptive efficiency  of the scalar advection problem on a cubed-sphere grid.}
\label{fig:scale_adaptive_sphere}
\end{figure}

\paragraph{Results}
\Fig{scale_adaptive_sphere} shows the adaptive efficiency as a function of the number of \gpp.
Adaptive efficiency generally
increases with more \gpp and reaches over 90\% with at least 100 \gpp.  Even though
the runs on the sphere domain have far fewer \gpp than on the replicated flat domain, the results in
\Fig{scale_adaptive_sphere} have the same characteristics as those seen in \Fig{scale_adaptive}.
In \Tab{sphere_details}, we show the cell processing rate for this problem.
This rate is about an order
of magnitude smaller than for the flat domain from the previous example.
This is due to the extra computational effort required to compute metric terms.

\begin{table}
\caption{Timing details from slotted-disk problem. Entries are taken along a diagonal
from \Tab{sphere_wallclock_gpp}. The rate in
the rightmost column is computed as total number of grid cells advanced per time per process.
All results were run using 16 ranks per  JUQUEEN node. }
\label{tab:sphere_details}
\begin{spheretable}{$32\times32$; Slotted-disk}
\rowsphere{mx=32,lmax=2, p=1,    pp=1, w=274.0, asteps=19200, gpp=96, a=265., gf=2.26, gc=0.00179}\\
\rowsphere{mx=32,lmax=3, p=4,    pp=1, w=366.,  asteps=24237, gpp=60, a=333., gf=6.65, gc=16.4}\\
\rowsphere{mx=32,lmax=4, p=16,   pp=1, w=508.,  asteps=30910, gpp=38, a=421., gf=10.8, gc=53.4}\\
\rowsphere{mx=32,lmax=5, p=64,   pp=1, w=1200., asteps=40996, gpp=25, a=819., gf=20.5, gc=268.}\\
\rowsphere{mx=32,lmax=6, p=256,  pp=1, w=1700., asteps=51210, gpp=16, a=688., gf=20.7, gc=862.}\\
\end{spheretable}
\end{table}

\ignore{
\begin{table}
\caption{Processing rates for cubed-sphere problem.  Rates are computed as
number of cells advanced per time, per process.}
\label{tab:sphere_rate}
\begin{center}
\begin{tabular}{
  S[table-format=4,table-column-width=1.cm]      
  *5{S[scientific-notation=true,                 
    table-number-alignment=right,
    table-text-alignment=center,
    round-precision=1,
    round-mode=places,
    fixed-exponent=4,
    table-figures-exponent=1,
    table-figures-decimal=1,
    input-symbols=\dash,
    table-column-width= 1.8cm]}
}
\toprule
{Ranks} &
\multicolumn{1}{c}{2}       &
\multicolumn{1}{c}{3}       &
\multicolumn{1}{c}{4}       &
\multicolumn{1}{c}{5}       &
\multicolumn{1}{c}{6}   \\
\midrule
    1  & 7.18e4  &  7.28e+4  &    \dash &    \dash &    \dash \\
    4  & 6.81e+04  &  6.78e+4  &    \dash &    \dash &    \dash \\
   16  & 6.40e+04  &  5.61e+4  &  6.23e+4 &    \dash &    \dash \\
   64  &  \dash    &    \dash  &  3.15e+4 &  3.51e+4 &    \dash \\
  256  &  \dash    &   \dash   &   \dash  &  2.15e+4 &  3.08e+4 \\
\bottomrule
\end{tabular}
\end{center}
\end{table}
}



\section{Conclusions and future work}

In this article, we describe our work in developing a forest-of-quadtrees
approach to block-structured adaptive mesh refinement using the algorithms
first put forth by Berger and Oliger in 1985.
We demonstrate that these ideas can be implemented using the scalable mesh
management library \pforest and hyperbolic solvers from the \clawpack library.

A particular focus of this article is on the implementation of the parallel
ghost filling algorithm and inter-grid indexing necessary to formulate unsplit
finite volume schemes on a \mblock adaptive hierarchy of non-overlapping grids.
We add a smooth refinement procedure to preserve moving features of the
solution on the finest levels.

We demonstrate our approach numerically by solving a scalar advection problem
on 1 to 64Ki MPI processes with good parallel scalability.
We also solve an advection problem on the cubed sphere that is composed of
multiple blocks.
In addition, we looked carefully at how AMR efficiency depends on the
granularity (\gpp) and fix-grid sizes.  We find that $32 \times 32$ grids strike
a favorable compromise between the flexible refinement offered by small grid
sizes and high arithmetic intensity offered by larger sizes.

While \forestclaw supports \mrate (locally adaptive) time stepping, we only
demonstrate global time stepping here.
A follow-up article will describe the \mrate time stepping capabilities
currently available in \forestclaw, report on detailed verification studies,
comparisons with other adaptive codes, and results from solving more complex
hyperbolic systems including the shallow-water and Euler equations.


\section{Acknowledgements}

Donna Calhoun would like to acknowledge the Isaac Newton Institute program
``Multiscale Numerics for the Ocean and Atmosphere'' for its support of much of
this work during the fall of 2012 and the National Science Foundation (NSF
DMS-1419108).
Carsten Burstedde is supported by the Hausdorff Center for Mathematics (HCM) at
Bonn University
funded by the German Research Foundation (DFG).

The authors gratefully acknowledge the Gau\ss{} Centre for Supercomputing (GCS)
for providing computing time through the John von Neumann Institute for
Computing (NIC) on the GCS share of the supercomputer JUQUEEN at J\"ulich
Supercomputing Centre (JSC).  GCS is the alliance of the three national
supercomputing centres HLRS (Universit\"at Stuttgart), JSC (Forschungszentrum
J\"ulich), and LRZ (Bayerische Akademie der Wissenschaften), funded by the
German Federal Ministry of Education and Research (BMBF) and the German State
Ministries for Research of Baden-W\"urttemberg (MWK), Bayern (StMWFK) and
Nordrhein-Westfalen (MIWF).


\end{document}